\pdfoutput=1
\documentclass[%
  reprint,
  amsmath,amssymb,amsthm,
  aps,
  prb,
  nobibnotes
]{revtex4-2}

\usepackage{graphicx}
\usepackage{microtype}
\usepackage{dcolumn}
\usepackage{bm}
\usepackage[usenames,dvipsnames]{xcolor}
\usepackage[framemethod=TikZ]{mdframed}
\usepackage{algorithm}      % for algorithm counter (we won't float it)
\usepackage{algpseudocode}  % for pseudocode
\usepackage{float}          % for [H] spec used in supplement algorithms
\usepackage{orcidlink}
\usepackage{import}
\usepackage{multirow}
\usepackage{braket}
\usepackage[caption=false]{subfig}
\usepackage{booktabs}
\usepackage{ytableau}
\usepackage{listings}
\usepackage{hyperref}
\usepackage{url}
\usepackage[
  left=0.72in,
  right=0.72in,
  top=0.72in,
  bottom=0.72in
]{geometry}

\newcommand\myshade{85}
\colorlet{mylinkcolor}{violet}
\colorlet{mycitecolor}{YellowOrange}
\colorlet{myurlcolor}{Aquamarine}
\hypersetup{
  linkcolor  = mylinkcolor!\myshade!black,
  citecolor  = mycitecolor!\myshade!black,
  urlcolor   = myurlcolor!\myshade!black,
  colorlinks = true
}

\newmdenv[
  linewidth=0.4pt,
  roundcorner=6pt,
  linecolor=black!20,
  innerleftmargin=8pt,innerrightmargin=8pt,
  innertopmargin=6pt,innerbottommargin=6pt,
  skipabove=6pt,skipbelow=6pt,
  backgroundcolor=white
]{algobox}
\newcommand{\algcaption}[1]{%
  \refstepcounter{algorithm}%
  \noindent\textbf{Algorithm~\thealgorithm: #1}\par\vspace{4pt}%
}

\newtheorem{definition}{Definition}

\newcommand{\adag}{a^{\dagger}}

\definecolor{codegreen}{rgb}{0,0.6,0}
\definecolor{codegray}{rgb}{0.5,0.5,0.5}
\definecolor{codepurple}{rgb}{0.58,0,0.82}
\definecolor{backcolour}{rgb}{0.95,0.95,0.92}
\lstdefinestyle{mystyle}{
  backgroundcolor=\color{backcolour},
  commentstyle=\color{codegreen},
  keywordstyle=\color{magenta},
  numberstyle=\tiny\color{codegray},
  stringstyle=\color{codepurple},
  basicstyle=\ttfamily\footnotesize,
  breakatwhitespace=false,
  breaklines=true,
  captionpos=b,
  keepspaces=true,
  numbers=left,
  numbersep=5pt,
  showspaces=false,
  showstringspaces=false,
  showtabs=false,
  tabsize=2
}
\newcommand{\beginsupplement}{%
  \onecolumngrid
  \clearpage
  \setcounter{equation}{0}%
  \setcounter{figure}{0}%
  \setcounter{table}{0}%
  \setcounter{algorithm}{0}%
  \renewcommand{\theequation}{S\arabic{equation}}%
  \renewcommand{\thefigure}{S\arabic{figure}}%
  \renewcommand{\thetable}{S\arabic{table}}%
  \renewcommand{\thealgorithm}{S\arabic{algorithm}}%
}

\begin{document}

\preprint{APS/123-QED}

\title{Universal initial state preparation for first quantized quantum simulations}

\author{Jack S. Baker$^{1,2}$ \orcidlink{0000-0001-6635-1397}}
\email{jack.baker@lge.com}
\author{Gaurav Saxena$^{1,2}$ \orcidlink{0000-0001-6551-1782}}
\author{Thi Ha Kyaw$^{1,2}$ \orcidlink{0000-0002-3557-2709}}
\email{thiha@lge.com}
\affiliation{$^1$LG Electronics Toronto AI Lab, Toronto, Ontario M5V 1M3, Canada}
\affiliation{$^2$Department of Chemistry, University of Toronto, Toronto, Ontario M5S 3H6, Canada}

\date{\today}

\begin{abstract}
Preparing symmetry-adapted initial states is a principal bottleneck in first-quantized quantum simulation. We present a universal approach that efficiently maps any polynomial-size superposition of occupation-number configurations to the first-quantized representation on a digital quantum computer. The method exploits the Jordan--Schwinger Lie algebra homomorphism, which identifies number-conserving second-quantized operators with their first-quantized action and induces an equivariant bijection between Fock occupations and $\mathfrak{su}(d)$ weight states within the Schur--Weyl decomposition. Operationally, we deterministically prepare an superposition of target Schur labels and apply the inverse quantum Schur transform. For $L$ configurations of $N$ particles over $d$ modes prepared to accuracy $\epsilon$, the most efficient variant of the algorithm runs with non-Clifford gate complexity $\mathrm{poly}(L, N, \log d, \log \epsilon^{-1})$. The protocol applies universally to fermions, bosons, and Green's paraparticles in arbitrary single-particle bases. Resource estimates establish practicality within leading first-quantized pipelines, with statistics-aware specializations promising further reductions.
\end{abstract}

\maketitle

Quantum computers offer a transformative new means of simulating complex many-body systems~\cite{feynman2018simulating,Georgescu2014}, with implications spanning chemistry~\cite{McArdle2020}, materials science~\cite{Alexeev2024}, and fundamental physics~\cite{Bauer2023}. Recent advances suggest that such simulations will attain genuine computational advantages over their classical counterparts once fault-tolerant quantum computers (FTQCs) are available~\cite{Daley2022,Lee2023,lanes2025frameworkquantumadvantage}. Indeed, FTQCs~\cite{shor1996fault,gottesman1998theory,Roffe_2019} capable of running powerful primitives such as the quantum Fourier transform (QFT) and subroutines that rely on it, such as quantum phase estimation (QPE)~\cite{kitaev1995quantummeasurementsabelianstabilizer}, are expected to be prerequisites for realizing this potential. Importantly, the success of many quantum-simulation techniques hinge on access to high-quality initial states and efficient protocols for their preparation. In particular, the success probability of QPE is proportional to the square of the overlap between the initial state and the target eigenstate, so even modest improvements in initial-state quality can substantially reduce the runtime or the number of required circuit repetitions. Furthermore, early fault-tolerant implementations of quantum simulation depend even more critically on this step~\cite{Katabarwa2024}, making robust state-preparation algorithms essential for achieving practical quantum advantage sooner.

State-preparation strategies differ markedly between the second- and first-quantization formalisms for many-body physics. In second quantization, the desired exchange symmetries are built into the creation and annihilation operators. In this setting, existing techniques can prepare physically motivated occupation-number states from classically computed approximate eigenstates ~\cite{Babbush2015,Kivlichan2018, tubman2018postponingorthogonalitycatastropheefficient, Fomichev2024, Berry2025} or from other physically motivated assumptions (see Ref. \cite{Fomichev2024} and references therein). By contrast, although first quantization offers substantial savings in qubit and gate counts when the number of particles $N$ is much smaller than the number of single-particle basis functions $d$~\cite{Su2021,Berry2024}, the required particle-exchange symmetries must be enforced explicitly in the many-body wavefunction thus complicating initial-state preparation~\cite{Berry2018,Ward2009, Huggins2025}. Furthermore, existing first-quantized routines are tied to particular bases like plane-waves \cite{Huggins2025} or real-space grids~\cite{Ward2009}, which are popular but not mandatory choices~\cite{Georges2025}, and/or require tedious circuit redesign to accommodate antisymmetrization/symmetrization for fermions and bosons~\cite{Ward2009,Berry2018,Huggins2025}. Crucially, existing approaches do not permit exploration of more exotic statistics. Beyond bosons and fermions, Green's \emph{parastatistics} admit intermediate exchange statistics (parabosons and parafermions)~\cite{Green1953}, which are relevant in quantum field theory~\cite{Bauer2023}, have been proposed to exist as quasiparticles in condensed matter systems \cite{Wang2025} driving new exotic behaviour and have recently been simulated in trapped-ion experiments~\cite{HuertaAlderete2018}. Taken together, these setbacks leave first-quantized initial-state preparation well behind its second-quantized counterpart, which (with trivial modification) may uniformly prepare arbitrary superpositions of particle configurations for fermions, bosons, and parastatistics in arbitrary single-particle bases. This long-standing bottleneck has impeded first-quantized quantum simulations across application domains ranging from electronic structure to fundamental particle physics.
\begin{figure*}
    \centering
    \includegraphics[width=\linewidth]{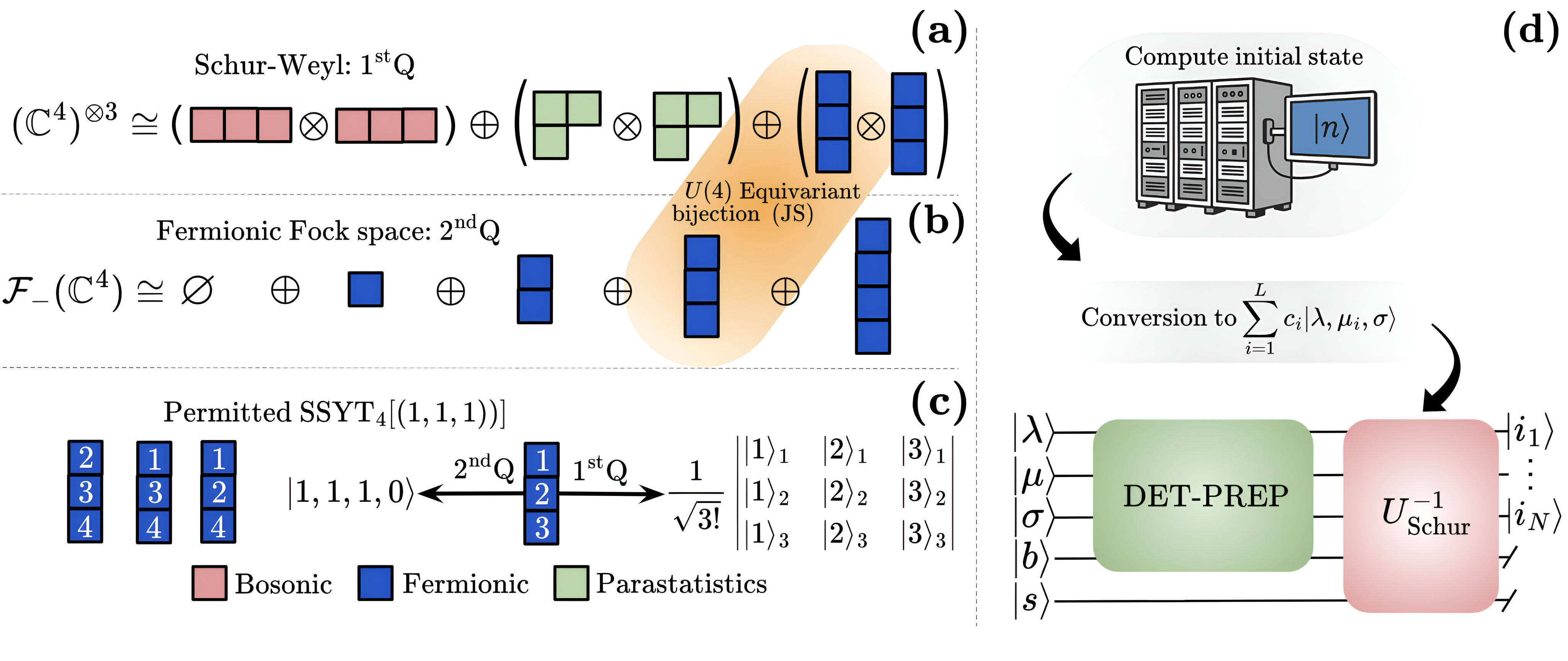}
    \caption{(a) Schur--Weyl decomposition of $(\mathbb{C}^{4})^{\otimes 3}$ into $U(4)$ irreps $V^{U(4)}_{\lambda}$ with $S_{3}$ multiplicities (in parentheses). Colors encode statistics: fermionic, bosonic, and parastatistics. (b) Fermionic Fock space $\mathcal{F}_{-}(\mathbb{C}^{4})=\bigoplus_{N=0}^{4}\wedge^{N}\mathbb{C}^{4}$. The shaded region highlights the $U(4)$-equivariant bijection induced by the Jordan--Schwinger (JS) realization between the fermionic Schur--Weyl sector in (a) and the corresponding Fock subspaces. (c) Permitted semistandard Young tableaux for $\lambda=(1,1,1)$ with $N=3$ and $d=4$. One tableau is identified with its Fock occupation $\ket{1,1,1,0}$ (2nd-Q) and with the first-quantized Slater determinant $\ket{1\wedge 2\wedge 3}=\tfrac{1}{\sqrt{3!}}\det\!\big(\ket{1}_r,\ket{2}_r,\ket{3}_r\big)_{r=1}^{3}$. (d) Universal first-quantized state-preparation pipeline: a classical high performance computer supplies $\sum_i c_i\ket{n}_i$ in 2nd-Q; coefficients are mapped to Schur labels $\sum_{i} c_{i}\ket{\lambda,\mu_i,\sigma}$; the deterministic unitary $\textsc{Det-Prep}=\textsc{ERASE}\cdot\textsc{WRITE}\cdot V_{\mathrm{amp}}$, acting on the label registers and an ancilla index bank $\ket{b}$, prepares the encoded Schur superposition with unit success probability and returns the bank to $\ket{0}$; an inverse quantum Schur transform $U_{\mathrm{Schur}}^{-1}$ produces the target first-quantized state on the system register.}
    \label{fig:hero}
\end{figure*}

In this communication, we close the gap between first- and second-quantized initial-state preparation techniques by providing a universal protocol in first quantization. By \emph{universality} we mean applicability to fermions, bosons, and paraparticles of arbitrary order (parastatistics), without time-consuming circuit redesign and equipped with compatibility with any single-particle basis. Our algorithm prepares, to accuracy $\epsilon$, an initial state that is a linear combination of $L$ configurations of $N$ particles over $d$ single-particle modes, in $\mathrm{poly}(L, N, d, \log \epsilon^{-1})$ time with the Bacon--Chuang--Harrow (BCH) varaint and in $\mathrm{poly}(L, N, \log d, \log \epsilon^{-1})$ time with the corrected Krovi--Burchardt (KB) variant. Our method is underpinned by a Lie-algebra homomorphism realized by the Jordan--Schwinger (JS) map~\cite{Jordan1935,Schwinger1952,Manko1994,dubus2024bosonsfermionsspinsmultimode}, which establishes a correspondence between particle-number-conserving second-quantized operators and their first-quantized counterparts. For given particle statistics, this correspondence yields a natural equivariant bijection between occupation-number (Fock) states and $\mathfrak{su}(d)$ weight states embedded within the Schur--Weyl decomposition of the first-quantized Hilbert space~\cite{ITZYKSON1966,Rowe2012,Mathur2002}. Given particle statistics and a target state in the occupation-number basis, we map it to the corresponding superposition of Schur-basis states via this bijection and prepare the encoded Schur labels using a deterministic, garbage-free write--erase routine that succeeds with probability one. Finally, applying an inverse quantum Schur transform ~\cite{bacon2005quantumschurtransformi,harrow2005applicationscoherentclassicalcommunication,Bacon2006,Kirby2018,Krovi2019,burchardt2025highdimensionalquantumschurtransforms} to the encoded state efficiently yields the desired first-quantized state in the computational basis. The conceptual foundation of our work and a visualization of our algorithm is shown in Fig.~\ref{fig:hero}. This communication leverages topics in group and representation theory. For self-containment, we provide the necessary background in the Supplemental Material. 

We begin by establishing that first-quantized operators and their particle-number-conserving second-quantized counterparts furnish equivalent representations of the same abstract Lie algebra. For ordinary bosons and fermions, a general particle-number-conserving second-quantized operator with up to $K$-body interactions can be expressed in normal-ordered form as
\begin{equation} O^{(K)} = \sum_{k=1}^{K} \sum_{\substack{p_{1},\dots ,p_{k}\\[2pt]q_{1},\dots ,q_{k}=1}}^{d} o^{(k)}_{p_{1}\dots p_{k},\,q_{1}\dots q_{k}}\, \adag_{p_{1}}\!\dots\adag_{p_{k}}\, a_{q_{k}}\!\dots a_{q_{1}}, \label{eq:secondQ_sumK} \end{equation}
where $p_i,q_i\in\{1,\ldots,d\}$ label single-particle modes and the sums run over all ordered $k$-tuples $(p_1,\ldots,p_k)$ and $(q_1,\ldots,q_k)$. The coefficients $o^{(k)}_{p_{1}\ldots p_{k},\,q_{1}\ldots q_{k}} = \big\langle \phi_{p_1}\cdots\phi_{p_k}\big| \hat{o}^{(k)} \big| \phi_{q_1}\cdots\phi_{q_k}\big\rangle = \int \mathrm{d}\mathbf{r}_1 \cdots \mathrm{d}\mathbf{r}_k\, \phi_{p_1}^*(\mathbf{r}_1)\cdots\phi_{p_k}^*(\mathbf{r}_k)\,\hat{o}^{(k)}\,\phi_{q_1}(\mathbf{r}_1)\cdots\phi_{q_k}(\mathbf{r}_k)$ are the $k$-particle matrix elements of the operator $\hat{o}^{(k)}$ in a chosen single-particle basis $\{\phi_i(\mathbf{r})\}_{i=1}^{d}$. Here $a_i$ and $a_i^\dagger$ denote annihilation and creation operators obeying the canonical bosonic or fermionic (anti)commutation relations; the extension to more general exchange statistics is given below. As a concrete example, setting $K=2$, imposing fermionic anticommutation relations, and identifying $\mathbf{o}^{(1)}=\mathbf{T}+\mathbf{V}_{\text{e--}n}$ (the sum of the electronic kinetic and electron--nuclear interaction matrices) and $\mathbf{o}^{(2)}=\mathbf{V}_{ee}$ (the antisymmetrized electron--electron Coulomb tensor) reduces Eq.~\eqref{eq:secondQ_sumK} to the familiar electronic-structure Hamiltonian.

To connect with the first-quantized picture while treating all exchange statistics on the same footing, we work with statistics-appropriate one-body generators $J_{p,q}$. For ordinary bosons and fermions, $J_{p,q}=\adag_{p}a_{q}$; for Green's paraparticles of arbitrary order, suitably normalized bilinears of the para-operators furnish the same generators~\cite{Green1953,Stoilova_2013, baker2026efficientquantumcircuitscoherent}. In every statistics sector these generators satisfy the $\mathfrak{gl}(d,\mathbb{C})$ commutation relations $[J_{p,q},J_{r,s}]=\delta_{qr}J_{p,s}-\delta_{ps}J_{r,q}$ and are normalized such that the diagonal generators coincide with the mode-occupation operators, $J_{p,p}=\hat{n}_p$. General number-conserving operators, including Eq.~\eqref{eq:secondQ_sumK} and its paraparticle counterparts, are then organized as polynomials of degree at most $K$ in the $J_{p,q}$. We now introduce the Jordan--Schwinger (JS) map ~\cite{Jordan1935,Schwinger1952,Manko1994,dubus2024bosonsfermionsspinsmultimode}
\begin{equation}
\Phi:\quad
X \;\longmapsto\; \sum_{p,q=1}^{d} X_{p,q}\,J_{p,q},
\label{eq:GeneralMap}
\end{equation}
for $X \in \mathfrak{gl}(d,\mathbb{C})$. By the commutation relations above, this map preserves the Lie bracket, $\Phi([X,Y])=[\Phi(X),\Phi(Y)]$, uniformly across the bosonic, fermionic, and parastatistical sectors; thus $\Phi$ is a Lie-algebra homomorphism. To restrict to $\mathfrak{su}(d)$, a convenient choice is the subspace generated by the Cartan--Weyl generators \cite{Knapp2023} $\mathcal{B}_{\text{CW}}=\{H_i \mid i=1,\ldots,d-1\}\,\cup\,\{E_{i,j} \mid i\neq j,\; i,j=1,\ldots,d\}$ with Cartan generators $H_i=E_{i,i}-E_{i+1,i+1}$ and root operators $E_{i,j}=(\delta_{ik}\delta_{jl})_{k,l=1}^{d}$. Under $\Phi$, the images of these generators are
\begin{align} \Lambda_{i,j} &= \begin{cases} J_{i,j}, & i \neq j, \ i,j \in \{1, \ldots, d \} \\[4pt] J_{i,i} - J_{i+1,i+1}, & i = j, \ i \in \{1, \ldots, d-1 \}. \end{cases} 
\label{eq:CW_JS_map} 
\end{align}
Extending to $\mathfrak{u}(d)\cong\mathfrak{su}(d)\oplus\mathfrak{u}(1)$ amounts to including the central element $I=\sum_{i=1}^{d} E_{i,i}$, which maps under $\Phi$ to the total particle-number operator $\hat{N}=\sum_{i=1}^{d} J_{i,i}=\sum_{i=1}^{d}\hat{n}_i$.

It is often convenient to represent the Cartan--Weyl generators as total operators $\Lambda_{i,j}=\sum_{k=1}^N Q_{i,j}^{(k)}$ with $Q_{i,j}^{(k)}=I^{\otimes (k-1)}\otimes Q_{i,j}\otimes I^{\otimes (N-k)}$, where $Q_{i,j}=E_{i,j}$ for $i\neq j$ and $Q_{i,i}=H_i$. With the total operator form and Eq.~\eqref{eq:CW_JS_map}, we can express Eq.~\eqref{eq:secondQ_sumK} as a linear combination of order-$K$ polynomials in $\Lambda_{i,j}$, plus an identity term. Since $E_{i,j}=|i\rangle\langle j|$, this yields
\begin{equation}
\begin{split}
\tilde{O}^{(K)} = {} &
\sum_{k=1}^{K}
\sum_{\substack{p_{1},\dots ,p_{k}\\ q_{1},\dots ,q_{k}=1}}^{d}
\sum_{1\le i_1<\dots<i_k\le N}
o^{(k)}_{p_1\dots p_k,\,q_1\dots q_k} \\
& \times
|p_1\rangle\langle q_1|_{(i_1)}\, |p_2\rangle\langle q_2|_{(i_2)} \cdots |p_k\rangle\langle q_k|_{(i_k)} ,
\end{split}
\label{eq:galerkin_generalK}
\end{equation}
where the product is the ordinary operator product of the embedded one-particle operators $|p\rangle\langle q|_{(i)} := I^{\otimes (i-1)} \otimes |p\rangle\langle q| \otimes I^{\otimes (N-i)}$, which act on disjoint tensor factors. Eqs.~\eqref{eq:secondQ_sumK} and \eqref{eq:galerkin_generalK} are thus two distinct realizations of the same underlying abstract structure: the universal enveloping algebra $\mathcal{U}[\mathfrak{u}(d)]$. The second-quantized form in Eq.~\eqref{eq:secondQ_sumK} acts on Fock space, whereas the first-quantized form in Eq.~\eqref{eq:galerkin_generalK} acts on the tensor product of single-particle spaces. It should also be noted that the form of Eq. \eqref{eq:galerkin_generalK} is applicable to simulation on real space grids, as is shown in Ref. \cite{feniou2025realspacechemistryquantumcomputers}.

Our findings allow us to demonstrate a key correspondence between these two Hilbert spaces. Consider the $i=j$ case of Eq.~\eqref{eq:CW_JS_map} and inspect the eigenstates of the right- and left-hand sides. The eigenstates of the former are Fock basis states $\ket{n}:=\ket{n_1,n_2,\ldots,n_{d}}$ satisfying $(J_{i,i} - J_{i+1,i+1})\ket{n}=(n_i-n_{i+1})\ket{n}$, since $J_{i,i}=\hat{n}_i$ in every statistics sector. The eigenstates of the latter are the \emph{weight states} of $\mathfrak{su}(d)$, denoted $\ket{\zeta,z}$, where $\zeta=[\zeta_1,\zeta_2,\ldots,\zeta_{d-1}]$ is the Dynkin label (i.e., label of the Dynkin diagram \cite{Knapp2023}) of the highest weight and $z=[z_1,z_2,\ldots,z_{d-1}]$ is the Dynkin label of a permissible weight, obeying $\Lambda_{i,i}\ket{\zeta,z}=z_i\ket{\zeta,z}$. Comparing eigenvalues yields
\begin{equation}
    z_i = n_i - n_{i+1}.
    \label{eq:weight_from_fock}
\end{equation}
Equation~\eqref{eq:weight_from_fock}, together with the fixed total-particle-number constraint $\sum_{i=1}^{d} n_i = N$, permits conversion between occupation numbers and Dynkin weights, and vice versa, therefore establishing an equivariant bijection.

The assignment of $\zeta$ is fixed by the particle statistics, with each labeling an irreducible representation (irrep) in the Schur--Weyl decomposition of the first quantized Hilbert space \cite{GoodmanWallach1998}. That is, for an $N$-particle system with $d$ single-particle states, the first-quantized Hilbert space $(\mathbb{C}^d)^{\otimes N}$ supports commuting actions of the unitary group $U(d)$ and the symmetric group $S_N$. Schur--Weyl duality states that this tensor product decomposes as
\begin{equation}
    (\mathbb{C}^d)^{\otimes N} \cong \bigoplus_{\lambda \vdash N,\;\ell(\lambda)\le d} V^{U(d)}_\lambda \otimes V^{S_N}_\lambda,
    \label{eq:sw_dual_main_text}
\end{equation}
where $\lambda$ runs over partitions of $N$ with at most $d$ parts. Here $V^{U(d)}_\lambda$ is an irrep of $U(d)$ with label $\lambda$, and $V^{S_N}_\lambda$ is the corresponding irrep of $S_N$. Each partition $\lambda$ is encoded by a Young diagram whose row lengths $\lambda=(\lambda_1,\ldots,\lambda_{d})$ determine the Dynkin label $\zeta=[\lambda_1-\lambda_2,\lambda_2-\lambda_3,\ldots,\lambda_{d-1}-\lambda_{d}]$ of $\mathfrak{su}(d)$. Because we work with finite-dimensional unitary representations of compact groups we may freely identify each $U(d)$ irrep with its differentiated $\mathfrak u(d)$ representation (labeled by $\mathfrak{su}(d)$ highest weights). Within each $U(d)$ irrep, the permissible weights are organized by Gelfand--Tsetlin (GT) patterns $\mu$, which form a multiplicity-free basis for $V^{U(d)}_\lambda$~\cite{gelfand1950finite}. We show in the Supplemental Material how to obtain $\mu$ from $z$. The additional label $\sigma$ indexes a basis vector of the corresponding $S_N$ irrep $V^{S_N}_\lambda$. Taken together, these labels yield an orthonormal Schur basis $\ket{\lambda,\mu,\sigma}_{\text{Sch}}$ of the first-quantized Hilbert space.

Within this framework, particle statistics appear as restrictions on admissible Young diagrams. For bosons, states lie entirely in the fully symmetric subspace corresponding to the single-row diagram $\lambda=(N)$. For fermions, states occupy the fully antisymmetric subspace corresponding to the single-column diagram $\lambda=(1^N)$. More generally, Green's parastatistics of order $p$ are realized by allowing Young diagrams with at most $p$ rows (parabosons) or at most $p$ columns (parafermions), each case giving rise to the corresponding family of highest weights.

For clarity, consider the minimal nontrivial case $N=2$ and $d=2$. The algebra reduces to $\mathfrak{su}(2)$, so weight states coincide with the familiar total–angular–momentum eigenstates $|\zeta=2J, z=2M\rangle$. In the forthcoming, consistent with our mode indexing, computational-basis labels  denote the single-particle modes $1,\ldots,d$ rather than binary values; thus $|12\rangle$ places the first particle in mode 1 and the second in mode 2, and the reader accustomed to qubit strings beginning at 0 should bear this convention in mind. Fermionic antisymmetry restricts to the single-column Young diagram ($\zeta=0$, i.e., $J=0$); the sole admissible Fock state $|n_1=1,n_2=1\rangle$ maps via Eq.~\eqref{eq:weight_from_fock} to $z=0$ ($M=0$), giving $|n_1=1,n_2=1\rangle \leftrightarrow |\zeta=0,z=0\rangle$, which in the computational basis is $(|12\rangle-|21\rangle)/\sqrt{2}$; a singlet state and a Slater determinant of the occupied modes. Bosons occupy the fully symmetric sector (single-row diagram, $\zeta=2$, $J=1$); the Fock states $|2,0\rangle$, $|0,2\rangle$, and $|1,1\rangle$ map to $|\zeta=2,z=2\rangle$, $|\zeta=2,z=-2\rangle$, and $|\zeta=2,z=0\rangle$, realized as $|11\rangle$, $|22\rangle$, and $(|12\rangle+|21\rangle)/\sqrt{2}$, respectively, which span the triplet states (a symmetrized product, i.e., a permanent). In both statistics the $S_N$ irrep is multiplicity-free, so we omit $\sigma$. Moreover, for $\mathfrak{su}(2)$ there is no weight-space degeneracy, so $z$ uniquely labels the irrep. Fig. \ref{fig:hero} (a-c) provides a higher dimensional example, focusing on the fermionic case in the equivalent language of semistandard Young tableux. A further example including treatment of weight and symmetric group multiplicities appears in the Supplemental Material.

In this setting, state preparation in first quantization reduces to preparing superpositions of Schur-basis states, which can be performed efficiently on a digital quantum computer using the inverse quantum Schur transform ~\cite{bacon2005quantumschurtransformi,harrow2005applicationscoherentclassicalcommunication,Bacon2006,Kirby2018,Krovi2019,burchardt2025highdimensionalquantumschurtransforms}. The forward transform implements the unitary change of basis $U_{\rm Schur}:(\mathbb{C}^{d})^{\otimes N}\!\to\!\bigoplus_{\lambda} V^{U(d)}_{\lambda}\!\otimes\! V^{S_N}_{\lambda}$.
Writing $S^{\lambda,\mu,\sigma}_{i_1,\ldots,i_N}:=\langle i_1,\ldots,i_N|U^{\dagger}_{\rm Schur}|\lambda,\mu,\sigma\rangle$, the inverse acts as
\begin{equation}
U_{\rm Schur}^{-1}\,|\lambda,\mu,\sigma\rangle = |\lambda,\mu,\sigma\rangle_{\text{Sch}}=\sum_{i_1,\ldots,i_N} S^{\lambda,\mu,\sigma}_{i_1,\ldots,i_N}\,|i_1,\ldots,i_N\rangle
\label{eq:IQST}
\end{equation}
where $|\lambda,\mu,\sigma\rangle := |\lambda\rangle \otimes |\mu\rangle \otimes |\sigma\rangle$ is a label state, with each of $\lambda$, $\mu$, and $\sigma$ encoded as dit strings, and $|i_1,\ldots,i_N\rangle$ denotes a qudit computational-basis state with $i_k\in\{1,2,\ldots,d\}$. Qudit operations and states can be encoded in qubits using standard techniques \cite{NielsenChuang2010}.
% -------------------- FLOATED ALGORITHM BOX --------------------
\begin{figure}[t]
\begin{algobox}
\algcaption{Universal initial-state preparation in first quantization}\label{alg:universal_prep}

\begin{enumerate}\itemsep3pt

\item Specify the target initial state as a normalized superposition of $L$ occupation-number basis states,
$\ket{\psi}=\sum_{i=1}^{L} c_i \ket{n}_i$, each with fixed particle number $N$.
Such a superposition may be chosen directly or obtained as the output of a classical algorithm.

\item Impose the particle statistics by selecting the appropriate Young diagram $\lambda$. Map each $\ket{n}_i$ to the Schur basis using Eq.~\eqref{eq:weight_from_fock} to obtain Dynkin labels $z$, convert these to GT patterns $\mu$ (see the Supplemental Material), and (for parastatistics) assign labels $\sigma$. This yields
$\ket{\psi }^{\lambda,\sigma}_{\text{Schur}}=\sum_{i=1}^L c_{i}\,\ket{\lambda,\mu_i,\sigma}$.

\item Classically encode the branch-dependent Schur data as label bitstrings $s_i$ in the convention of the chosen inverse-Schur realization: the GT pattern in a naive or compressed encoding for the Bacon--Chuang--Harrow (BCH) transform, or the compressed label triple of the Krovi--Burchardt (KB) transform (see the Supplemental Material).

\item Apply the deterministic preparation $\textsc{Det-Prep}$ of Eq.~\eqref{eq:detprep}: garbage-free amplitude loading $V_{\mathrm{amp}}$ onto the ancilla bank $b$, label writing \textsc{WRITE}, and deterministic index erasure \textsc{ERASE}. The procedure succeeds with probability $1$ and returns all ancillas to $\ket{0}$.

\item Apply the inverse quantum Schur transform $U_{\rm Schur}^{-1}$ (Eq.~\eqref{eq:IQST}).
\end{enumerate}

\noindent\textbf{Output:} Target initial state in first quantization on the system register.\\
\textbf{Time complexity:} $\mathrm{poly}(L, N, d, \log \epsilon^{-1})$ with the BCH realization; $\mathrm{poly}(L, N, \log d, \log \epsilon^{-1})$ with the KB realization.
\end{algobox}
\end{figure}
% -----------------------------------------------------------------
\begin{figure*}
    \centering
    \includegraphics[width=\linewidth]{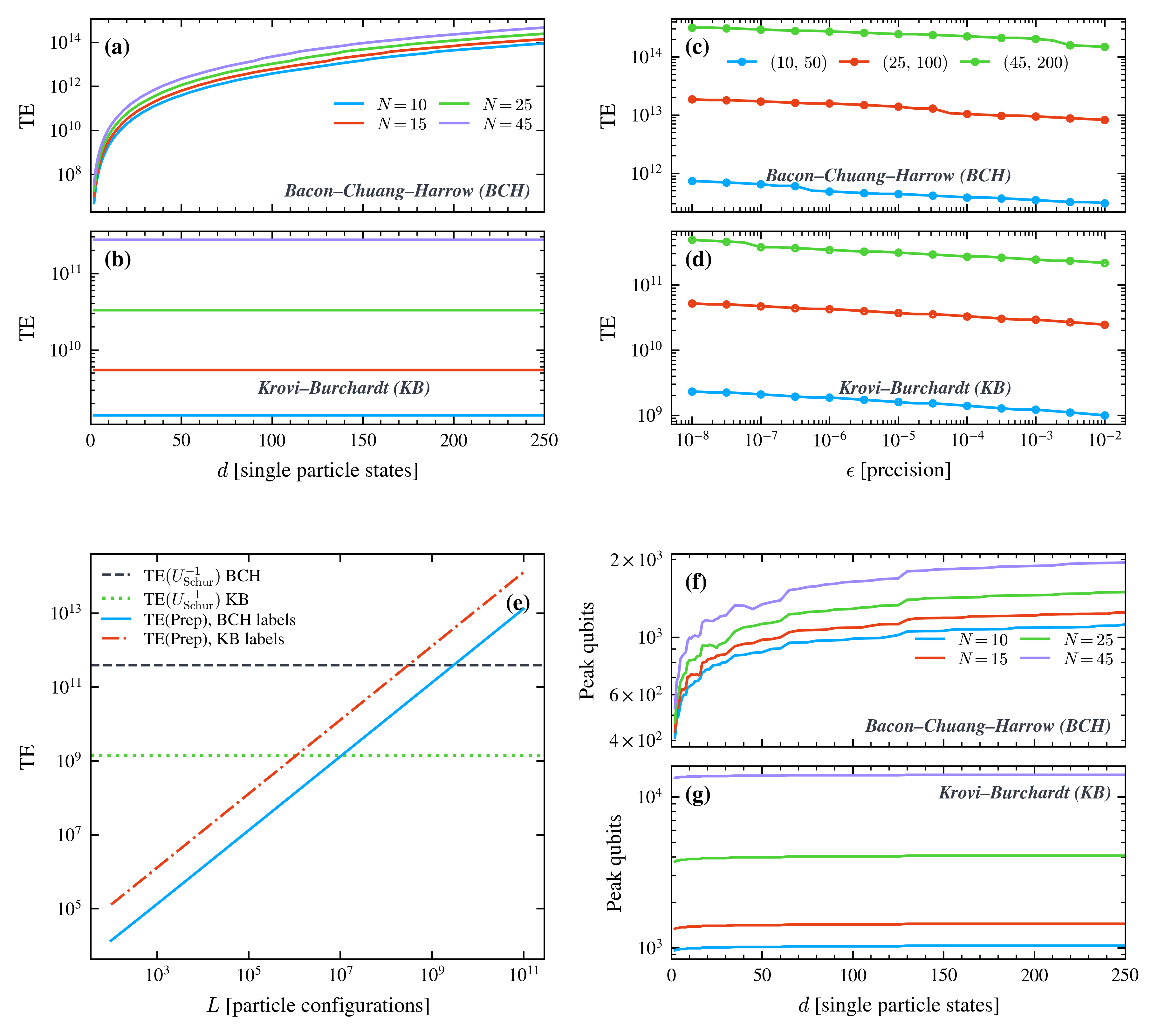}
    \caption{Resource estimates for universal first-quantized state preparation (algorithm \ref{alg:universal_prep}) utilizing the Bacon--Chuang--Harrow (BCH) and corrected Krovi--Burchardt (KB) inverse quantum Schur transform variants. Panels (a), (b), (f) and (g) assume an equal superposition of $L=50$ configurations at fixed precision $\epsilon=10^{-4}$, prepared deterministically with the write--erase procedure of the Supplemental Material; the error budget is split evenly between the preparation and inverse-Schur stages. (a), (b) Toffoli-equivalent gate count, $\mathrm{TE}=\#\mathrm{Toffoli}+(\mathrm{T\!-\!count})/7$~\cite{TEcostnote}, versus single-particle dimension $d$ for the BCH and KB constructions, respectively. (c), (d) $\mathrm{TE}$ versus target precision $\epsilon$ for BCH and KB (legends list $(N,d)$ tuples). (e) Crossover in $\mathrm{TE}$ as a function of the number of configurations $L$: the deterministic label preparation $\mathrm{TE}(\mathrm{Prep})$, instantiated with the compressed GT label register of the BCH pipeline (BCH labels) and with the compressed label triple of the KB pipeline (KB labels), compared against the inverse Schur-transform cost $\mathrm{TE}(U_{\mathrm{Schur}}^{-1})$ for both constructions. We set $N=10$, $d=50$, and $\epsilon=10^{-4}$. (f), (g) Peak qubit count versus $d$ for BCH and KB, respectively, at $\epsilon=10^{-4}$.}
\label{fig:res_est}
\end{figure*}
Superposition states can be prepared by applying $U_{\rm Schur}^{-1}$ to arbitrary normalized superpositions of Schur labels, $|\psi \rangle_{\text{Schur}}^{\lambda,\sigma}=\sum_{i=1}^L c_i\,|\lambda,\mu_i,\sigma\rangle$, $c_i\in\mathbb{C}$, where we fix $\lambda$ and $\sigma$, thereby fixing the particle statistics and selecting a canonical symmetric group copy. Indeed, the JS homomorphism is insensitive to the $\sigma$ label, so any valid $\sigma$ may be used here; different $\sigma$ merely label separate valid orthonormal bases in the irreducible subspace. The registers $\ket{\lambda}$ and $\ket{\sigma}$ are branch-independent and are written by Clifford operations; the branch-dependent data are held in a label register whose width $n_{\mathrm{lab}}$ depends on the chosen realization of $U_{\rm Schur}^{-1}$, holding the GT pattern in a naive or compressed encoding for BCH, or the compressed label triple for KB, in which case $n_{\mathrm{lab}}=\mathcal{O}(N\log d + N^{2}\log N)$ with only logarithmic dependence on $d$. The label superposition is produced by the deterministic unitary $\textsc{Det-Prep}$, which acts on the label register together with an ancilla bank of $b=\lceil\log_2 L\rceil$ index qubits:
\begin{align}
\ket{0^{b}}\!\ket{0^{n_{\mathrm{lab}}}}
&\xrightarrow{V_{\mathrm{amp}}}
\sum_{i=1}^{L} c_i \ket{i}\!\ket{0^{n_{\mathrm{lab}}}} \notag\\
&\xrightarrow{\textsc{WRITE}}
\sum_{i=1}^{L} c_i \ket{i}\!\ket{s_i} \notag\\
&\xrightarrow{\textsc{ERASE}}
\ket{0^{b}}\otimes\sum_{i=1}^{L} c_i \ket{s_i}.
\label{eq:detprep}
\end{align}
Here $V_{\mathrm{amp}}$ loads the amplitudes $c_i$ onto the ancilla bank through advanced quantum read-only memory (QROAM) angle lookups, \textsc{WRITE} copies the classically precomputed label bitstring $s_i$ into the label register conditioned on the index, and \textsc{ERASE} returns the ancilla bank to $\ket{0^{b}}$ exactly by recomputing the index from the label. The procedure therefore succeeds with probability one, at a Toffoli-equivalent cost of $\Theta(L\, n_{\mathrm{lab}})$, which is optimal in $L$ up to the label width since any preparation circuit must access all $L$ classical coefficients; further details are provided in the Supplemental Material.

A high-level algorithm for universal state preparation in first quantization is presented in Algorithm~\ref{alg:universal_prep} and further visualized in Fig. \ref{fig:hero}(d). The runtime complexity is attained by combining clean-ancilla advanced QROAM~\cite{Berry2019qubitizationof} for the amplitude and label lookups of $\textsc{Det-Prep}$, an erase cascade linear in $L$, and either the BCH \cite{bacon2005quantumschurtransformi, Bacon2006} or the high-dimensional Krovi inverse quantum Schur transform~\cite{Krovi2019}, with time complexity $\mathrm{poly}(N,d,\log \epsilon^{-1})$ and $\mathrm{poly}(N,\log d,\log \epsilon^{-1})$, respectively; combined with the preparation stage this yields the end-to-end scalings of Algorithm~\ref{alg:universal_prep}. The correctness of Krovi's original algorithm was recently questioned, and a corrected version, hereafter Krovi--Burchardt (KB), has been developed~\cite{burchardt2025highdimensionalquantumschurtransforms}. The correction preserves the crucial $\mathrm{poly}(\log d)$ scaling, derived from the compression of ``type vectors" in the preparation stage \cite{Krovi2019, burchardt2025highdimensionalquantumschurtransforms}, so the efficiency of high-dimensional transforms remains; it is this corrected construction that we implement below. As is standard in FTQC implementation, $\text{poly}(\log \epsilon^{-1})$ scaling originates from the approximation of arbitrary single qubit rotations following the Solovay-Kitaev theorem \cite{Kitaev1997QCAE}.

Figure~\ref{fig:res_est} summarizes non-Clifford resource estimates for concrete realizations of Algorithm~\ref{alg:universal_prep}. We consider preparing an equal superposition of $L$ number-basis configurations using the deterministic write--erase procedure, with the amplitude and label tables loaded via clean-ancilla QROAM~\cite{Berry2019qubitizationof}, and benchmark the two realizations of the inverse quantum Schur transform $U_{\mathrm{Schur}}^{-1}$: the BCH construction~\cite{bacon2005quantumschurtransformi} and the corrected KB construction~\cite{Krovi2019,burchardt2025highdimensionalquantumschurtransforms}. A detailed account of the resource estimations is provided in the Supplemental Material.

Panels~\ref{fig:res_est}(a) and \ref{fig:res_est}(b) contrast usage of the two Schur-transform varaints at fixed precision. The BCH cost climbs steeply with the single-particle dimension, reflecting its $\widetilde{\mathcal{O}}(Nd^{4})$ scaling, whereas the KB cost is essentially independent of $d$, confirming its $\widetilde{\mathcal{O}}(N^{4})$ scaling with only polylogarithmic dependence on the basis size. Across the surveyed range the KB construction lowers the Toffoli-equivalent (TE) count (see Fig.~\ref{fig:res_est} caption) by two to four orders of magnitude, and the advantage widens as $d$ grows. Panels~\ref{fig:res_est}(c) and \ref{fig:res_est}(d) show that both realizations depend only weakly on the target precision $\epsilon$, as expected from the logarithmic overhead of rotation synthesis. Panels~\ref{fig:res_est}(f) and \ref{fig:res_est}(g) reveal the accompanying cost: the BCH qubit footprint remains near $10^{3}$ and grows only mildly with $d$, whereas the KB footprint, though essentially independent of $d$, grows quadratically with $N$ through its compressed Gelfand--Tsetlin register and exceeds $10^{4}$ qubits at $N=45$. The choice between the two constructions is therefore a genuine space--time trade-off governed by $(N,d)$: KB is decisively preferable whenever logical qubits are plentiful and $d\gg N$, while BCH remains attractive on qubit-constrained architectures. Crucially, the end-to-end TE counts sit within one order of magnitude of, and with the KB realization frequently well below, those reported for full QPE-based ground-state preparation in first quantization~\cite{Georges2025}, establishing that our procedure is practical within state-of-the-art simulation pipelines. Panel~\ref{fig:res_est}(e) compares the cost of the deterministic label preparation, instantiated with the label register appropriate to each pipeline, against the fixed cost of $U_{\mathrm{Schur}}^{-1}$ under both constructions. The preparation cost grows strictly linearly in $L$, dominated by the erase cascade, and the KB-label curve sits a constant factor above its BCH counterpart, reflecting the wider compressed label triple against the (optimally) compressed GT register alone. For moderate and practically relevant $L$, the inverse Schur transform dominates the TE budget under either construction. The KB transform lowers this ceiling by more than two orders of magnitude, and since its labels are also wider, the crossover beyond which the preparation stage dominates shifts to smaller $L$ on both accounts. The schur label preparation stage is thus never the leading expense of a first-quantized workflow until the number of configurations becomes very large, at which point the label-erasure cascade, rather than the Schur transform, sets the budget.

To summarize, we have presented a universal, fault-tolerant protocol for symmetry-adapted state preparation in first quantization. Exploiting the Jordan--Schwinger homomorphism, we have identified number-conserving second-quantized operators with their first-quantized action and obtained an equivariant bijection between Fock occupations and $\mathfrak{su}(d)$ weight states within the Schur--Weyl decomposition. This enables the preparation of arbitrary polynomial-size superpositions of particle configurations via a deterministic, write--erase preparation of encoded Schur labels, succeeding with unit probability, followed by an inverse quantum Schur transform, with time complexity $\mathrm{poly}(L, N, d, \log \epsilon^{-1})$ for the BCH variant and $\mathrm{poly}(L, N, \log d, \log \epsilon^{-1})$ for the KB variant. Resource estimates employing both inverse Schur transforms place the Toffoli-equivalent cost within, and in the KB case frequently below, the budgets of leading first-quantized pipelines, with the two constructions offering a complementary space--time trade-off across the $(N,d)$ landscape. Beyond state preparation, our framework suggests several avenues to further advance quantum simulations in first quantization: (i) a universal, coherent quantization transform that toggles on-device between second and first quantization; (ii) block-diagonalization and potential fast-forwarding of first-quantized Hamiltonians using the quantum Schur transform (e.g., \cite{burkat2025quantumpaldustransformefficient,Gu2021}); (iii) Schur-basis algorithms exploiting the Wigner--Eckart theorem, for example to factor matrix elements and shrink measurement budgets and (iv) asymptotic improvements on algorithm \ref{alg:universal_prep} by dropping universality in favor of particle statistics dependent operations. These advances broaden the scope of first-quantized simulation across chemistry, materials, and fundamental physics.

\section*{Acknowledgements}
We are grateful for informative conversations with Benoît Dubus, Tobias Haas, Nicolas J. Cerf regarding their work on the extension of the Jordan-Schwinger map to many modes \cite{dubus2024bosonsfermionsspinsmultimode}. We extend our thanks to Kevin Ferreira, Yipeng Ji and Paria Nejat of the LG Electronics Toronto AI Lab and to Sean Kim of LG Electronics, AI Lab, for their ongoing support of our research.

\bibliography{ms.bib}% Produces the bibliography via BibTeX.

@incollection{feynman2018simulating,
  title={Simulating physics with computers},
  author={Feynman, Richard P},
  booktitle={Feynman and computation},
  pages={133--153},
  year={2018},
  publisher={cRc Press}
}

@article{Daley2022,
  title = {Practical quantum advantage in quantum simulation},
  volume = {607},
  ISSN = {1476-4687},
  url = {http://dx.doi.org/10.1038/s41586-022-04940-6},
  DOI = {10.1038/s41586-022-04940-6},
  number = {7920},
  journal = {Nature},
  publisher = {Springer Science and Business Media LLC},
  author = {Daley,  Andrew J. and Bloch,  Immanuel and Kokail,  Christian and Flannigan,  Stuart and Pearson,  Natalie and Troyer,  Matthias and Zoller,  Peter},
  year = {2022},
  month = jul,
  pages = {667–676}
}

@article{Lee2023,
  title = {Evaluating the evidence for exponential quantum advantage in ground-state quantum chemistry},
  volume = {14},
  ISSN = {2041-1723},
  url = {http://dx.doi.org/10.1038/s41467-023-37587-6},
  DOI = {10.1038/s41467-023-37587-6},
  number = {1},
  journal = {Nature Communications},
  publisher = {Springer Science and Business Media LLC},
  author = {Lee,  Seunghoon and Lee,  Joonho and Zhai,  Huanchen and Tong,  Yu and Dalzell,  Alexander M. and Kumar,  Ashutosh and Helms,  Phillip and Gray,  Johnnie and Cui,  Zhi-Hao and Liu,  Wenyuan and Kastoryano,  Michael and Babbush,  Ryan and Preskill,  John and Reichman,  David R. and Campbell,  Earl T. and Valeev,  Edward F. and Lin,  Lin and Chan,  Garnet Kin-Lic},
  year = {2023},
  month = apr 
}

@misc{lanes2025frameworkquantumadvantage,
      title={A Framework for Quantum Advantage}, 
      author={Olivia Lanes and Mourad Beji and Antonio D. Corcoles and Constantin Dalyac and Jay M. Gambetta and Loic Henriet and Ali Javadi-Abhari and Abhinav Kandala and Antonio Mezzacapo and Christopher Porter and Sarah Sheldon and John Watrous and Christa Zoufal and Alexandre Dauphin and Borja Peropadre},
      year={2025},
      eprint={2506.20658},
      archivePrefix={arXiv},
      primaryClass={quant-ph},
      url={https://arxiv.org/abs/2506.20658}, 
}

@misc{tubman2018postponingorthogonalitycatastropheefficient,
      title={Postponing the orthogonality catastrophe: efficient state preparation for electronic structure simulations on quantum devices}, 
      author={Norm M. Tubman and Carlos Mejuto-Zaera and Jeffrey M. Epstein and Diptarka Hait and Daniel S. Levine and William Huggins and Zhang Jiang and Jarrod R. McClean and Ryan Babbush and Martin Head-Gordon and K. Birgitta Whaley},
      year={2018},
      eprint={1809.05523},
      archivePrefix={arXiv},
      primaryClass={quant-ph},
      url={https://arxiv.org/abs/1809.05523}, 
}

@article{Berry2025,
  title = {Rapid Initial-State Preparation for the Quantum Simulation of Strongly Correlated Molecules},
  volume = {6},
  ISSN = {2691-3399},
  url = {http://dx.doi.org/10.1103/PRXQuantum.6.020327},
  DOI = {10.1103/prxquantum.6.020327},
  number = {2},
  journal = {PRX Quantum},
  publisher = {American Physical Society (APS)},
  author = {Berry,  Dominic W. and Tong,  Yu and Khattar,  Tanuj and White,  Alec and Kim,  Tae In and Low,  Guang Hao and Boixo,  Sergio and Ding,  Zhiyan and Lin,  Lin and Lee,  Seunghoon and Chan,  Garnet Kin-Lic and Babbush,  Ryan and Rubin,  Nicholas C.},
  year = {2025},
  month = may 
}

@misc{feniou2025realspacechemistryquantumcomputers,
      title={Real-Space Chemistry on Quantum Computers: A Fault-Tolerant Algorithm with Adaptive Grids and Transcorrelated Extension}, 
      author={César Feniou and Christopher Cherfan and Julien Zylberman and Baptiste Claudon and Jean-Philip Piquemal and Emmanuel Giner},
      year={2025},
      eprint={2507.20583},
      archivePrefix={arXiv},
      primaryClass={quant-ph},
      url={https://arxiv.org/abs/2507.20583}, 
}

@article{Huggins2025,
  title = {Efficient State Preparation for the Quantum Simulation of Molecules in First Quantization},
  volume = {6},
  ISSN = {2691-3399},
  url = {http://dx.doi.org/10.1103/PRXQuantum.6.020319},
  DOI = {10.1103/prxquantum.6.020319},
  number = {2},
  journal = {PRX Quantum},
  publisher = {American Physical Society (APS)},
  author = {Huggins,  William J. and Leimkuhler,  Oskar and Stetina,  Torin F. and Whaley,  K. Birgitta},
  year = {2025},
  month = apr 
}

@article{Babbush2015,
  title = {Chemical basis of Trotter-Suzuki errors in quantum chemistry simulation},
  volume = {91},
  ISSN = {1094-1622},
  url = {http://dx.doi.org/10.1103/PhysRevA.91.022311},
  DOI = {10.1103/physreva.91.022311},
  number = {2},
  journal = {Physical Review A},
  publisher = {American Physical Society (APS)},
  author = {Babbush,  Ryan and McClean,  Jarrod and Wecker,  Dave and Aspuru-Guzik,  Alán and Wiebe,  Nathan},
  year = {2015},
  month = feb 
}

@article{Kivlichan2018,
  title = {Quantum Simulation of Electronic Structure with Linear Depth and Connectivity},
  volume = {120},
  ISSN = {1079-7114},
  url = {http://dx.doi.org/10.1103/PhysRevLett.120.110501},
  DOI = {10.1103/physrevlett.120.110501},
  number = {11},
  journal = {Physical Review Letters},
  publisher = {American Physical Society (APS)},
  author = {Kivlichan,  Ian D. and McClean,  Jarrod and Wiebe,  Nathan and Gidney,  Craig and Aspuru-Guzik,  Alán and Chan,  Garnet Kin-Lic and Babbush,  Ryan},
  year = {2018},
  month = mar 
}

@article{Wang2025,
  title = {Particle exchange statistics beyond fermions and bosons},
  volume = {637},
  ISSN = {1476-4687},
  url = {http://dx.doi.org/10.1038/s41586-024-08262-7},
  DOI = {10.1038/s41586-024-08262-7},
  number = {8045},
  journal = {Nature},
  publisher = {Springer Science and Business Media LLC},
  author = {Wang,  Zhiyuan and Hazzard,  Kaden R. A.},
  year = {2025},
  month = jan,
  pages = {314–318}
}

@book{NielsenChuang2010,
  author    = {Michael A. Nielsen and Isaac L. Chuang},
  title     = {Quantum Computation and Quantum Information},
  edition   = {10th Anniversary Edition},
  year      = {2010},
  publisher = {Cambridge University Press},
  address   = {Cambridge},
  isbn      = {978-1-107-00217-3},
  note      = {See Secs.~4.5.1--4.5.2 for two-level unitaries and universality (decomposition of general $d\times d$ unitaries into qubit operations).}
}

@article{Kitaev1997QCAE,
  author  = {Kitaev, A. Yu.},
  title   = {Quantum computations: algorithms and error correction},
  journal = {Russian Mathematical Surveys},
  year    = {1997},
  volume  = {52},
  number  = {6},
  pages   = {1191--1249},
  doi     = {10.1070/RM1997v052n06ABEH002155},
  url     = {https://www.mathnet.ru/eng/rm892}
}

@article{Roffe_2019,
   title={Quantum error correction: an introductory guide},
   volume={60},
   ISSN={1366-5812},
   url={http://dx.doi.org/10.1080/00107514.2019.1667078},
   DOI={10.1080/00107514.2019.1667078},
   number={3},
   journal={Contemporary Physics},
   publisher={Informa UK Limited},
   author={Roffe, Joschka},
   year={2019},
   month=jul, pages={226–245} }

@misc{kitaev1995quantummeasurementsabelianstabilizer,
      title={Quantum measurements and the Abelian Stabilizer Problem}, 
      author={A. Yu. Kitaev},
      year={1995},
      eprint={quant-ph/9511026},
      archivePrefix={arXiv},
      primaryClass={quant-ph},
      url={https://arxiv.org/abs/quant-ph/9511026}, 
}

@article{Gu2021,
  title = {Fast-forwarding quantum evolution},
  volume = {5},
  ISSN = {2521-327X},
  url = {http://dx.doi.org/10.22331/q-2021-11-15-577},
  DOI = {10.22331/q-2021-11-15-577},
  journal = {Quantum},
  publisher = {Verein zur Forderung des Open Access Publizierens in den Quantenwissenschaften},
  author = {Gu,  Shouzhen and Somma,  Rolando D. and Şahinoğlu,  Burak},
  year = {2021},
  month = nov,
  pages = {577}
}

@article{Fomichev2024,
  title = {Initial State Preparation for Quantum Chemistry on Quantum Computers},
  volume = {5},
  ISSN = {2691-3399},
  url = {http://dx.doi.org/10.1103/PRXQuantum.5.040339},
  DOI = {10.1103/prxquantum.5.040339},
  number = {4},
  journal = {PRX Quantum},
  publisher = {American Physical Society (APS)},
  author = {Fomichev,  Stepan and Hejazi,  Kasra and Zini,  Modjtaba Shokrian and Kiser,  Matthew and Fraxanet,  Joana and Casares,  Pablo Antonio Moreno and Delgado,  Alain and Huh,  Joonsuk and Voigt,  Arne-Christian and Mueller,  Jonathan E. and Arrazola,  Juan Miguel},
  year = {2024},
  month = dec 
}

@article{Katabarwa2024,
  title = {Early Fault-Tolerant Quantum Computing},
  volume = {5},
  ISSN = {2691-3399},
  url = {http://dx.doi.org/10.1103/PRXQuantum.5.020101},
  DOI = {10.1103/prxquantum.5.020101},
  number = {2},
  journal = {PRX Quantum},
  publisher = {American Physical Society (APS)},
  author = {Katabarwa,  Amara and Gratsea,  Katerina and Caesura,  Athena and Johnson,  Peter D.},
  year = {2024},
  month = jun 
}

@article{Su2021,
  title = {Fault-Tolerant Quantum Simulations of Chemistry in First Quantization},
  volume = {2},
  ISSN = {2691-3399},
  url = {http://dx.doi.org/10.1103/PRXQuantum.2.040332},
  DOI = {10.1103/prxquantum.2.040332},
  number = {4},
  journal = {PRX Quantum},
  publisher = {American Physical Society (APS)},
  author = {Su,  Yuan and Berry,  Dominic W. and Wiebe,  Nathan and Rubin,  Nicholas and Babbush,  Ryan},
  year = {2021},
  month = nov 
}

@inproceedings{gelfand1950finite,
  title={Finite-dimensional representations of the group of unimodular matrices},
  author={Gelfand, Israel M and Tsetlin, Michael L},
  booktitle={Dokl. Akad. Nauk SSSR},
  volume={71},
  pages={825},
  year={1950}
}

@article{Alexeev2024,
  title = {Quantum-centric supercomputing for materials science: A perspective on challenges and future directions},
  volume = {160},
  ISSN = {0167-739X},
  url = {http://dx.doi.org/10.1016/j.future.2024.04.060},
  DOI = {10.1016/j.future.2024.04.060},
  journal = {Future Generation Computer Systems},
  publisher = {Elsevier BV},
  author = {Alexeev,  Yuri and Amsler,  Maximilian and Barroca,  Marco Antonio and Bassini,  Sanzio and Battelle,  Torey and Camps and others},
  year = {2024},
  month = nov,
  pages = {666–710}
}

@article{Bauer2023,
  title = {Quantum simulation of fundamental particles and forces},
  volume = {5},
  ISSN = {2522-5820},
  url = {http://dx.doi.org/10.1038/s42254-023-00599-8},
  DOI = {10.1038/s42254-023-00599-8},
  number = {7},
  journal = {Nature Reviews Physics},
  publisher = {Springer Science and Business Media LLC},
  author = {Bauer,  Christian W. and Davoudi,  Zohreh and Klco,  Natalie and Savage,  Martin J.},
  year = {2023},
  month = jun,
  pages = {420–432}
}

@article{Georgescu2014,
  title = {Quantum simulation},
  volume = {86},
  ISSN = {1539-0756},
  url = {http://dx.doi.org/10.1103/RevModPhys.86.153},
  DOI = {10.1103/revmodphys.86.153},
  number = {1},
  journal = {Reviews of Modern Physics},
  publisher = {American Physical Society (APS)},
  author = {Georgescu,  I. M. and Ashhab,  S. and Nori,  Franco},
  year = {2014},
  month = mar,
  pages = {153–185}
}

@article{McArdle2020,
  title = {Quantum computational chemistry},
  author = {McArdle, Sam and Endo, Suguru and Aspuru-Guzik, Al\'an and Benjamin, Simon C. and Yuan, Xiao},
  journal = {Rev. Mod. Phys.},
  volume = {92},
  issue = {1},
  pages = {015003},
  numpages = {51},
  year = {2020},
  month = {Mar},
  publisher = {American Physical Society},
  doi = {10.1103/RevModPhys.92.015003},
  url = {https://link.aps.org/doi/10.1103/RevModPhys.92.015003}
}

@article{Georges2025,
  title = {Quantum simulations of chemistry in first quantization with any basis set},
  volume = {11},
  ISSN = {2056-6387},
  url = {http://dx.doi.org/10.1038/s41534-025-00987-1},
  DOI = {10.1038/s41534-025-00987-1},
  number = {1},
  journal = {npj Quantum Information},
  publisher = {Springer Science and Business Media LLC},
  author = {Georges,  Timothy N. and Bothe,  Marius and S\"{u}nderhauf,  Christoph and Berntson,  Bjorn K. and Izsák,  Róbert and Ivanov,  Aleksei V.},
  year = {2025},
  month = apr 
}

@article{Berry2024,
  title = {Quantum simulation of realistic materials in first quantization using non-local pseudopotentials},
  volume = {10},
  ISSN = {2056-6387},
  url = {http://dx.doi.org/10.1038/s41534-024-00896-9},
  DOI = {10.1038/s41534-024-00896-9},
  number = {1},
  journal = {npj Quantum Information},
  publisher = {Springer Science and Business Media LLC},
  author = {Berry,  Dominic W. and Rubin,  Nicholas C. and Elnabawy,  Ahmed O. and Ahlers,  Gabriele and DePrince,  A. Eugene and Lee,  Joonho and Gogolin,  Christian and Babbush,  Ryan},
  year = {2024},
  month = dec 
}

@article{Berry2019qubitizationof,
  doi = {10.22331/q-2019-12-02-208},
  url = {https://doi.org/10.22331/q-2019-12-02-208},
  title = {Qubitization of {A}rbitrary {B}asis {Q}uantum {C}hemistry {L}everaging {S}parsity and {L}ow {R}ank {F}actorization},
  author = {Berry, Dominic W. and Gidney, Craig and Motta, Mario and McClean, Jarrod R. and Babbush, Ryan},
  journal = {{Quantum}},
  issn = {2521-327X},
  publisher = {{Verein zur F{\"{o}}rderung des Open Access Publizierens in den Quantenwissenschaften}},
  volume = {3},
  pages = {208},
  month = dec,
  year = {2019}
}

@article{Jordan1935,
  title = {Der Zusammenhang der symmetrischen und linearen Gruppen und das Mehrk\"{o}rperproblem},
  volume = {94},
  ISSN = {1434-601X},
  url = {http://dx.doi.org/10.1007/BF01330618},
  DOI = {10.1007/bf01330618},
  number = {7–8},
  journal = {Zeitschrift f\"{u}r Physik},
  publisher = {Springer Science and Business Media LLC},
  author = {Jordan,  P.},
  year = {1935},
  month = jul,
  pages = {531–535}
}

@article{Manko1994,
  title = {A Generalization of the Jordan-Schwinger Map: The Classical Verion and its q deformation},
  volume = {09},
  ISSN = {1793-656X},
  url = {http://dx.doi.org/10.1142/S0217751X94002260},
  DOI = {10.1142/s0217751x94002260},
  number = {31},
  journal = {International Journal of Modern Physics A},
  publisher = {World Scientific Pub Co Pte Lt},
  author = {Man’ko,  V.I. and Marmo,  G. and Vitale,  P. and Zaccaria,  F.},
  year = {1994},
  month = dec,
  pages = {5541–5561}
}

@misc{dubus2024bosonsfermionsspinsmultimode,
      title={From bosons and fermions to spins: A multi-mode extension of the Jordan-Schwinger map}, 
      author={Benoît Dubus and Tobias Haas and Nicolas J. Cerf},
      year={2024},
      eprint={2411.04918},
      archivePrefix={arXiv},
      primaryClass={quant-ph},
      url={https://arxiv.org/abs/2411.04918}, 
}

@techreport{Schwinger1952,
    url = {https://www.ifi.unicamp.br/%7Ecabrera/teaching/paper_schwinger.pdf},
    institution = {Harvard University, Nuclear Development Associates, Inc., United States Department of Energy},
    volume = {32},
    pages = {229-279},
    author = {Schwinger, Julian},
    title = {{On angular momentum}},
    year = {1952},
    number = {NYO-3071}
}

@article{Ward2009,
  title = {Preparation of many-body states for quantum simulation},
  volume = {130},
  ISSN = {1089-7690},
  url = {http://dx.doi.org/10.1063/1.3115177},
  DOI = {10.1063/1.3115177},
  number = {19},
  journal = {The Journal of Chemical Physics},
  publisher = {AIP Publishing},
  author = {Ward,  Nicholas J. and Kassal,  Ivan and Aspuru-Guzik,  Alán},
  year = {2009},
  month = may 
}

@article{Berry2018,
  title = {Improved techniques for preparing eigenstates of fermionic Hamiltonians},
  volume = {4},
  ISSN = {2056-6387},
  url = {http://dx.doi.org/10.1038/s41534-018-0071-5},
  DOI = {10.1038/s41534-018-0071-5},
  number = {1},
  journal = {npj Quantum Information},
  publisher = {Springer Science and Business Media LLC},
  author = {Berry,  Dominic W. and Kieferová,  Mária and Scherer,  Artur and Sanders,  Yuval R. and Low,  Guang Hao and Wiebe,  Nathan and Gidney,  Craig and Babbush,  Ryan},
  year = {2018},
  month = may 
}

@misc{bacon2005quantumschurtransformi,
      title={The Quantum Schur Transform: I. Efficient Qudit Circuits}, 
      author={Dave Bacon and Isaac L. Chuang and Aram W. Harrow},
      year={2005},
      eprint={quant-ph/0601001},
      archivePrefix={arXiv},
      primaryClass={quant-ph},
      url={https://arxiv.org/abs/quant-ph/0601001}, 
}

@article{Bacon2006,
  title = {Efficient Quantum Circuits for Schur and Clebsch-Gordan Transforms},
  volume = {97},
  ISSN = {1079-7114},
  url = {http://dx.doi.org/10.1103/PhysRevLett.97.170502},
  DOI = {10.1103/physrevlett.97.170502},
  number = {17},
  journal = {Physical Review Letters},
  publisher = {American Physical Society (APS)},
  author = {Bacon,  Dave and Chuang,  Isaac L. and Harrow,  Aram W.},
  year = {2006},
  month = oct 
}

@article{Kirby2018,
  title = {A practical quantum algorithm for the Schur transform},
  volume = {18},
  ISSN = {1533-7146},
  url = {http://dx.doi.org/10.26421/QIC18.9-10-1},
  DOI = {10.26421/qic18.9-10-1},
  number = {9 & 10},
  journal = {Quantum Information and Computation},
  publisher = {Rinton Press},
  author = {Kirby,  William M. and Strauch,  Frederick W.},
  year = {2018},
  month = aug,
  pages = {721–742}
}

@article{kawano2016qftsn,
  author  = {Kawano, Yasuhito and Sekigawa, Hiroshi},
  title   = {Quantum {Fourier} transform over symmetric groups---improved result},
  journal = {Journal of Symbolic Computation},
  volume  = {75},
  pages   = {219--243},
  year    = {2016},
  doi     = {10.1016/j.jsc.2015.11.016}
}

@article{Green1953,
  author = {Green, H.~S.},
  title = {A generalized method of field quantization},
  journal = {Phys. Rev.},
  volume = {90},
  pages = {270--273},
  year = {1953}
}

@article{HuertaAlderete2018,
  author = {Huerta Alderete, C. and Rodr{\'\i}guez-Lara, B.~M.},
  title = {Simulating para-Fermi oscillators},
  journal = {Sci. Rep.},
  volume = {8},
  pages = {11572},
  year = {2018}
}

@misc{burchardt2025highdimensionalquantumschurtransforms,
      title={High-dimensional quantum Schur transforms}, 
      author={Adam Burchardt and Jiani Fei and Dmitry Grinko and Martin Larocca and Maris Ozols and Sydney Timmerman and Vladyslav Visnevskyi},
      year={2025},
      eprint={2509.22640},
      archivePrefix={arXiv},
      primaryClass={quant-ph},
      url={https://arxiv.org/abs/2509.22640}, 
}

@misc{harrow2005applicationscoherentclassicalcommunication,
      title={Applications of coherent classical communication and the Schur transform to quantum information theory}, 
      author={Aram W. Harrow},
      year={2005},
      eprint={quant-ph/0512255},
      archivePrefix={arXiv},
      primaryClass={quant-ph},
      url={https://arxiv.org/abs/quant-ph/0512255}, 
}

@inproceedings{shor1996fault,
  title={Fault-tolerant quantum computation},
  author={Shor, Peter W},
  booktitle={Proceedings of 37th conference on foundations of computer science},
  pages={56--65},
  year={1996},
  organization={IEEE}
}

@inproceedings{Beals1997,
  series = {STOC ’97},
  title = {Quantum computation of Fourier transforms over symmetric groups},
  url = {http://dx.doi.org/10.1145/258533.258548},
  DOI = {10.1145/258533.258548},
  booktitle = {Proceedings of the twenty-ninth annual ACM symposium on Theory of computing  - STOC ’97},
  publisher = {ACM Press},
  author = {Beals,  Robert},
  year = {1997},
  pages = {48–53},
  collection = {STOC ’97}
}

@article{gottesman1998theory,
  title={Theory of fault-tolerant quantum computation},
  author={Gottesman, Daniel},
  journal={Physical Review A},
  volume={57},
  number={1},
  pages={127},
  year={1998},
  publisher={APS}
}

@misc{baker2026efficientquantumcircuitscoherent,
      title={Efficient Quantum Circuits for Coherent Conversion Between General First- and Second-Quantized Many-Body Representations}, 
      author={Jack S. Baker and Gaurav Saxena and Thi Ha Kyaw},
      year={2026},
      eprint={2606.25029},
      archivePrefix={arXiv},
      primaryClass={quant-ph},
      url={https://arxiv.org/abs/2606.25029}, 
}

@misc{TEcostnote,
  note = {We adopt the conventional circuit-level decomposition of one
  Toffoli into seven $T$ gates when defining the Toffoli-equivalent metric
  used here. This should be regarded only as a simple normalization
  convention rather than an architecture-independent physical cost ratio.
  More optimized implementations can realize a Toffoli using four $T$ gates,
  while dedicated Toffoli-state distillation can make a
  Toffoli magic state cost approximately twice that of a $T$-state in
  appropriate fault-tolerant architectures.
  Conversely, when $T$-state factories optimized for moderate magic-state
  demand are employed, a four-$T$ decomposition
  provides a better comparison. We retain the naive seven-$T$
  conversion throughout solely to provide a uniform and readily reproducible
  resource metric across all circuits considered.}
}

@article{Stoilova_2013,
doi = {10.1088/1751-8113/46/47/475202},
url = {https://doi.org/10.1088/1751-8113/46/47/475202},
year = {2013},
month = {nov},
publisher = {IOP Publishing},
volume = {46},
number = {47},
pages = {475202},
author = {Stoilova, N I},
title = {The parastatistics Fock space and explicit Lie superalgebra representations},
journal = {Journal of Physics A: Mathematical and Theoretical}
}

@misc{burkat2025quantumpaldustransformefficient,
      title={The Quantum Paldus Transform: Efficient Circuits with Applications}, 
      author={Jędrzej Burkat and Nathan Fitzpatrick},
      year={2025},
      eprint={2506.09151},
      archivePrefix={arXiv},
      primaryClass={quant-ph},
      url={https://arxiv.org/abs/2506.09151}, 
}

@book{Knapp2023,
  author    = {Anthony W. Knapp},
  title     = {Lie Groups: Beyond an Introduction},
  edition   = {Digital Second Edition},
  year      = {2023},
  publisher = {Self published},
  doi       = {10.3792/euclid/9798989504206},
  isbn      = {979-8-9895042-0-6},
  url       = {https://www.math.stonybrook.edu/~aknapp/download/Beyond2.pdf},
  note      = {Cartan--Weyl description for \emph{SU}(d): see Section II.1 (Classical root-space decompositions)}
}

@book{GoodmanWallach1998,
  author    = {Roe Goodman and Nolan R. Wallach},
  title     = {Representations and Invariants of the Classical Groups},
  series    = {Encyclopedia of Mathematics and its Applications},
  volume    = {68},
  publisher = {Cambridge University Press},
  address   = {Cambridge},
  year      = {1998},
  isbn      = {0521582733},
  note      = {For a Schur--Weyl duality treatment, see Ch.~10 ``Commuting Algebras on Tensor Spaces''.}
}

@article{Rowe2012,
  title = {Dual pairing of symmetry and dynamical groups in physics},
  volume = {84},
  ISSN = {1539-0756},
  url = {http://dx.doi.org/10.1103/RevModPhys.84.711},
  DOI = {10.1103/revmodphys.84.711},
  number = {2},
  journal = {Reviews of Modern Physics},
  publisher = {American Physical Society (APS)},
  author = {Rowe,  D. J. and Carvalho,  M. J. and Repka,  J.},
  year = {2012},
  month = may,
  pages = {711–757}
}

@article{ITZYKSON1966,
  title = {Unitary Groups: Representations and Decompositions},
  volume = {38},
  ISSN = {0034-6861},
  url = {http://dx.doi.org/10.1103/RevModPhys.38.95},
  DOI = {10.1103/revmodphys.38.95},
  number = {1},
  journal = {Reviews of Modern Physics},
  publisher = {American Physical Society (APS)},
  author = {Itzykson, C. and Nauenberg, M.},
  year = {1966},
  month = jan,
  pages = {95–120}
}

@article{Mathur2002,
  title = {SU(N) coherent states},
  volume = {43},
  ISSN = {1089-7658},
  url = {http://dx.doi.org/10.1063/1.1513651},
  DOI = {10.1063/1.1513651},
  number = {11},
  journal = {Journal of Mathematical Physics},
  publisher = {AIP Publishing},
  author = {Mathur,  Manu and Mani,  H. S.},
  year = {2002},
  month = nov,
  pages = {5351–5364}
}

@book{Fulton,
	author = {Fulton, William and Harris, Joe},
	title = {{Representation Theory}},
	journal = {SpringerLink},
	isbn = {978-1-4612-0979-9},
	publisher = {Springer},
	address = {New York, NY, USA},
	url = {https://link.springer.com/book/10.1007/978-1-4612-0979-9},
    year = {2004}
}

@book{Sengupta,
	author = {Sengupta, Ambar N.},
	title = {{Representing Finite Groups}},
	journal = {SpringerLink},
	isbn = {978-1-4614-1231-1},
	publisher = {Springer},
	address = {New York, NY, USA},
	url = {https://link.springer.com/book/10.1007/978-1-4614-1231-1},
    year = {2012}
}

@article{BaconChuangHarrow2006,
  author    = {Dave Bacon and Isaac L. Chuang and Aram W. Harrow},
  title     = {Efficient Quantum Circuits for Schur and Clebsch--Gordan Transforms},
  journal   = {Phys. Rev. Lett.},
  volume    = {97},
  number    = {17},
  pages     = {170502},
  year      = {2006},
  eprint    = {quant-ph/0601001}
}

@article{Barenco1995,
  author    = {Adriano Barenco and Charles H. Bennett and Richard Cleve and David P. DiVincenzo and Norman Margolus and Peter Shor and Tycho Sleator and John A. Smolin and Harald Weinfurter},
  title     = {Elementary gates for quantum computation},
  journal   = {Phys. Rev. A},
  volume    = {52},
  number    = {5},
  pages     = {3457--3467},
  year      = {1995}
}

@misc{Cuccaro2004,
  author    = {Steven A. Cuccaro and Thomas G. Draper and Samuel A. Kutin and David Petrie Moulton},
  title     = {A new quantum ripple-carry addition circuit},
  howpublished = {arXiv:quant-ph/0410184},
  year      = {2004}
}

@article{Gidney2018,
  author    = {Craig Gidney},
  title     = {Halving the cost of quantum addition},
  journal   = {Quantum},
  volume    = {2},
  pages     = {74},
  year      = {2018},
  eprint    = {arXiv:1709.06648}
}

@book{BrentZimmermann,
  author    = {Richard P. Brent and Paul Zimmermann},
  title     = {Modern Computer Arithmetic},
  publisher = {Cambridge University Press},
  year      = {2010}
}

@article{Volder1959,
  author    = {Jack E. Volder},
  title     = {The CORDIC trigonometric computing technique},
  journal   = {IRE Trans. Electronic Computers},
  volume    = {EC-8},
  number    = {3},
  pages     = {330--334},
  year      = {1959}
}

@article{Walther1971,
  author    = {J. S. Walther},
  title     = {A Unified Algorithm for Elementary Functions},
  journal   = {Spring Joint Computer Conference (AFIPS)},
  volume    = {38},
  pages     = {379--385},
  year      = {1971}
}

@article{Meher2009,
  author    = {Parhi K. Meher and Javier Valls and Tso-Bing Juang and K. Sridharan and K. Maharatna},
  title     = {50 Years of CORDIC: Algorithms, Architectures, and Applications},
  journal   = {IEEE Trans. Circuits and Systems I: Regular Papers},
  volume    = {56},
  number    = {9},
  pages     = {1893--1907},
  year      = {2009}
}

@article{Krovi2019,
  title = {An efficient high dimensional quantum Schur transform},
  volume = {3},
  ISSN = {2521-327X},
  url = {http://dx.doi.org/10.22331/q-2019-02-14-122},
  DOI = {10.22331/q-2019-02-14-122},
  journal = {Quantum},
  publisher = {Verein zur Forderung des Open Access Publizierens in den Quantenwissenschaften},
  author = {Krovi,  Hari},
  year = {2019},
  month = feb,
  pages = {122}
}

@article{BocharovRoettelerSvore2015,
  author  = {Bocharov, Alex and Roetteler, Martin and Svore, Krysta M.},
  title   = {Efficient Synthesis of Universal Repeat-Until-Success Quantum Circuits},
  journal = {Physical Review Letters},
  volume  = {114},
  number  = {8},
  pages   = {080502},
  year    = {2015},
  doi     = {10.1103/PhysRevLett.114.080502},
  eprint  = {1404.5320},
  archivePrefix = {arXiv}
}

@article{Biane1998,
  author    = {Philippe Biane},
  title     = {Representations of unitary groups and free probability},
  journal   = {Advances in Mathematics},
  year      = {1998},
  volume    = {138},
  number    = {1},
  pages     = {126--181},
  doi       = {10.1006/aima.1998.1755}
}

@article{BerryGidneyMottaMcCleanBabbush2019,
  title   = {Qubitization of Arbitrary Basis Quantum Chemistry to Exponential Precision},
  author  = {Berry, Dominic W. and Gidney, Craig and Motta, Mario and McClean, Jarrod R. and Babbush, Ryan},
  journal = {Quantum},
  volume  = {3},
  pages   = {208},
  year    = {2019},
  doi     = {10.22331/q-2019-12-02-208},
  note    = {Advanced QROM/QROAM, clean-ancilla constructions and coherent alias sampling used for PREP.}
}

@article{CollinsSniady2006,
  author    = {Beno{\^i}t Collins and Piotr {\'S}niady},
  title     = {Integration with respect to the Haar measure on unitary, orthogonal and symplectic group},
  journal   = {Communications in Mathematical Physics},
  year      = {2006},
  volume    = {264},
  number    = {3},
  pages     = {773--795},
  doi       = {10.1007/s00220-006-1554-3}
}

\beginsupplement

\noindent\textbf{\large Supplemental Material}\\ \\
\noindent\textit{Universal initial state preparation for first quantized quantum simulations}\\
Jack S. Baker$^{1}$, Gaurav Saxena$^{1}$, and Thi Ha Kyaw$^{1}$\\
$^{1}$LG Electronics Toronto AI Lab, Toronto, Ontario M5V 1M3, Canada
\vspace{6pt}

% Optional: a hyperlinked ToC in the supplement
\setcounter{tocdepth}{2}
\tableofcontents
\bigskip

\section{Overview}
This document contains supplemental material pertaining to the article
``Universal initial state preparation for first quantized quantum
simulations''. We begin in Section~\ref{sec:theory_primer} with a short
primer on the group- and representation-theoretic concepts underlying the
main article and this supplement. Section~\ref{sec:u3-example} then presents
a complete $U(3)$ example for $N=3$ particles in $d=3$ modes, listing every
Schur-basis state together with its Fock occupation, $\mathfrak{su}(3)$
Dynkin weight, Gelfand--Tsetlin (GT) label, and normalized
computational-basis expansion (Table~\ref{table:all_schur_states}); this
example exhibits the weight degeneracies and symmetric-group multiplicities
absent from the minimal $\mathfrak{su}(2)$ case of the main text.
Section~\ref{sec:dynkin-to-gt} supplies the classical preprocessing routine
invoked in the main text: a deterministic polynomial-time algorithm,
\textsc{DynkinToGT}, that maps any permissible Dynkin weight of an irrep to
a valid GT pattern of the chosen Young diagram, with explicit recovery of
standard weights and enforcement of the interlacing constraints
(Algorithm~\ref{alg:DynkinToGT}).

The resource-estimation core of the supplement develops closed-form
Toffoli-equivalent (TE) and qubit models for the two realizations of the
quantum Schur transform employed in the main text. For the
Bacon--Chuang--Harrow (BCH) construction, we detail register encodings,
two-level compilation of the reduced-Wigner operators, global error
budgeting for rotation synthesis, and the online evaluation of
reduced-Wigner coefficients via streaming reversible arithmetic with CORDIC
angle generation. A parallel model is developed in
Section~\ref{sec:krovi_resource} for the corrected Krovi--Burchardt (KB)
high-dimensional Schur transform, covering its three stages: the
preprocessing isometry, the quantum Fourier transform over the symmetric
group (for which we analyze and compare both the Beals and the
Kawano--Sekigawa constructions), and the compression isometry built from
$F$-moves, whose rotation angles we provision by online evaluation of the
closed-form $F$-symbols.

Section~\ref{sec:det-prep} then presents the first stage of our pipeline in
full: the deterministic, garbage-free preparation of a normalized
superposition of $L$ classically specified Schur-label bitstrings via the
write--erase construction $\textsc{Det-Prep}$, including a proof of the
exact index erasure, complete primitive-level costing under clean-ancilla
QROAM conventions, and the associated error budget. Finally,
Section~\ref{sec:total-cost} assembles the end-to-end accounting: the
classical preprocessing stage, the total TE cost and peak qubit footprint
of the full pipeline under both the BCH and KB realizations, their
asymptotic complexities, and a comparison of the space--time trade-off the
two variants realize across the $(N,d)$ landscape.

\section{Mathematical Preliminaries \label{sec:theory_primer}}
This section provides a brief overview of groups, group representations, and the Schur-Weyl duality, thus providing a short primer to help navigate the main article and this supplement. 
\begin{definition} A group $G$ is a set together with a law of composition and having the following properties:
\begin{enumerate}
    \item The law of composition obeys associativity, i.e., $(ab)c = a(bc),\,\forall\,a,b,c\in G$.
    \item $G$ contains an identity element $e$ such that $ea = ae = a, \, \forall\, a\in G$.
    \item Every element $a\in G$ must have an inverse $b\in G$ such that $ab=ba=e$.
\end{enumerate}
\end{definition}
A group is called \textit{abelian} if the law of composition is commutative.
The \textit{order} of a group is the number of elements it contains. If the order is finite (infinite), $G$ is said to be a finite (infinite) group.
\textit{Lie groups} are an example of infinite groups that are also continuous. Formally, they are defined as groups that are also smooth differentiable manifolds.
Every Lie group gives rise to a Lie algebra, which is a vector space $\mathfrak{g}$ equipped with an operation called the Lie bracket, which is an alternating bilinear map $\mathfrak{g}\times \mathfrak{g}\to \mathfrak{g}$ that satisfies the Jacobi identity.

The following are some of the groups used in the main text:
\begin{enumerate}
    \item $GL(d)$ denotes the $d \times d$ general linear group, the group of all $d\times d$ invertible matrices. We denote the group by $GL(d,\mathbb{R})$ or $GL(d,\mathbb{C})$ when we have to indicate that we are working with real or complex matrices, respectively.
    \item $U(d)$ denotes the unitary group, the group of all $d\times d$ unitary matrices. It is also a subgroup of $GL(d,\mathbb{C})$. 
    \item $SU(d)$ or the special unitary group, denotes a subgroup of $U(d)$ which consists of $d\times d$ unitary matrices with determinant equal to 1. 
    \item $S_n$ denotes the symmetric group, the group of permutations of the set of indices $\{1,2,\ldots,n\}$.
\end{enumerate}

Groups, as abstractly defined above, play an important role in many physical theories, especially where symmetries are present. 
To take advantage of this mathematical framework and understand more properties of the systems involved, we need to establish a correspondence between the symmetries of the physical systems and mathematical structures while faithfully preserving the underlying physics. This correspondence is known as a homomorphism, and the group structure and the symmetry properties of the physical system will be represented using matrices. We give formal definitions below.

\begin{definition}
    Given two groups $G_1$ and $G_2$, a group homomorphism is a map $f:G_1\to G_2$ that preserves the group structure. That is for any $g,h\in G_1$, the map $f$ satisfies 

\begin{equation}
    f(gh) = f(g)f(h)
\end{equation}
\end{definition}

\begin{definition}
    A matrix representation of a group $G$ is a homomorphism $f:G\to GL(d)$, mapping every element $g\in G$ to a matrix $f(g)\in GL(d)$.
\end{definition}
\noindent We will call the \textit{matrix representation} of $G$ simply as the \textit{representation} of $G$.
For a given representation $f:G\to GL(d)$, we call a subspace $\mathcal{H}(d')$ of the Hilbert space $\mathcal{H}(d)$ (where $d'\leq d$), a $G$-\textit{invariant subspace}, if for all $v\in \mathcal{H}(d')$ and all $g\in G$, the vector $f(g)\cdot v$ is also in $\mathcal{H}(d')$.
A representation is called \textit{irreducible}, or an \textit{irrep}, if the only G-invariant subspaces are the empty subspace and the entire subspace $\mathcal{H}(d)$.
For finite groups, Maschke's theorem states that every representation on a nonzero, finite-dimensional complex vector space is a direct sum of irreducible representations.
So, if $\mathcal{H}(d)$ is a direct sum of $G$-invariant subspaces, $\mathcal{H}(d_1)$ and $\mathcal{H}(d_2)$, the representation $f$ on $\mathcal{H}(d)$ is given by the direct sum of its restrictions to $\mathcal{H}(d_1)$ and $\mathcal{H}(d_2)$, and is written as $f=f_1\oplus f_2$, where $f_1$ and $f_2$ are $f$'s restrictions on $\mathcal{H}(d_1)$ and $\mathcal{H}(d_2)$, respectively. Moreover, the matrix representation $f(g)$, denoted F, will have a block form 
\begin{equation}
F = \begin{pmatrix}
F_1 & 0 \\
0 & F_2
\end{pmatrix}
\end{equation}
where $F_1$ and $F_2$ are matrix representations of $f_1$ and $f_2$, respectively.
A direct sum of irreducible (also called simple) representations is also known as a completely reducible (or a semisimple) representation.
Interested readers may find more insights in this vast subject in many standard textbooks such as~\cite{Fulton, Sengupta}. Next, we briefly discuss the Schur-Weyl duality that is central to the results of the main article.

For a given semisimple representation, we can express it in terms of unique or non-isomorphic irreducible representations by using the isotypic components. The isotypic component is the direct sum of all isomorphic subrepresentations.
Let $\{W_1, W_2,\ldots,W_k\}$ be a complete list of unique irreducible representations of $G$. We can define the isotypic component $F^{(i)}$ corresponding to each $W_i$ as 
\begin{equation}
F^{(i)}\cong W_i\otimes \mathbb{C}^{w_i}
\end{equation} 
which represents that $W_i$ has multiplicity $w_i$ in the decomposition of the representation $F$.
Then, the representation $F$ can be decomposed as
\begin{equation}
    F\cong \underset{i}{\oplus} F^{(i)}
\end{equation}
which is known as the isotypic decomposition of $F$.
Such an isotypic decomposition can be found for the representation of both the symmetric and the unitary groups.
Denoting the representation of the unitary group $U(d)$ as $V^{U(d)}$ and the representation of the symmetric group $S_n$ on the space $(\mathbb{C}^d)^{\otimes n}$ as $V^{S_n}$, the isotypic decomposition of the representation of their combined action (i.e., the group $U(d)\times S_n$) can be written as
\begin{equation}
V^{U(d)}V^{S_n}\cong \underset{i}{\oplus}\underset{j}{\oplus} V^{U(d)}_i\otimes V_j^{S_n}\otimes \mathbb{C}^{v_{i,j}}
\end{equation}
where $v_{i,j}$ denotes the multiplicity of the irrep $V^{U(d)}_i\otimes V_j^{S_n}$, and $V^{U(d)}_i$ and $V_j^{S_n}$ denote the non-isomorphic irreps of $U(d)$ and $S_n$, respectively.
Further, due to the commuting properties of the two groups, it can be shown that the multiplicities $v_{i,j}$ are either zero or one; thus, the above equation can be simplified as
\begin{equation}
    V^{U(d)}V^{S_n}\cong \underset{\lambda}{\oplus} V^{U(d)}_{\lambda}\otimes V_{\lambda}^{S_n}
\end{equation}
where $\lambda$ runs over some unspecified set.
The Schur-Weyl duality provides a characterization of the above $\lambda$ in terms of the Young diagrams with $n$ boxes and at most $d$ rows.
A detailed discussion and proof can be found in~\cite{Fulton}.

\section{Concrete example in $U(3)$}\label{sec:u3-example}

In the main text we illustrated the equivariant bijection between Fock and Schur bases in the simplest setting of two particles in two modes, where the $\mathfrak{su}(2)$ Dynkin weight states coincide with the familiar total angular momentum eigenstates. We now turn to the next nontrivial case, three particles in three modes. This example exhibits genuinely higher-rank features, most notably the degeneracy of $\mathfrak{su}(3)$ Dynkin weights, which motivates working in the Gelfand--Tsetlin (GT) basis \cite{gelfand1950finite} to resolve degeneracies by distinct GT patterns. It also allows us to display an intermediate Young diagram (neither fully symmetric nor fully antisymmetric), corresponding to parastatistics in the Jordan--Schwinger picture.

For $N=3$ and $d=3$, Schur--Weyl duality gives
\begin{equation}
(\mathbb{C}^3)^{\otimes 3}
\;\cong\;
V^{U(3)}_{(3,0,0)}\!\otimes V^{S_3}_{(3, 0, 0)}
\;\oplus\;
V^{U(3)}_{(2,1,0)}\!\otimes V^{S_3}_{(2,1, 0)}
\;\oplus\;
V^{U(3)}_{(1,1,1)}\!\otimes V^{S_3}_{(1,1,1)}.
\label{eq:schur-weyl-u3}
\end{equation}
Equivalently, in Young-diagrammatic form,
\begin{equation}
(\mathbb{C}^3)^{\otimes 3}
\;\cong\;
\left(\ydiagram{3}\,\otimes\,\ydiagram{3} \right)
\;\oplus\;
\left(\ydiagram{2,1}\,\otimes\,\ydiagram{2,1} \right)
\;\oplus\;
\left(\ydiagram{1,1,1}\,\otimes\,\ydiagram{1,1,1}\, \right).
\label{eq:young-u3}
\end{equation}
The $U(3)$ irrep dimensions are $\dim V_{(3,0,0)}=10$, $\dim V_{(2,1,0)}=8$, and $\dim V_{(1,1,1)}=1$, while the corresponding $S_3$ irreps have dimensions $1$, $2$, and $1$, respectively, so that $10\cdot1+8\cdot2+1\cdot1=27=3^3$. The shapes $(3, 0,  0)$, $(2,1, 0)$, and $(1,1,1)$ encode the fully symmetric (bosonic), mixed-symmetry (parastatistics with minimal order $p=2$), and fully antisymmetric (fermionic) sectors.

Next, let us define the concept of GT patterns. A GT pattern for $U(d)$ is a triangular array of integers with interlacing rows. Specializing to $U(3)$ with highest weight $\lambda=(\lambda_1,\lambda_2,\lambda_3)$, each basis vector is uniquely labeled by
\begin{equation}
\mu \;=\; 
\begin{array}{ccc}
\lambda_1 & \lambda_2 & \lambda_3 \\
& m_1 & m_2 \\
&& k
\end{array}
\qquad\text{with}\qquad
\lambda_1 \ge m_1 \ge \lambda_2 \ge m_2 \ge \lambda_3,\quad m_1 \ge k \ge m_2.
\label{eq:gt-u3-shape}
\end{equation}
We use the compressed notation $(x,y;\,k)\equiv(m_1,m_2;\,k)$ in Table~\ref{table:all_schur_states}. Given $(x,y;\,k)$, the associated $U(3)$ weight components (row-sum differences) are
\begin{equation}
\omega_1 = k,\qquad
\omega_2 = (x{+}y)-k,\qquad
\omega_3 = (\lambda_1{+}\lambda_2{+}\lambda_3)-(x{+}y),
\label{eq:omega-from-gt}
\end{equation}
and the corresponding $\mathfrak{su}(3)$ Dynkin weight is
\begin{equation}
z=(z_1,z_2)=(\omega_1-\omega_2,\;\omega_2-\omega_3)
=\big(2k-(x{+}y),\;2(x{+}y)-k-(\lambda_1{+}\lambda_2{+}\lambda_3)\big).
\label{eq:dynkin-from-gt}
\end{equation}
In our case $\lambda_1{+}\lambda_2{+}\lambda_3=3$, so
\begin{equation}
z=\big(2k-(x{+}y),\;2(x{+}y)-k-3\big).
\label{eq:dynkin-from-gt-u3}
\end{equation}
The inverse problem of reconstructing $(x,y;k)$ from a given $z$ is degenerate and is treated later in Section~\ref{sec:dynkin-to-gt}. The Fock-to-Dynkin map used to assign $z$ to each row is recalled in the main text: $z_i=n_i-n_{i+1}$ for occupations $\lvert n_1,n_2,n_3\rangle$.
\begin{table*}
\caption{\textbf{Schur data for $N{=}3$, $d{=}3$, grouped by $U(3)$ irrep (and $S_3$ copy where applicable).} Columns: Fock content $\lvert n_1,n_2,n_3\rangle$, Dynkin weight $z=(z_1,z_2)$, GT label $(x,y;k)$ in compressed notation, and normalized three-qutrit computational-basis expansion. Computational-basis labels follow the main-text convention $i_k\in\{1,2,3\}$, with label $p$ denoting single-particle mode $p$. Statistics: $(3,0,0)$ bosonic; $(2,1,0)$ mixed symmetry (parastatistics, minimal order $p\ge2$); $(1,1,1)$ fermionic. Here $\sigma$ indexes the $S_3$ copy; we use the subgroup-adapted Young–Yamanouchi basis, with $T_2$ antisymmetric under the adjacent swap of equal symbols and $T_1$ its orthogonal complement.}
\vspace{-0.5em}
\small
\begin{ruledtabular}
\begin{tabular}{l l l l}
Fock $\lvert n_1,n_2,n_3\rangle$& Dynkin $z$ & GT $(x,y;k)$ & \multicolumn{1}{c}{$|\lambda, \mu, \sigma \rangle_{\text{Sch}}$ in comp. basis (norm.)} \\ \hline
\multicolumn{4}{l}{\textit{$\lambda=(3,0,0)$ — fully symmetric (bosonic), $\dim=10$}}\\ \hline
$\lvert 0,0,3\rangle$ & $(0,-3)$ & $(0,0;0)$ & $\lvert 333\rangle$ \\
$\lvert 0,1,2\rangle$ & $(-1,-1)$ & $(1,0;0)$ & $\dfrac{1}{\sqrt{3}}\!\left(\lvert 233\rangle{+}\lvert 323\rangle{+}\lvert 332\rangle\right)$ \\
$\lvert 1,0,2\rangle$ & $(1,-2)$ & $(1,0;1)$ & $\dfrac{1}{\sqrt{3}}\!\left(\lvert 133\rangle{+}\lvert 313\rangle{+}\lvert 331\rangle\right)$ \\
$\lvert 0,2,1\rangle$ & $(-2,1)$ & $(2,0;0)$ & $\dfrac{1}{\sqrt{3}}\!\left(\lvert 223\rangle{+}\lvert 232\rangle{+}\lvert 322\rangle\right)$ \\
$\lvert 1,1,1\rangle$ & $(0,0)$ & $(2,0;1)$ & $\dfrac{1}{\sqrt{6}}\!\!\sum_{\pi\in S_3}\!\lvert \pi(1)\,\pi(2)\,\pi(3)\rangle$ \\
$\lvert 2,0,1\rangle$ & $(2,-1)$ & $(2,0;2)$ & $\dfrac{1}{\sqrt{3}}\!\left(\lvert 113\rangle{+}\lvert 131\rangle{+}\lvert 311\rangle\right)$ \\
$\lvert 0,3,0\rangle$ & $(-3,3)$ & $(3,0;0)$ & $\lvert 222\rangle$ \\
$\lvert 1,2,0\rangle$ & $(-1,2)$ & $(3,0;1)$ & $\dfrac{1}{\sqrt{3}}\!\left(\lvert 122\rangle{+}\lvert 212\rangle{+}\lvert 221\rangle\right)$ \\
$\lvert 2,1,0\rangle$ & $(1,1)$ & $(3,0;2)$ & $\dfrac{1}{\sqrt{3}}\!\left(\lvert 112\rangle{+}\lvert 121\rangle{+}\lvert 211\rangle\right)$ \\
$\lvert 3,0,0\rangle$ & $(3,0)$ & $(3,0;3)$ & $\lvert 111\rangle$ \\ \hline
\multicolumn{4}{l}{\textit{$\lambda=(2,1,0)$, $\sigma=T_2$ — mixed symmetry (parastatistics $p\ge2$), $\dim=8$}}\\ \hline
$\lvert 1,2,0\rangle$ & $(-1,2)$ & $(2,1;1)$ & $\dfrac{1}{\sqrt{2}}\!\left(\lvert 221\rangle{-}\lvert 212\rangle\right)$ \\
$\lvert 2,1,0\rangle$ & $(1,1)$ & $(2,1;2)$ & $\dfrac{1}{\sqrt{2}}\!\left(\lvert 112\rangle{-}\lvert 121\rangle\right)$ \\
$\lvert 0,2,1\rangle$ & $(-2,1)$ & $(2,0;0)$ & $\dfrac{1}{\sqrt{2}}\!\left(\lvert 223\rangle{-}\lvert 232\rangle\right)$ \\
$\lvert 2,0,1\rangle$ & $(2,-1)$ & $(2,0;2)$ & $\dfrac{1}{\sqrt{2}}\!\left(\lvert 113\rangle{-}\lvert 131\rangle\right)$ \\
$\lvert 0,1,2\rangle$ & $(-1,-1)$& $(1,0;0)$ & $\dfrac{1}{\sqrt{2}}\!\left(\lvert 233\rangle{-}\lvert 323\rangle\right)$ \\
$\lvert 1,0,2\rangle$ & $(1,-2)$ & $(1,0;1)$ & $\dfrac{1}{\sqrt{2}}\!\left(\lvert 133\rangle{-}\lvert 313\rangle\right)$ \\
$\lvert 1,1,1\rangle$ & $(0,0)$ & $(2,0;1)$ & $\dfrac{1}{\sqrt{2}}\!\left(\lvert 123\rangle{-}\lvert 231\rangle\right)$ \\
$\lvert 1,1,1\rangle$ & $(0,0)$ & $(1,1;1)$ & $\dfrac{1}{\sqrt{2}}\!\left(\lvert 132\rangle{-}\lvert 321\rangle\right)$ \\ \hline
\multicolumn{4}{l}{\textit{$\lambda=(2,1,0)$, $\sigma=T_1$ — mixed symmetry (parastatistics $p\ge2$), $\dim=8$}}\\ \hline
$\lvert 1,2,0\rangle$ & $(-1,2)$ & $(2,1;1)$ & $\dfrac{1}{\sqrt{6}}\!\left(\lvert 221\rangle{+}\lvert 212\rangle{-}2\lvert 122\rangle\right)$ \\
$\lvert 2,1,0\rangle$ & $(1,1)$ & $(2,1;2)$ & $\dfrac{1}{\sqrt{6}}\!\left(\lvert 112\rangle{+}\lvert 121\rangle{-}2\lvert 211\rangle\right)$ \\
$\lvert 0,2,1\rangle$ & $(-2,1)$ & $(2,0;0)$ & $\dfrac{1}{\sqrt{6}}\!\left(\lvert 223\rangle{+}\lvert 232\rangle{-}2\lvert 322\rangle\right)$ \\
$\lvert 2,0,1\rangle$ & $(2,-1)$ & $(2,0;2)$ & $\dfrac{1}{\sqrt{6}}\!\left(\lvert 113\rangle{+}\lvert 131\rangle{-}2\lvert 311\rangle\right)$ \\
$\lvert 0,1,2\rangle$ & $(-1,-1)$& $(1,0;0)$ & $\dfrac{1}{\sqrt{6}}\!\left(\lvert 233\rangle{+}\lvert 323\rangle{-}2\lvert 332\rangle\right)$ \\
$\lvert 1,0,2\rangle$ & $(1,-2)$ & $(1,0;1)$ & $\dfrac{1}{\sqrt{6}}\!\left(\lvert 133\rangle{+}\lvert 313\rangle{-}2\lvert 331\rangle\right)$ \\
$\lvert 1,1,1\rangle$ & $(0,0)$ & $(2,0;1)$ & $\dfrac{1}{\sqrt{6}}\!\left(\lvert 123\rangle{+}\lvert 231\rangle{-}2\lvert 312\rangle\right)$ \\
$\lvert 1,1,1\rangle$ & $(0,0)$ & $(1,1;1)$ & $\dfrac{1}{\sqrt{6}}\!\left(\lvert 132\rangle{+}\lvert 321\rangle{-}2\lvert 213\rangle\right)$ \\ \hline
\multicolumn{4}{l}{\textit{$\lambda=(1,1,1)$ — fully antisymmetric (fermionic), $\dim=1$}}\\ \hline
$\lvert 1,1,1\rangle$ & $(0,0)$ & $(1,1;1)$ & $\dfrac{1}{\sqrt{6}}\!\left(\lvert 123\rangle{-}\lvert 132\rangle{-}\lvert 213\rangle{+}\lvert 231\rangle{+}\lvert 312\rangle{-}\lvert 321\rangle\right)$ \\
\end{tabular}
\label{table:all_schur_states}
\end{ruledtabular}
\end{table*}
Table~\ref{table:all_schur_states} lists all Schur-basis states $\lvert\lambda,\mu,\sigma\rangle_{\text{Schur}}$ for $N=3$, $d=3$, grouped by $U(3)$ irrep (and by the two standard copies $\sigma=T_1,T_2$ when $\lambda=(2,1,0)$). For each row we display the Fock content, its Dynkin weight $z$, the GT label $(x,y;k)$ and the corresponding normalized three-qutrit computational-basis superposition. In the bosonic sector $\lambda=(3,0,0)$ one obtains the permanents of the occupied modes; in the fermionic sector $\lambda=(1,1,1)$ the unique state is the Slater determinant. Intermediate shapes realize \emph{immanants} of the single-particle orbital matrix, interpolating between these two extremes. The zero-weight subspace for $\lambda=(2,1,0)$ is two-fold degenerate; GT labels $(2,0;1)$ and $(1,1;1)$ resolve this degeneracy, and the $S_3$ multiplicity label $\sigma\in\{T_1,T_2\}$ distinguishes the two standard copies. While these choices leave the underlying Fock vector $\lvert1,1,1\rangle$ unchanged, they are essential inputs to the inverse quantum Schur transform used later.

\section{From Dynkin Weights to Gelfand--Tsetlin Patterns}\label{sec:dynkin-to-gt}

We describe a deterministic, polynomial-time procedure that maps a permissible Dynkin weight $z$ of an $\mathfrak{su}(d)$ irrep with highest weight $\zeta$ (equivalently, Young diagram $\lambda$ with at most $d$ rows) to a valid Gelfand--Tsetlin (GT) pattern of shape $\lambda$. The procedure returns one GT pattern among the (possibly many) patterns associated with a degenerate weight, which is sufficient for our purposes.

Let $\zeta=(\zeta_1,\dots,\zeta_{d-1})$ denote the $\mathfrak{su}(d)$ Dynkin labels of the highest weight, and let $\lambda=(\lambda_1,\dots,\lambda_d)$ be the corresponding partition (Young diagram row lengths), related by
\begin{equation}
\lambda_d = 0,\qquad
\lambda_i = \zeta_i + \lambda_{i+1}\quad (i=d{-}1,\dots,1).
\label{eq:lambda-from-zeta}
\end{equation}
Given a Dynkin weight $z=(z_1,\dots,z_{d-1})$ within the irrep, write the corresponding standard weight components as $\boldsymbol{\omega}=(\omega_1,\dots,\omega_d)$, defined by $z_i=\omega_i-\omega_{i+1}$ together with the total-sum constraint $\sum_{i=1}^d \omega_i=\sum_{i=1}^d \lambda_i$. Solving,
\begin{equation}
\omega_d \;=\; \frac{1}{d}\!\left(\sum_{i=1}^{d}\lambda_i - \sum_{i=1}^{d-1} i\,z_i\right),\qquad
\omega_i \;=\; z_i+\omega_{i+1}\ (i=d{-}1,\dots,1).
\label{eq:omega-recovery}
\end{equation}
For a GT pattern with top row $\lambda$ and rows $x_{r,1}\ge\cdots\ge x_{r,r}$ ($r=1,\dots,d$), the interlacing constraints are
\begin{equation}
x_{r+1,j}\ \ge\ x_{r,j}\ \ge\ x_{r+1,j+1}\qquad (r=1,\dots,d{-}1;\ j=1,\dots,r),
\label{eq:interlace}
\end{equation}
and the row-sum differences encode the weight:
\begin{equation}
\sum_{j=1}^{d}x_{d,j}-\sum_{j=1}^{d-1}x_{d-1,j}=\omega_d,\quad
\sum_{j=1}^{d-1}x_{d-1,j}-\sum_{j=1}^{d-2}x_{d-2,j}=\omega_{d-1},\ \dots,\ 
x_{1,1}=\omega_1.
\label{eq:row-sum-diffs}
\end{equation}

\begin{algorithm}[H]
\caption{\textsc{DynkinToGT}$(\texttt{Highest},\,z)$: produce one GT pattern for weight $z$}
\label{alg:DynkinToGT}
\begin{algorithmic}[1]
\Require Either $\texttt{Highest}=\zeta\in\mathbb{Z}_{\ge0}^{d-1}$ (Dynkin) or $\texttt{Highest}=\lambda\in\mathbb{Z}_{\ge0}^{d}$ (partition, nonincreasing); a permissible Dynkin weight $z\in\mathbb{Z}^{d-1}$.
\Ensure A GT pattern $\{x_{r,j}\}$ with top row $\lambda$ that realizes weight $z$.
\State \textbf{if} $\texttt{Highest}$ has length $d{-}1$ \textbf{then} compute $\lambda$ from $\zeta$ via $\lambda_d\!\gets\!0$, $\lambda_i\!\gets\!\zeta_i+\lambda_{i+1}$ \textbf{else} $\lambda\!\gets\!\texttt{Highest}$.
\State Sort $\lambda$ nonincreasing; set $T\gets \sum_{i=1}^{d}\lambda_i$.
\State \textbf{(recover standard weights)} $\omega_d \gets \big(T - \sum_{i=1}^{d-1} i\,z_i\big)/d$; assert $\omega_d\in\mathbb{Z}$. For $i=d{-}1$ down to $1$, set $\omega_i \gets z_i+\omega_{i+1}$.
\State \textbf{(target row sums)} Set $S_1\gets T$ and for $r=2$ to $d$ let $S_r \gets S_{r-1} - \omega_{d-r+2}$.
\State Initialize top row $x_{d,j}\gets \lambda_j$ for $j=1,\dots,d$ and append row $d$ to output.
\For{$r\gets d{-}1$ down to $1$} \Comment{Build row $r$ from row $r{+}1$}
  \State \textbf{baseline} $x_{r,j}\gets x_{r+1,j+1}$ for $j=1,\dots,r$.
  \State $\Delta \gets S_r - \sum_{j=1}^{r} x_{r,j}$.
  \For{$j\gets 1$ \textbf{to} $r$}
    \State $u \gets x_{r+1,j} - x_{r+1,j+1}$ \Comment{max allowed increment at position $j$}
    \State $\mathrm{inc} \gets \min\{u,\ \Delta\}$; $x_{r,j}\gets x_{r,j} + \mathrm{inc}$; $\Delta\gets \Delta - \mathrm{inc}$.
    \If{$\Delta=0$} \textbf{break} \EndIf
  \EndFor
  \State assert $\Delta=0$; append row $r$ to output.
\EndFor
\State \Return GT rows $\{x_{d,1{:}d}\,,\,x_{d-1,1{:}d-1}\,,\,\dots\,,\,x_{1,1}\}$.
\end{algorithmic}
\end{algorithm}

The construction enforces interlacing at every step. The baseline row obtained by shifting the row above to the right is the pointwise minimal interlacing choice; the remaining shortfall to the target sum is a single nonnegative integer $\Delta$, which is distributed left-to-right subject to the local capacities $x_{r+1,j}-x_{r+1,j+1}$. Each increment respects $x_{r+1,j}\ge x_{r,j}\ge x_{r+1,j+1}$, and once $\Delta$ is exhausted the row sum equals $S_r$. Induction from the top row shows that all rows interlace and the row-sum differences match $\boldsymbol{\omega}$, hence the resulting GT pattern realizes the desired Dynkin weight $z$. When $z$ is degenerate, this rule deterministically selects one of the admissible patterns.

\section{Non-Clifford resource estimation for the BCH Schur transform}
We now derive a non-clifford resource model used for resource estimations in the main text. The first routine we cover is the inverse quantum Schur transform, following the Bacon-Chuang-Harrow (BCH) variant~\cite{bacon2005quantumschurtransformi, BaconChuangHarrow2006}. The transform maps between the computational basis on $(\mathbb{C}^d)^{\otimes N}$ and the Schur basis $\{\lvert \lambda,\mu,\sigma\rangle\}$ by a cascade of $N{-}1$ Clebsch--Gordan (CG) steps. We formalize the forward pass (computational $\to$ Schur); the inverse (Schur $\to$ computational), needed for state preparation, is assumed to consume the same quantum computational resources. A high level algorithm for the forward transform is given in Algorithm \ref{alg:BCH-forward}.

\begin{algorithm}[H]
\caption{\textsc{BCH-Schur-Transform}$(d,N,\varepsilon)$: forward (computational $\to$ Schur)}
\label{alg:BCH-forward}
\begin{algorithmic}[1]
\Require Qudit registers $I_1,\ldots,I_N \in \{1,\ldots,d\}$ (computational basis).
\Require Target overall diamond-norm error $\varepsilon\in(0,1/2)$ for the full transform. This $\varepsilon$ also determines the online arithmetic precision used to compute rotation angles.
\Ensure Schur registers $\lvert \lambda\rangle$ (Young diagram), $\lvert \mu\rangle$ (Gelfand--Tsetlin pattern), and $\lvert \sigma\rangle$ (Young--Yamanouchi multiplicity for $S_N$) stored as the uncompressed path $(j_1,\ldots,j_{N-1})$.
\State Initialize $U(d)$ irrep label $\lvert \lambda_1\rangle\gets\lvert(1,0,\ldots,0)\rangle$.
\State Initialize $U(d)$ internal state $\lvert q_1\rangle\gets\lvert I_1\rangle$.
\For{$t=1$ {\bf to} $N{-}1$}
  \State Compute allowed add-one-box rows of $\lambda_t$; let $m_t\in\{1,\ldots,d\}$ be the count.
  \State Apply the $U_d$ CG transform
  \begin{equation}
  U_{\mathrm{CG}}^{(\lambda_t)}:\ 
  \lvert \lambda_t\rangle\,\lvert q_t\rangle\,\lvert I_{t+1}\rangle
  \mapsto
  \sum_{j_t=1}^{m_t}
  U_{\mathrm{CG}}^{(\lambda_t)}(j_t)\;
  \lvert j_t\rangle\,\lvert \lambda_{t+1}\rangle\,\lvert q_{t+1}\rangle,
  \label{eq:cg-map}
  \end{equation}
  with $\lambda_{t+1}=\lambda_t+e_{j_t}$ (add one box in row $j_t$).
\EndFor
\State Output $\lvert \lambda\rangle\gets\lvert \lambda_N\rangle$, $\lvert \mu\rangle\gets\lvert q_N\rangle$, and the path $\lvert \sigma\rangle\equiv\lvert j_1,\ldots,j_{N-1}\rangle$ (uncompressed).
\end{algorithmic}
\end{algorithm}

We now give the resource model and a stepwise Toffoli-equivalent (TE) accounting that depends explicitly on $(d,N,\varepsilon)$. Throughout, TE equals the number of Toffolis plus the number of synthesized $T$ gates divided by $7$ (the number of T-gates used in the standard ancilla-free decomposition of a Toffoli gate).

\subsection{Register encodings \label{subsec:registers}}

Let $n_d:=\lceil\log_2 d\rceil$. A direct encoding uses
\begin{equation}
n_\lambda^{\mathrm{naive}}=d\,\lceil\log_2(N{+}1)\rceil,\qquad
n_\mu^{\mathrm{naive}}=\tfrac{d(d-1)}{2}\,\lceil\log_2(N{+}1)\rceil,\qquad
n_\sigma=(N{-}1)\,\lceil\log_2 d\rceil,
\label{eq:registers-naive}
\end{equation}
and $N\,n_d$ qubits to hold the $N$ output qudits when running the inverse transform. To reduce control width, we adopt compressed encodings. If $p_d(N)$ is the number of partitions of $N$ with at most $d$ parts, then
\begin{equation}
n_\lambda^{\mathrm{comp}}=\lceil\log_2 p_d(N)\rceil.
\label{eq:nlambda-comp-again}
\end{equation}
Algorithm~\ref{alg:pdN} gives a simple classical dynamic program for evaluating $p_d(N)$.

\begin{algorithm}[H]
\caption{Exact count of partitions with at most $d$ parts}
\label{alg:pdN}
\begin{algorithmic}[1]
\Require Nonnegative integers $N,d$
\Ensure $p_d(N)$
\State \textbf{let} \texttt{ways}[0..$N$] $\gets$ 0 \Comment{$\texttt{ways}[n]$ will hold the number of ways to sum to $n$}
\State \texttt{ways}[0] $\gets$ 1 \Comment{Empty partition}
\For{$i \gets 1$ \textbf{to} $d$} \Comment{Allow parts of size $i$}
  \For{$n \gets i$ \textbf{to} $N$}
    \State \texttt{ways[$n$]} $\gets$ \texttt{ways[$n$]} $+$ \texttt{ways[$n-i$]}
  \EndFor
\EndFor
\State \Return \texttt{ways[$N$]}
\Statex
\Statex \textbf{Time/Space:} $O(Nd)$ integer operations, $O(N)$ memory.
\end{algorithmic}
\end{algorithm}

For the $U(d)$-irrep label, the Weyl dimension formula gives
\begin{equation}
\dim_{U(d)}(\lambda)=\prod_{1\le i<j\le d}\frac{\lambda_i-\lambda_j+j-i}{\,j-i\,}.
\label{eq:weyl}
\end{equation}
Therefore a qubit register covering all shapes has size
\begin{equation}
n_\mu^{\mathrm{comp}}
=\Big\lceil \log_2 \max_{\lambda\vdash N,\ \ell(\lambda)\le d}\ \dim_{U(d)}(\lambda)\Big\rceil .
\label{eq:nmu-comp-again}
\end{equation}

While one can enumerate all $\lambda$ and evaluate Eq.~\ref{eq:weyl} in polynomial time, a concise and asymptotically faithful proxy is obtained by evaluating the Weyl formula on the \emph{balanced} (almost rectangular) diagram \cite{Biane1998, CollinsSniady2006}, whose parts differ by at most one. Writing
\begin{equation}
N=qd+r,\quad 0\le r<d,\qquad 
\lambda^{\mathrm{bal}}(N,d)=\big(\underbrace{q{+}1,\ldots,q{+}1}_{r},\ \underbrace{q,\ldots,q}_{d-r}\big),
\label{eq:balanced-shape}
\end{equation}
the Weyl dimension reduces to a closed form that depends only on $(r,d)$:
\begin{equation}
\dim_{U(d)}\!\big(\lambda^{\mathrm{bal}}\big)
=\prod_{i=1}^{r}\ \prod_{j=r+1}^{d}\ \frac{j-i+1}{\,j-i\,},
\label{eq:weyl-balanced}
\end{equation}
and we take
\begin{equation}
n_\mu^{\mathrm{bal}}(N,d)=\Big\lceil \log_2 \dim_{U(d)}\!\big(\lambda^{\mathrm{bal}}(N,d)\big)\Big\rceil .
\label{eq:nmu-balanced}
\end{equation}

\subsection{Step 0: top level}

The transform is a cascade of $(N{-}1)$ CG steps. A conservative upper bound is given by considering the worst case TE count at each $U_d$ CG step:
\begin{equation}
\mathrm{TE}\!\left(\mathrm{Schur}(d,N,\varepsilon)\right)\ \le\ (N{-}1)\ \mathrm{TE}\!\left(U_d\text{-CG}(d,N,\varepsilon)\right).
\label{eq:schur-top}
\end{equation}
A high level quantum circuit for the transformation is shown in Fig. \ref{fig:BCH_high_level}
\begin{figure}
    \centering
    \includegraphics[width=0.6\linewidth]{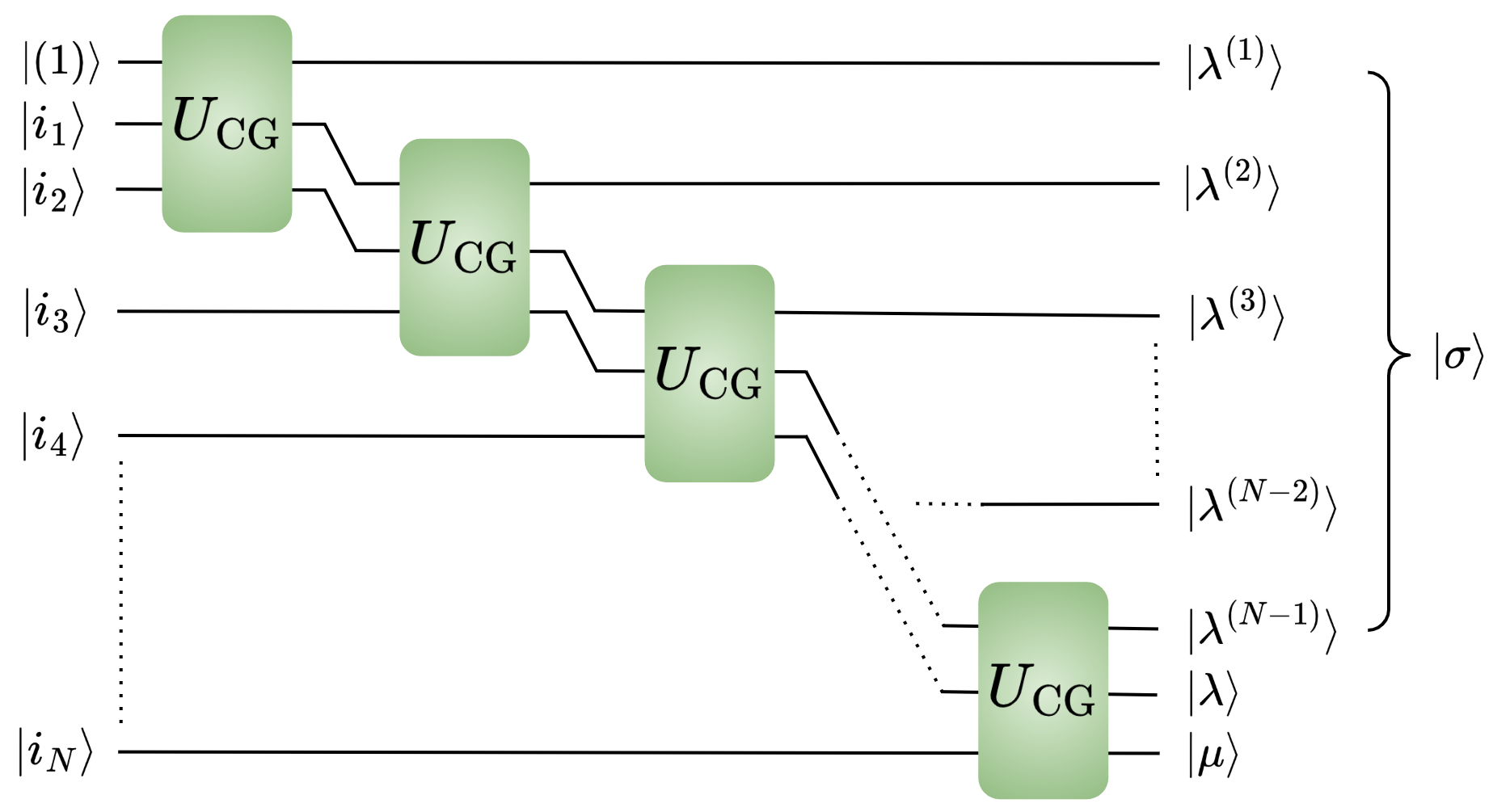}
    \caption{Abstract quantum circuit diagram implementing the BCH quantum Schur transform. A cascade of $N - 1$ $U_{\text{CG}}$ circuits is applied, resulting in a history (corresponding to $|\sigma \rangle$), young diagram $|\lambda \rangle$ and GT pattern register $| \mu \rangle$.}
    \label{fig:BCH_high_level}
\end{figure}

\subsection{Step 1: $U_d$ CG as a recursion in the rank}

Each $U_d$ CG factors into levels $s=d,d{-}1,\ldots,2$ (processed in that order). At level $s$ there is a label-controlled reduced-Wigner operator $\widehat T^{[s]}$ acting on an $s$-dimensional add-row register, followed by a recursive call to $U_{s-1}$. Thus
\begin{equation}
\mathrm{TE}\!\left(U_d\text{-CG}(d,N,\varepsilon)\right)=\sum_{s=2}^{d}\mathrm{TE}_s(d,N,\varepsilon).
\label{eq:ud-sum}
\end{equation}

\subsection{Step 2: two-level compilation, global error budget, and synthesis tolerances}

At fixed level $s$, the $s\times s$ unitary $\widehat T^{[s]}$ on the add-row register is compiled into
\begin{equation}
M_s=\binom{s}{2},
\label{eq:twolevel-count}
\end{equation}
two-level operations (for example via Givens/QR decomposition). We encode the $s$-level register into $n_s=\lceil\log_2 s\rceil$ qubits and realize each two-level operation along a Gray path of Hamming length $h\le n_s$, worst case $h=n_s$, which yields per two-level operation
\begin{equation}
E_s=2n_s-1
\label{eq:enables}
\end{equation}
address-selective, label-controlled single-qubit rotations. Let $k_{\lambda\mu}$ be the control width from the $(\lambda,\mu)$ registers; let $k_{\mathrm{addr}}(s)=n_s-1$ be the address-control width; and set
\begin{equation}
k_{\mathrm{tot}}(s)=k_{\lambda\mu}+k_{\mathrm{addr}}(s).
\label{eq:ktot}
\end{equation}
We implement $k$-controlled $X$ gates with the clean-ancilla linear construction of~\cite{Barenco1995}, which has Toffoli cost
\begin{equation}
C_{\mathrm{Tof}}(k;a)\le 2k-3\quad \text{for } a\ge k-2\ \text{ clean ancillas}.
\label{eq:mcx}
\end{equation}

We split the global error budget $\varepsilon$ between rotation synthesis and online arithmetic (detailed later) as
\begin{equation}
\varepsilon \;=\; \varepsilon_{\mathrm{rot}} \;+\; \varepsilon_{\mathrm{arith}}, 
\qquad \varepsilon_{\mathrm{rot}}=\tfrac{1}{2}\varepsilon,\ \ \varepsilon_{\mathrm{arith}}=\tfrac{1}{2}\varepsilon .
\label{eq:eps-split}
\end{equation}
Across one CG step, the number of controlled single-qubit rotation invocations is
\begin{equation}
K_{\mathrm{rot}}^{\mathrm{(CG)}} \;=\; \sum_{s=2}^{d} \binom{s}{2}\,E_s ,
\label{eq:Krot-per-CG}
\end{equation}
and across the $(N{-}1)$ CG steps of the full transform,
\begin{equation}
K_{\mathrm{rot}} \;=\; (N-1)\sum_{s=2}^{d}\binom{s}{2}\,E_s .
\label{eq:Krot}
\end{equation}
By submultiplicativity and a telescoping expansion, a per-invocation synthesis tolerance $\delta_{\mathrm{rot}}$ accumulates at most additively, so we set
\begin{equation}
\delta_{\mathrm{rot}} \;=\; \varepsilon_{\mathrm{rot}}/K_{\mathrm{rot}} .
\label{eq:delta-rot}
\end{equation}
We adopt direct single-qubit rotation synthesis~\cite{BocharovRoettelerSvore2015} with
\begin{equation}
T_{\mathrm{dir}}(\delta)=\alpha_{\mathrm{dir}}\log_2(1/\delta)+\beta_{\mathrm{dir}},
\qquad \alpha_{\mathrm{dir}}=1.149,\ \ \beta_{\mathrm{dir}}=9.2,
\label{eq:direct-single}
\end{equation}
so a controlled rotation at tolerance $\delta_{\mathrm{rot}}$ costs
\begin{equation}
T_{\mathrm{cR}}^{\mathrm{dir}}(\delta_{\mathrm{rot}})
 \;=\; 2\,T_{\mathrm{dir}}(\delta_{\mathrm{rot}}/2)
 \;=\; 2\!\left[\alpha_{\mathrm{dir}}\log_2\!\frac{2}{\delta_{\mathrm{rot}}}+\beta_{\mathrm{dir}}\right].
\label{eq:direct-controlled}
\end{equation}
Each enable therefore consists of: flag-set via one $C_{\mathrm{Tof}}\!\big(k_{\mathrm{tot}}(s);a\big)$, a controlled rotation contributing $T_{\mathrm{cR}}^{\mathrm{dir}}(\delta_{\mathrm{rot}})$ $T$ gates (counted as $T/7$ in TE), and flag-clear via the same multi-control. The TE per two-level operation is
\begin{equation}
\widetilde T_{2\mathrm{lvl}}(s)=E_s\left[\,2\,C_{\mathrm{Tof}}\!\big(k_{\mathrm{tot}}(s);a\big)+\frac{T_{\mathrm{cR}}^{\mathrm{dir}}(\delta_{\mathrm{rot}})}{7}\right],
\label{eq:twolevel-te}
\end{equation}
and the level-$s$ compilation contribution is
\begin{equation}
\mathrm{TE}_s^{(\mathrm{comp})}=M_s\,\widetilde T_{2\mathrm{lvl}}(s)=\binom{s}{2}\,(2n_s-1)\!\left[\,2\,C_{\mathrm{Tof}}\!\big(k_{\mathrm{tot}}(s);a\big)+\frac{T_{\mathrm{cR}}^{\mathrm{dir}}(\delta_{\mathrm{rot}})}{7}\right].
\label{eq:level-compile}
\end{equation}

\subsection{Step 3: online evaluation of reduced-Wigner coefficients and angle generation}

The arithmetic that computes rotation angles must respect the remaining error budget. We budget per invocation an angle-quantization allowance
\begin{equation}
\varepsilon_\theta \;=\; \varepsilon_{\mathrm{arith}}/K_{\mathrm{rot}},
\label{eq:eps-theta}
\end{equation}
which enforces worst-case additive accumulation over all invocations. Angles are represented with $f$ fractional bits so that $\lvert\theta-\hat\theta\rvert\le \pi\,2^{-f}\le \varepsilon_\theta$, i.e.
\begin{equation}
f \;=\; \left\lceil \log_2\!\Big(\pi/\varepsilon_\theta\Big)\right\rceil
 \;=\; \left\lceil \log_2\!\Big(\tfrac{\pi\,K_{\mathrm{rot}}}{\varepsilon_{\mathrm{arith}}}\Big)\right\rceil .
\label{eq:f-fractional}
\end{equation}
Intermediate products and divisions during the evaluation of the reduced-Wigner block require guard bits to cover products of $O(s)$ factors; we take
\begin{equation}
g(s) \;=\; \left\lceil \log_2 s\right\rceil + 3,
\label{eq:guards}
\end{equation}
and bound the integer dynamic range of shifted GT differences by
\begin{equation}
i(N,s) \;=\; \left\lceil \log_2(2N+2s+1)\right\rceil + 2 .
\label{eq:int-range}
\end{equation}
The working word size (signed two's-complement) used throughout rank $s$ is
\begin{equation}
w(s) \;=\; i(N,s) \;+\; f \;+\; g(s).
\label{eq:w-of-s}
\end{equation}

We now define the reduced-Wigner elements evaluated online. With the shifted labels
\begin{equation}
\tilde\mu_i=\mu_i+(s-i),\ \ i\in\{1,\ldots,s\},\qquad
\tilde\mu'_j=\mu'_j+(s-1-j),\ \ j\in\{1,\ldots,s{-}1\},
\label{eq:shifts}
\end{equation}
and the sign $S(j{-}j')=+1$ if $j\ge j'$ and $-1$ otherwise, the entries of the $s\times(s{-}1)$ reduced-Wigner isometry $\widehat T^{[s]}$ are (BCH Eq.~(55) \cite{bacon2005quantumschurtransformi})
\begin{equation}
\widehat T^{\mu,j,\mu',j'}=
\begin{cases}
\displaystyle S(j{-}j')\left[
\frac{\Big(\prod_{i\neq j}\big(\tilde\mu_i-\tilde\mu'_{j'}+1\big)\Big)\Big(\prod_{i\neq j'}\big(\tilde\mu'_i-\tilde\mu_j+1\big)\Big)}
{\Big(\prod_{i\neq j}\big(\tilde\mu_i-\tilde\mu_j+1\big)\Big)\Big(\prod_{i\neq j'}\big(\tilde\mu'_i-\tilde\mu'_{j'}+1\big)\Big)}
\right]^{\!1/2}, & j'\in\{1,\ldots,s{-}1\},\\[4mm]
\displaystyle S(j{-}s)\left[
\frac{\prod_{i\neq j}\big(\tilde\mu_i-\tilde\mu_j\big)}
{\prod_{i=1}^{s-1}\big(\tilde\mu'_i-\tilde\mu_j\big)}
\right]^{\!1/2}, & j'=0.
\end{cases}
\label{eq:T-hat-def}
\end{equation}

We use standard reversible primitives, counting TE as number of Toffolis since these routines use only Toffoli and Clifford. CDKM ripple-carry add/subtract~\cite{Cuccaro2004} has cost
\begin{equation}
A(w)=2w-1 \quad \text{TE}, \qquad \text{ancilla: } 0,
\label{eq:A-w}
\end{equation}
schoolbook shift-and-add multiplication with $w$ single-controlled adds~\cite{Gidney2018} has
\begin{equation}
M(w)=2w^2+w \quad \text{TE}, \qquad \text{ancilla: } 0,
\label{eq:M-w}
\end{equation}
Newton reciprocal with a unit-cost seed~\cite{BrentZimmermann} uses
\begin{equation}
I_{\mathrm{rec}}(w)=\left\lceil \log_2 w\right\rceil + 2, 
\qquad C_{\mathrm{recip}}(w)=I_{\mathrm{rec}}(w)\,\big(2\,M(w)+3\,A(w)\big),
\label{eq:recip}
\end{equation}
and Newton square-root uses
\begin{equation}
I_{\sqrt{\phantom{.}}}(w)=\left\lceil \log_2 w\right\rceil + 2, 
\qquad C_{\sqrt{\phantom{.}}}(w)=I_{\sqrt{\phantom{.}}}(w)\,\Big(C_{\mathrm{recip}}(w)+M(w)+2A(w)\Big).
\label{eq:sqrt}
\end{equation}
Angles are obtained by reversible CORDIC in vectoring mode with $f$ iterations~\cite{Volder1959,Walther1971,Meher2009},
\begin{equation}
C_{\mathrm{CORDIC}}(w)=3\,f\,A(w).
\label{eq:cordic-cost}
\end{equation}

We evaluate $\widehat T^{[s]}$ once per rank using a streaming, space--time balanced plan that avoids large caches. First, we form the cross-row differences
\begin{equation}
\mathcal{D}^{(1)}_{i,t}=\tilde\mu_i-\tilde\mu'_t \ \ (i\in[s],\, t\in[s-1]),
\qquad
\mathcal{D}^{(2)}_{u,v}=\tilde\mu'_u-\tilde\mu_v \ \ (u\in[s-1],\, v\in[s]),
\label{eq:diffs-np}
\end{equation}
together with the within-row/within-column differences required in the denominators of Eq.~\ref{eq:T-hat-def}. This costs
\begin{equation}
C_{\mathrm{diff}}^{[s]} \;=\; 2\,s(s-1)\,A\!\big(w(s)\big).
\label{eq:diff-cost-np}
\end{equation}
Next, for each of the $s(s-1)$ entries (including the $j'=0$ column), we assemble the required leave-one-out products using rolling prefix/suffix accumulators rather than stored arrays. In this streaming variant, each entry consumes eight multiplies, one reciprocal, one square-root, and a constant number of adds/subtracts, i.e.
\begin{equation}
C_{\mathrm{entry}}^{\mathrm{stream}}(w)=8\,M(w)+C_{\mathrm{recip}}(w)+C_{\sqrt{\phantom{.}}}(w)+5\,A(w),
\qquad
C_{\mathrm{entries}}^{[s]}=s(s-1)\,C_{\mathrm{entry}}^{\mathrm{stream}}\!\big(w(s)\big).
\label{eq:entry-cost-stream}
\end{equation}
Finally, the compiled two-level sequence requires $\binom{s}{2}$ angles,
\begin{equation}
C_{\mathrm{angles}}^{[s]} \;=\; \binom{s}{2}\,C_{\mathrm{CORDIC}}\!\big(w(s)\big)
\;=\; \frac{3}{2}\,s(s-1)\,f\,A\!\big(w(s)\big).
\label{eq:angles-cost-np}
\end{equation}
The per-rank online evaluation cost is
\begin{equation}
C_{\mathrm{eval}}^{[s]}(s;N,\varepsilon)
= 2s(s-1)\,A\!\big(w(s)\big)
   + s(s-1)\,C_{\mathrm{entry}}^{\mathrm{stream}}\!\big(w(s)\big)
   + \tfrac{3}{2}\,s(s-1)\,f\,A\!\big(w(s)\big).
\label{eq:Ceval-s}
\end{equation}

\subsection{Per-rank and overall toffoli-equivalent cost}

The level-$s$ total is the sum of the compilation and online components,
\begin{equation}
\mathrm{TE}_s(d,N,\varepsilon) \;=\;
\binom{s}{2}\,(2n_s-1)\!\left[\,2\,C_{\mathrm{Tof}}\!\big(k_{\mathrm{tot}}(s);a\big)
+\frac{T_{\mathrm{cR}}^{\mathrm{dir}}(\delta_{\mathrm{rot}})}{7}\right]
\;+\;
C_{\mathrm{eval}}^{[s]}(s;N,\varepsilon),
\label{eq:level-total-np}
\end{equation}
and
\begin{equation}
\mathrm{TE}\!\left(U_d\text{-CG}(d,N,\varepsilon)\right)=\sum_{s=2}^{d}\mathrm{TE}_s(d,N,\varepsilon),
\qquad
\mathrm{TE}\!\left(\mathrm{Schur}(d,N,\varepsilon)\right)\le (N{-}1)\sum_{s=2}^{d}\mathrm{TE}_s(d,N,\varepsilon).
\label{eq:CG-and-Schur-np}
\end{equation}

\subsection{Ancilla accounting for arithmetic}

We use a streaming strategy for memory usage to avoid the need for a large cache. A conservative bound for the arithmetic work space at rank $s$ is
\begin{equation}
a_{\mathrm{arith}}^{[s]} \;\le\; 12\,w(s) \quad \text{qubits}
\label{eq:anc-arith-lean}
\end{equation}
as inferred by the number of aforementioned arithmetic operations required to obtain a single rotation angle. These arithmetic ancillas are separate from the clean-ancilla bank used for multi-controls in Eq.~\ref{eq:mcx}.

\subsection{Total qubit accounting}
\label{subsec:total-qubits}

We separate persistent \emph{system} registers from transient \emph{ancilla}. The system footprint (present throughout the transform) is
\begin{equation}
Q_{\mathrm{sys}}(d,N)
\;=\;
N\,n_d \;+\; n_\lambda \;+\; n_\mu \;+\; n_\sigma,
\label{eq:Qsys-final}
\end{equation}
with register sizes taken from Sec. \ref{subsec:registers}

At rank $s$ the work registers comprise three parts. (i) The $s$-level add-row register has size $n_s=\lceil\log_2 s\rceil$. (ii) Multi-controlled operations use a clean-ancilla bank for the linear-cost construction in Eq.~\ref{eq:mcx}. The minimum number that attains the $2k{-}3$ Toffoli scaling is
\begin{equation}
a_{\mathrm{mcx}}^{\min}(s)
\;=\;
\max\{0,\,k_{\mathrm{tot}}(s)-2\}
\;=\;
\max\{0,\,k_{\lambda\mu}+n_s-3\}.
\label{eq:anc-mcx-min}
\end{equation}
If a fixed bank $a_{\mathrm{mcx}}^{\mathrm{prov}}$ is provisioned and reused across all enables, the effective allocation at rank $s$ is
\begin{equation}
a_{\mathrm{mcx}}^{\mathrm{use}}(s)
\;=\;
\max\!\big\{\,a_{\mathrm{mcx}}^{\mathrm{prov}},\; a_{\mathrm{mcx}}^{\min}(s)\,\big\}.
\label{eq:anc-mcx-prov}
\end{equation}
(iii) The streaming arithmetic that evaluates $\widehat T^{[s]}$ (Step~3) requires only rolling accumulators and a constant number of temporaries; with the working word size $w(s)$ from Eq.~\ref{eq:w-of-s} we bounded the arithmetic qubits in Eq. \ref{eq:anc-arith-lean}

The transient ancilla at rank $s$ is then
\begin{equation}
Q_{\mathrm{anc}}^{[s]}
\;=\;
n_s
\;+\;
a_{\mathrm{mcx}}^{\mathrm{use}}(s)
\;+\;
a_{\mathrm{arith}}^{[s]},
\label{eq:Qanc-level-final}
\end{equation}
and the overall peak qubit demand of the quantum Schur transform is
\begin{equation}
Q_{\mathrm{total}}(d,N,\varepsilon;\,a_{\mathrm{mcx}}^{\mathrm{prov}})
\;=\;
Q_{\mathrm{sys}}(d,N)
\;+\;
\max_{2\le s\le d}\; Q_{\mathrm{anc}}^{[s]} .
\label{eq:Qtotal-final}
\end{equation}
All ancillas are returned to $\lvert 0\rangle$ by construction (uncomputation of arithmetic and flag-clears for multi-controls), so Eq.~\ref{eq:Qtotal-final} is also the peak footprint for the inverse transform used in state preparation.

\section{Non-Clifford resource estimation for the corrected Krovi Schur transform}
\label{sec:krovi_resource}

As an alternative to the Bacon--Chuang--Harrow (BCH) construction analyzed in
the preceding section, we develop a non-Clifford resource model for the
high-dimensional quantum Schur transform introduced by Krovi~\cite{Krovi2019}
and recently revised by Burchardt \emph{et al.}~\cite{burchardt2025highdimensionalquantumschurtransforms},
which fixes a crucial error in the original algorithm. In the regime $d\geq N$
relevant to first-quantized simulation, this construction replaces the
$\widetilde{\mathcal{O}}(Nd^{4})$ scaling of the BCH transform by
$\widetilde{\mathcal{O}}(N^{4})$, with only polylogarithmic dependence on the
single-particle dimension $d$. The reduction is achieved by a preprocessing
isometry that compresses the $d$-alphabet input into an $N$-alphabet
representation, after which the remaining stages operate on registers whose
widths depend only on $N$.

Following Theorem~1 and Fig.~2 of
Ref.~\cite{burchardt2025highdimensionalquantumschurtransforms}, the forward
transform factorizes as
\begin{equation}
  U_{\mathrm{Schur}}
  \;=\;
  V^{\dagger}\cdot \mathrm{QFT}_{S_{N}}\cdot P,
  \label{eq:krovi_factorization}
\end{equation}
where $P$ is the preprocessing isometry, $\mathrm{QFT}_{S_{N}}$ is the quantum
Fourier transform over the symmetric group, and $V^{\dagger}$ is the
compression isometry that rotates the split basis produced by
$\mathrm{QFT}_{S_{N}}$ into the combined Young--Yamanouchi and
Gelfand--Tsetlin bases via a sequence of $F$-moves. Because each stage is
self-inverse in resource, the inverse transform $U_{\mathrm{Schur}}^{-1}$ has
identical Toffoli-equivalent cost and peak qubit footprint to the forward
transform analysed here.

Throughout this section, Toffoli equivalents (TE) are defined as in the
preceding section,
\begin{equation}
  \mathrm{TE}
  \;=\;
  \#\mathrm{Toffoli} + (\text{T-count})/7,
\end{equation}
and the global diamond-norm error budget $\varepsilon$ is partitioned evenly
across the three stages,
\begin{equation}
  \varepsilon_{P}
  \;=\; \varepsilon_{Q}
  \;=\; \varepsilon_{V}
  \;=\; \tfrac{1}{3}\,\varepsilon.
  \label{eq:krovi_eps_split}
\end{equation}
Within each stage, the per-rotation synthesis tolerance is fixed so that the
worst-case additive diamond-norm accumulation respects the stage budget; in
Stage~3, which requires coherent arithmetic to produce its rotation angles,
the stage budget is further split evenly between rotation synthesis and
arithmetic, mirroring the convention of the BCH analysis. Arbitrary
single-qubit rotations at tolerance $\delta$ are synthesized using the direct
construction of Ref.~\cite{BocharovRoettelerSvore2015} with T-count
\begin{equation}
  T_{\mathrm{dir}}(\delta)
  \;=\;
  \alpha_{\mathrm{dir}}\,\log_{2}(1/\delta) + \beta_{\mathrm{dir}},
  \qquad
  \alpha_{\mathrm{dir}}=1.149,
  \quad
  \beta_{\mathrm{dir}}=9.2,
\end{equation}
and controlled rotations at tolerance $\delta$ cost
\begin{equation}
  T_{\mathrm{cR}}(\delta)
  \;=\;
  2\,T_{\mathrm{dir}}(\delta/2).
\end{equation}
Multi-controlled NOT gates are implemented with the clean-ancilla linear
construction of Ref.~\cite{Barenco1995} at Toffoli cost
\begin{equation}
  C_{\mathrm{Tof}}(k;a)
  \;=\;
  2k-3
  \qquad\text{for } a\geq k-2 \text{ clean ancillas},
  \quad k\geq 2.
\end{equation}
Integer addition and subtraction are performed with CDKM ripple-carry
adders~\cite{Cuccaro2004} at cost
\begin{equation}
  A(w)
  \;=\;
  2w-1
  \qquad\text{Toffolis},
\end{equation}
and comparators at width $w$ are implemented as an adder followed by sign
extraction, so that a guard computed and later uncomputed contributes
$2\,A(w)$ Toffolis. Once a one-bit guard flag is available, a flag-controlled
SWAP of a $w$-wide register pair is priced, consistently with the convention
already used for subroutine $B$ below, at $2w\,C_{\mathrm{Tof}}(3;a)$
Toffolis, a flag-controlled increment or decrement of a $w$-wide register at
$w-1$ Toffolis, and a flag-controlled adder at $A(w)$.

\subsection{Register encodings and persistent state}
\label{subsec:krovi_encodings}

Let $n_{d}:=\lceil\log_{2}d\rceil$ and $n_{N}:=\lceil\log_{2}N\rceil$. We use
the dense qubit encodings of each dit employed throughout
Ref.~\cite{burchardt2025highdimensionalquantumschurtransforms}. The registers
involved in the transform are summarised in
Table~\ref{tab:krovi_registers}. The input $\ket{x}$ is consumed by the
preprocessing isometry $P$, which produces the triple
$(\ket{p},\ket{\mu},\ket{tY_{\mu}})$; the coset register $\ket{tY_{\mu}}$ is
then consumed by $\mathrm{QFT}_{S_{N}}$, which produces the partition label
$\ket{\lambda}$, the standard Young tableau $\ket{T}$, and the
Young--Yamanouchi path $\ket{\sigma}$ labelling the $S_{N}$ irrep copy;
finally $V^{\dagger}$ consumes $\ket{T}$ and produces the compressed
Gelfand--Tsetlin pattern $\ket{\widetilde{M}}$. The persistent output
registers of the full transform are therefore $\ket{p}$, $\ket{\mu}$,
$\ket{\lambda}$, $\ket{\widetilde{M}}$, and $\ket{\sigma}$.

\begin{table}[t]
\centering
\small
\begin{tabular}{lll}
\hline
Register & Width (qubits) & Role \\
\hline
$\ket{x}$             & $N n_{d}$                   & input computational state  \\
$\ket{p}$             & $N n_{d}$                   & alphabet map (decoding vector)  \\
$\ket{\mu}$           & $N n_{N}$                   & composition (type vector)  \\
$\ket{tY_{\mu}}$      & $N n_{N}$                   & coset representative of $S_{N}/Y_{\mu}$  \\
$\ket{T}$             & $(N-1)n_{N}$                & standard Young tableau (path)  \\
$\ket{\widetilde{M}}$ & $\tfrac{N(N+1)}{2}\,n_{N}$  & compressed Gelfand--Tsetlin pattern  \\
$\ket{\lambda}$       & $n_{\lambda}$               & Young diagram (partition of $N$)  \\
$\ket{\sigma}$        & $(N-1)n_{N}$                & $S_{N}$ Young--Yamanouchi path  \\
\hline
\end{tabular}
\caption{Register encodings for the corrected Krovi Schur transform. Here
$n_{\lambda}=\lceil\log_{2}p(N)\rceil$, with $p(N)$ the number of
integer partitions of $N$. All registers are dit-packed into qubits.}
\label{tab:krovi_registers}
\end{table}

\subsection{Stage 1: Preprocessing isometry \texorpdfstring{$P$}{P}}
\label{subsec:krovi_stage1}

The preprocessing isometry decomposes recursively as
$P=\mathrm{Prep}_{1}\cdot\mathrm{Prep}_{2}\cdots\mathrm{Prep}_{N}$, where
each $\mathrm{Prep}_{n}$ absorbs the $n$-th input qudit $\ket{x_{n}}$
into the triple produced from the first $n-1$ qudits (Fig.~4 of
Ref.~\cite{burchardt2025highdimensionalquantumschurtransforms}). Internally,
$\mathrm{Prep}_{n}$ consists of six subcircuits $A,B,C,D,E,H$ executed in
sequence (Fig.~5 of the same reference), which we analyse in turn. Throughout
this subsection, $n$ denotes the current recursion index, so that all
per-step quantities carry an implicit dependence on $n$.

\paragraph{Subroutine $A$ (compute $e_{n}$ and $c_{n}$).}
Figure~6a of Ref.~\cite{burchardt2025highdimensionalquantumschurtransforms}
shows $A$ as the cascade
$A_{1}A_{2}\cdots A_{n-1}A'_{1}A'_{2}\cdots A'_{n-1}$. Each $A_{k}$ is a
comparator-controlled incrementer of width $n_{d}$ on the position register
$\ket{\cdot}_{p}$: it increments by one if $p'_{k}<x_{n}$. Each $A'_{k}$ is
an equality-controlled bit flip of the uniqueness register
$\ket{\cdot}_{u}$. Implementing each comparator as a CDKM subtractor followed
by sign extraction, and implementing the controlled $(+1)$ on the $n_{d}$-wide
position register by the standard Toffoli-chain incrementer, we obtain
\begin{align}
  \mathrm{TE}(A_{k})
  &\;=\; 2\,A(n_{d}) + (n_{d}-1)
  \;=\; 5n_{d}-3,
  \\
  \mathrm{TE}(A'_{k})
  &\;=\; 2\,A(n_{d}) + 1
  \;=\; 4n_{d}-1.
\end{align}
The factor of two in each line accounts for forward computation and
uncomputation of the comparator. Summing yields
\begin{equation}
  \mathrm{TE}(A\mid n)
  \;=\; (n-1)\bigl(9n_{d}-4\bigr).
  \label{eq:TE_A_per_n}
\end{equation}

\paragraph{Subroutine $B$ (update $\ket{p'}\ket{x_{n}}\to\ket{p}$).}
Figure~6b shows $B$ as $B'_{1}B'_{2}\cdots B'_{n-1}\,B_{n}B_{n-1}\cdots B_{2}$.
Each $B_{k}$ is a controlled SWAP of two $n_{d}$-wide data registers, guarded
by the conjunction $\{c_{n}=0\}\wedge\{k>n-e_{n}\}$. The guard is a width-$n_{N}$
comparator plus a one-bit check; with the flag-controlled SWAP convention
fixed above,
\begin{equation}
  \mathrm{TE}(B_{k})
  \;=\; 2\,A(n_{N}) + 2\,n_{d}\cdot C_{\mathrm{Tof}}(3;a)
  \;=\; 4n_{N}-2 + 6n_{d}.
\end{equation}
Each $B'_{k}$ is a guarded subtractor
$\ket{x_{n}}\to\ket{x_{n}-p'_{k}}$ of width $n_{d}$, conditioned on
$\{c_{n}=1\}\wedge\{e_{n}=k\}$, at cost
\begin{equation}
  \mathrm{TE}(B'_{k})
  \;=\; A(n_{d}) + 2\,A(n_{N})
  \;=\; 2n_{d}+4n_{N}-3.
\end{equation}
Summing,
\begin{equation}
  \mathrm{TE}(B\mid n)
  \;=\; (n-1)\bigl(8n_{d}+8n_{N}-5\bigr).
  \label{eq:TE_B_per_n}
\end{equation}

\paragraph{Subroutine $C$ (update $\ket{\mu'}\ket{0}\to\ket{\mu}$).}
Figure~6c shows $C$ as $C_{0}C_{n}C_{n-1}\cdots C_{2}\,C'_{1}C'_{2}\cdots C'_{n-1}$.
The gate $C_{0}$ is a singly controlled $X$ acting on the lowest bit of
$\ket{\mu_{n}}$ and is therefore Clifford; it contributes no Toffoli
equivalents. Each $C_{k}$ is a controlled SWAP of $n_{N}$-wide registers
guarded by $\{c_{n}=0\}\wedge\{e_{n}>n-k\}$, priced identically to the
$B_{k}$ gates but at data width $n_{N}$,
\begin{equation}
  \mathrm{TE}(C_{k})
  \;=\; 2\,A(n_{N}) + 2\,n_{N}\cdot C_{\mathrm{Tof}}(3;a)
  \;=\; 10n_{N}-2.
\end{equation}
Each $C'_{k}$ is a guarded increment of $\ket{\mu_{k}}$ conditioned on
$\{c_{n}=1\}\wedge\{e_{n}=k\}$, at cost
\begin{equation}
  \mathrm{TE}(C'_{k})
  \;=\; 2\,A(n_{N}) + (n_{N}-1)
  \;=\; 5n_{N}-3.
\end{equation}
Summing,
\begin{equation}
  \mathrm{TE}(C\mid n)
  \;=\; (n-1)\bigl(15n_{N}-5\bigr).
  \label{eq:TE_C_per_n}
\end{equation}

\paragraph{Subroutine $D$ (uncompute $c_{n}$).}
Figure~6d shows $D_{n}D_{n-1}\cdots D_{1}$, where each $D_{k}$ is a single
controlled-$X$ on $\ket{\cdot}_{u}$ guarded by
$\{e_{n}=k\}\wedge\{\mu_{k}\neq 1\}$, at cost
\begin{equation}
  \mathrm{TE}(D_{k})
  \;=\; A(n_{N}) + 1
  \;=\; 2n_{N}.
\end{equation}
Summing,
\begin{equation}
  \mathrm{TE}(D\mid n)
  \;=\; 2n\,n_{N}.
  \label{eq:TE_D_per_n}
\end{equation}

\paragraph{Subroutine $E$ (prepare coset state on transversal register).}
Subroutine $E$ (Fig.~7a of
Ref.~\cite{burchardt2025highdimensionalquantumschurtransforms}) is the only
(explicit) source of T-gates in Stage~1. It is the cascade
$E_{1}\cdots E_{n}\,E'_{1}\cdots E'_{n-1}\,E''_{1}\cdots E''_{n-1}$. Each
$E_{k}=E_{k,0}E_{k,1}\cdots E_{k,n}$ is a sequence of $n+1$
controlled two-level Givens rotations on the transversal register, with
angles determined entirely by the integer $\mu_{k}\in\{1,\ldots,N\}$
(Eqs.~(158)--(159) of
Ref.~\cite{burchardt2025highdimensionalquantumschurtransforms}). The total
number of synthesized single-qubit rotations contributed by $E$ at step $n$
is therefore $n(n+1)$, and the total across the preprocessing cascade is the
exact count
\begin{equation}
  K_{E}
  \;=\;
  \sum_{n=1}^{N} n(n+1)
  \;=\;
  \frac{N(N+1)(N+2)}{3}.
  \label{eq:KE_definition}
\end{equation}
Because the rotation angles depend only on the integer $\mu_{k}\leq N$, the
angle table has at most $N$ distinct entries; we classically precompute this
table once and load the relevant angle on each invocation via a clean-ancilla
QROAM lookup of size $N$. We assign a per-rotation synthesis tolerance
\begin{equation}
  \delta_{E}
  \;=\;
  \frac{\varepsilon_{P}}{K_{E}}
  \;=\;
  \frac{\varepsilon}{3K_{E}},
  \label{eq:delta_E}
\end{equation}
so that additive diamond-norm accumulation across the stage does not exceed
$\varepsilon_{P}$. The corresponding QROAM word width is
\begin{equation}
  R_{E}
  \;=\;
  \left\lceil \log_{2}\!\bigl(\pi K_{E}/\varepsilon_{P}\bigr)\right\rceil
  + 1,
  \label{eq:R_E}
\end{equation}
where the additional bit accommodates the sign of the rotation angle.

Each invocation of a controlled Givens $E_{k,j}$ consists of the QROAM lookup
delivering the $R_{E}$-bit angle, the controlled rotation itself synthesised
to tolerance $\delta_{E}$, and flag-set and flag-clear multi-controls of
effective width $k_{E}=n_{N}+n_{d}$ on the composite index $(\mu_{k},e_{n})$.
Denoting the near-optimal QROAM blocking parameter by $k^{\star}$ in
accordance with the convention of the preceding section, the per-rotation
cost is
\begin{equation}
  \mathrm{TE}_{E,\mathrm{rot}}
  \;=\;
  \mathrm{TE}_{\mathrm{QROAM}}(N,R_{E})
  + \frac{T_{\mathrm{cR}}(\delta_{E})}{7}
  + 2\,C_{\mathrm{Tof}}(k_{E};a),
  \label{eq:TE_E_rot}
\end{equation}
where we use the explicit lookup cost
\begin{equation}
  \mathrm{TE}_{\mathrm{QROAM}}(L,R)
  \;=\;
  \left\lceil \frac{L}{2^{k^{\star}}}\right\rceil
  + R(k^{\star}-1)
  + \left\lceil \frac{L}{2^{k^{\star}}}\right\rceil
  + k^{\star}
  \label{eq:qroam_lookup_krovi}
\end{equation}
from the preceding section, with $k^{\star}$ chosen as the near-optimal
blocking parameter for the $(L,R)$ in question. Besides the Givens rotations,
$E$ contains $n-1$ controlled integer shifts $E'_{k}$ and $n-1$
comparator-conditioned cyclic shifts $E''_{k}$. Each $E'_{k}$ adds $\mu_{k}$
to the transversal entry conditioned on the position check $e_{n}>k$; the
guard is a comparator computed and uncomputed at $2\,A(n_{N})$, and the
conditional addition is a flag-controlled adder $A(n_{N})$, giving
$\mathrm{TE}(E'_{k})=3\,A(n_{N})=6n_{N}-3$. Each $E''_{k}$ is a cyclic
increment of the register $\ket{\sigma'(k)}$ conditioned on the comparison
$\sigma'(k)\geq\ell$ against the last transversal entry (Eq.~(164) of
Ref.~\cite{burchardt2025highdimensionalquantumschurtransforms}); the
comparator costs $2\,A(n_{N})$ and the conditional increment $n_{N}-1$,
giving $\mathrm{TE}(E''_{k})=5n_{N}-3$. Per step,
\begin{equation}
  \mathrm{TE}(E_{\mathrm{shifts}}\mid n)
  \;=\;
  (n-1)\bigl[(6n_{N}-3)+(5n_{N}-3)\bigr]
  \;=\;
  (n-1)(11n_{N}-6),
  \label{eq:TE_Eshift_per_n}
\end{equation}
and summing over $n=1,\ldots,N$, the total $E$ contribution is
\begin{equation}
  \mathrm{TE}_{E}^{\mathrm{total}}
  \;=\;
  K_{E}\,\mathrm{TE}_{E,\mathrm{rot}}
  + \frac{N(N-1)}{2}\,(11n_{N}-6).
  \label{eq:TE_E_total}
\end{equation}

\paragraph{Subroutine $H$ (uncompute $e_{n}$).}
Figure~7b of Ref.~\cite{burchardt2025highdimensionalquantumschurtransforms}
realises $H$ as $\mathrm{SUM}^{\dagger}\cdot H'_{n}\cdots H'_{1}\cdot\mathrm{SUM}$.
At recursion step $n$ the SUM gate acts on the $n$ populated entries of the
composition register, transforming $\ket{\mu}$ into the vector of prefix sums
$\ket{\widetilde{\mu}}$ via $n-1$ CDKM adders of width $n_{N}$, at per-step
cost
\begin{equation}
  \mathrm{TE}_{\mathrm{SUM}}(n)
  \;=\;
  (n-1)\,A(n_{N})
  \;=\;
  (n-1)(2n_{N}-1).
  \label{eq:TE_SUM}
\end{equation}
Each controlled shift $H'_{k}$ is a comparison of $\widetilde{\mu}_{k}$ with
the transversal entry $\sigma(n)$ followed by a controlled decrement of the
position register, giving
\begin{equation}
  \mathrm{TE}(H'_{k})
  \;=\;
  2\,A(n_{N}) + (n_{d}-1)
  \;=\;
  4n_{N}+n_{d}-3.
\end{equation}
The per-step cost of $H$ is therefore
\begin{equation}
  \mathrm{TE}(H\mid n)
  \;=\;
  2\,\mathrm{TE}_{\mathrm{SUM}}(n)
  + n\bigl(4n_{N}+n_{d}-3\bigr).
  \label{eq:TE_H_per_n}
\end{equation}

\paragraph{Stage-1 total.}
Combining Eqs.~\eqref{eq:TE_A_per_n}--\eqref{eq:TE_D_per_n},
\eqref{eq:TE_E_total} and \eqref{eq:TE_H_per_n}, and summing the per-step
contributions over $n=1,\ldots,N$, the Toffoli-equivalent cost of the
preprocessing isometry is
\begin{equation}
  \boxed{\;
  \mathrm{TE}_{P}
  \;=\;
  \mathrm{TE}_{ABCD}
  + \mathrm{TE}_{E}^{\mathrm{total}}
  + \mathrm{TE}_{H}^{\mathrm{total}},
  \;}
  \label{eq:TE_P_boxed}
\end{equation}
where
\begin{align}
  \mathrm{TE}_{ABCD}
  &\;=\;
  \sum_{n=1}^{N}\Bigl[\mathrm{TE}(A\mid n)+\mathrm{TE}(B\mid n)+\mathrm{TE}(C\mid n)+\mathrm{TE}(D\mid n)\Bigr]
  \nonumber\\
  &\;=\;
  \frac{N(N-1)}{2}\,\bigl(17n_{d}+23n_{N}-14\bigr)
  + N(N+1)\,n_{N},
  \label{eq:TE_ABCD_total}\\[2pt]
  \mathrm{TE}_{H}^{\mathrm{total}}
  &\;=\;
  N(N-1)(2n_{N}-1)
  + \frac{N(N+1)}{2}\,\bigl(4n_{N}+n_{d}-3\bigr).
  \label{eq:TE_H_total}
\end{align}
The peak qubit demand during Stage~1 consists of the persistent output
registers $\ket{p}$, $\ket{\mu}$, and $\ket{tY_{\mu}}$ together with the
transient workspace that is recycled between successive $\mathrm{Prep}_{n}$
calls. The persistent output width is $Nn_{d}+2Nn_{N}$. The transient
workspace comprises the position register $\ket{\cdot}_{p}$ of width $n_{d}$,
the uniqueness qubit $\ket{\cdot}_{u}$, a QROAM bank of size
$Q_{\mathrm{QROAM}}(N,R_{E})$ for the $E$-angle table, a clean-ancilla
multi-control bank of size $\max(0,k_{E}-2)$, and
$\lceil\log_{2}(1/\delta_{E})\rceil$ rotation-synthesis qubits. The Stage-1
peak is therefore
\begin{equation}
  \boxed{\;
  Q_{P}^{\mathrm{peak}}
  \;=\;
  Nn_{d} + 2Nn_{N}
  + n_{d} + 1
  + Q_{\mathrm{QROAM}}(N,R_{E})
  + \max(0,k_{E}-2)
  + \lceil\log_{2}(1/\delta_{E})\rceil.
  \;}
  \label{eq:Q_P_boxed}
\end{equation}

\subsection{Stage 2: Symmetric-group quantum Fourier transform}
\label{subsec:krovi_stage2}

Stage~2 performs a quantum Fourier transform over the symmetric group
$S_{N}$ on the coset register $\ket{tY_{\mu}}$ produced by the
preprocessing isometry $P$, returning the partition label $\ket{\lambda}$,
the standard Young tableau $\ket{T}$, and the Young--Yamanouchi path
$\ket{\sigma}$ that labels the $S_{N}$ irrep copy. We consider two efficient
constructions for the QFT over $S_N$.

Section~\ref{subsubsec:beals_qft} describes the original Beals
construction~\cite{Beals1997}, which is structurally simpler but has a larger
gate count. Section~\ref{subsubsec:ks_qft} describes the Kawano--Sekigawa
(KS) construction~\cite{kawano2016qftsn}, which exploits the tighter
Bratteli-diagram bound $e_{\lambda}<\sqrt{2m}+1$ (Lemma~5 of
Ref.~\cite{kawano2016qftsn}) and is the construction invoked in Lemma~2 of
Ref.~\cite{burchardt2025highdimensionalquantumschurtransforms}. Correctly
unrolling the KS recursion (see below) yields $\Theta(N^{3})$ synthesised
rotations, so that after multi-control inflation the KS variant attains the
$\widetilde{\mathcal{O}}(N^{4})$ elementary-gate scaling reported in
Refs.~\cite{kawano2016qftsn,burchardt2025highdimensionalquantumschurtransforms},
whereas the Beals variant carries $\Theta(N^{4})$ rotations and costs
$\widetilde{\mathcal{O}}(N^{5})$; the two variants therefore differ by a
factor of $\Theta(N)$, as made precise in
Section~\ref{subsubsec:qft_comparison}.

Throughout Stage~2, Toffoli equivalents and the synthesis budget follow
the conventions fixed at the top of this section: arbitrary single-qubit
rotations are synthesised with the direct construction of
Ref.~\cite{BocharovRoettelerSvore2015} and a per-rotation tolerance
$\delta$ is obtained by dividing the stage budget $\varepsilon_{Q}$
evenly across the synthesised rotations. For either QFT construction we
write the synthesised rotation count as $K_{Q}$, with a superscript
indicating the variant when disambiguation is needed. The classical
rotation-angle table, of size $T_{Q}$, is loaded on demand via QROAM
with cost given by Eq.~\eqref{eq:qroam_lookup_krovi}; because repeated
invocations of a primitive draw on the same table, the table sizes below
depend only on the set of distinct angles and not on primitive
multiplicities.

% -----------------------------------------------------------------
\subsubsection{Beals's construction}
\label{subsubsec:beals_qft}
% -----------------------------------------------------------------

Beals adapts the Clausen / Diaconis--Rockmore recursion to the quantum
setting. The subgroup tower
$\{1\}\subset S_{2}\subset\cdots\subset S_{N}$ is walked bottom up: at
level $m$ we have the subgroup $H=S_{m-1}$ of index $m$ in $S_{m}$ with
coset representatives $\{\tau_{1},\ldots,\tau_{m}\}$ where $\tau_{k}$
maps the point $1$ to $k$. Beals's Section~6 fixes
\begin{equation}
  \tau_{k} \;=\; (k\; k{-}1)\,(k{-}1\; k{-}2)\,\cdots\,(2\; 1),
  \label{eq:beals_coset_reps}
\end{equation}
so that each $\tau_{k}$ is a product of exactly $k-1$ adjacent
transpositions. The level-$m$ step of the forward transform acts on the
register triple $(\ket{\lambda},\ket{i},\ket{j})$, with
$\lambda\in\widehat{S}_{m}$ labelling an irrep in Young Orthogonal Form
and $(i,j)$ indexing a matrix entry of the corresponding irrep. Writing
$A_{\rho}$ for the matrix whose $(i,j)$ entry is the amplitude of the
computational basis state $\ket{\rho,i,j}$, Beals's inductive step is,
for each $k=1,\ldots,m$: first, right-multiplication by
$\rho(\tau_{k}^{-1})$, implemented as a product of $k-1$ elementary
transpositions $\rho((j,j{+}1))$, each block-diagonal in Young Orthogonal
Form with $1\times1$ and $2\times2$ blocks whose non-trivial entries are
$\pm 1/\delta$ and $\sqrt{1-1/\delta^{2}}$ for an integer axial distance
$\delta\in\{2,\ldots,m\}$; second, the addition step (Beals Lemma~5.1,
applied in reverse), in which a controlled isometry $T$ takes the tape
contents $(\sigma,i,j,-1)$ to the branching-rule superposition over
$\rho\in\widehat{S}_{m}$ whose restriction to $H$ contains $\sigma$,
\begin{equation}
  T:\;\ket{\sigma,i,j,-1}
  \;\longmapsto\;
  \sum_{\rho:\rho|_{H}\supset\sigma}
  \sqrt{\tfrac{d_{\rho}\,|H|}{d_{\sigma}\,|G|}}\;
  \ket{\rho,i',j'},
  \label{eq:beals_T_iso}
\end{equation}
with the addition step realised as the sandwich $T^{-1}\cdot S\cdot T$
for a classical reversible permutation $S$; and third,
right-multiplication by $\rho(\tau_{k})$, again a product of $k-1$
elementary transpositions.

The branching isometry $T$ of Eq.~\eqref{eq:beals_T_iso} acts on an
add-row register of width $\lceil\log_{2}m\rceil$ and maps one basis
state into a superposition over at most $m$ children of $\sigma$. Via
Givens/QR compilation, each $T$ requires at most $\binom{m}{2}\le m^{2}/2$
two-level rotations. Each sandwich invokes $T$ and its inverse once, so
a single coset contributes $m^{2}$ $A$-type two-level rotations. Beals's
recursion $F_{m}=(F_{m-1}\otimes I)\cdot[\text{level-}m\text{ work}]$ is
linear, so each level's work is executed exactly once, and the exact
rotation counts are
\begin{align}
  \mathcal{R}_{K}^{\mathrm{Beals}}
  &\;=\;
  2\sum_{m=2}^{N}m(m-1)
  \;=\;
  \frac{2N(N-1)(N+1)}{3},
  \label{eq:beals_RK}\\
  \mathcal{R}_{A}^{\mathrm{Beals}}
  &\;=\;
  \sum_{m=2}^{N} m^{3}
  \;=\;
  \left(\tfrac{N(N+1)}{2}\right)^{\!2}-1,
  \label{eq:beals_RA}
\end{align}
where $\mathcal{R}_{K}^{\mathrm{Beals}}$ counts the label-controlled
adjacent transpositions arising from the two coset multiplications and
$\mathcal{R}_{A}^{\mathrm{Beals}}$ counts the $A$-type two-level Givens
rotations contributed by the branching isometries. The total number of
synthesised rotations is
\begin{equation}
  K_{Q}^{\mathrm{Beals}}
  \;:=\;
  \mathcal{R}_{K}^{\mathrm{Beals}} + \mathcal{R}_{A}^{\mathrm{Beals}}
  \;=\;
  \Theta(N^{4}).
  \label{eq:KQ_beals}
\end{equation}

Each $K$-type angle is parameterised by a single integer axial distance
$\delta\in\{2,\ldots,N\}$, giving a $K$-angle table of size $N-1$. Each
$A$-type angle is a branching-rule amplitude
$\sqrt{d_{\rho}|H|/(d_{\sigma}|G|)}$ indexed by an edge $(\sigma,\rho)$
of the Bratteli diagram; summing over levels the combined angle table
satisfies
\begin{equation}
  T_{Q}^{\mathrm{Beals}}
  \;\leq\;
  (N-1) + \sum_{m=1}^{N}p(m)\,m,
  \label{eq:TQ_beals}
\end{equation}
with $p(m)$ the integer partition function. The per-rotation synthesis
tolerance is
\begin{equation}
  \delta_{Q}^{\mathrm{Beals}}
  \;=\;
  \frac{\varepsilon_{Q}}{K_{Q}^{\mathrm{Beals}}},
  \qquad
  R_{Q}^{\mathrm{Beals}}
  \;=\;
  \left\lceil\log_{2}\!\bigl(\pi K_{Q}^{\mathrm{Beals}}/\varepsilon_{Q}\bigr)\right\rceil
  + 1,
  \label{eq:beals_tolerances}
\end{equation}
so that worst-case additive diamond-norm accumulation respects the stage
budget. The effective multi-control width is
$k_{Q}=(N-1)n_{N}+n_{\lambda}+n_{N}$ as for KS below, since both
constructions control on the same partition-label, SYT-path, and
coset-index registers. The per-rotation Toffoli-equivalent cost is
\begin{equation}
  \mathrm{TE}_{Q,\mathrm{rot}}^{\mathrm{Beals}}
  \;=\;
  \mathrm{TE}_{\mathrm{QROAM}}(T_{Q}^{\mathrm{Beals}},R_{Q}^{\mathrm{Beals}})
  + \frac{T_{\mathrm{cR}}(\delta_{Q}^{\mathrm{Beals}})}{7}
  + 2\,C_{\mathrm{Tof}}(k_{Q};a)
  + 3\,A(n_{N}).
  \label{eq:TE_Q_rot_beals}
\end{equation}
Beals's inter-level basis extensions and permutations have the same form
as the $T_{m}$, $P_{m}$ primitives of KS (analysed below); we denote
their combined Toffoli cost $\mathrm{TE}_{TP}$, with the explicit formula
deferred to Eqs.~\eqref{eq:TE_Tm_krovi}--\eqref{eq:TE_TP_combined}.
Summing,
\begin{equation}
  \boxed{\;
  \mathrm{TE}_{Q}^{\mathrm{Beals}}
  \;=\;
  K_{Q}^{\mathrm{Beals}}\,\mathrm{TE}_{Q,\mathrm{rot}}^{\mathrm{Beals}}
  + \mathrm{TE}_{TP}.
  \;}
  \label{eq:TE_Q_beals_boxed}
\end{equation}
The Stage-2 peak qubit footprint under the Beals variant is
\begin{equation}
  Q_{Q}^{\mathrm{peak,Beals}}
  \;=\;
  Nn_{d} + Nn_{N} + 2(N-1)n_{N} + n_{\lambda}
  + Q_{\mathrm{QROAM}}(T_{Q}^{\mathrm{Beals}},R_{Q}^{\mathrm{Beals}})
  + \max(0,k_{Q}-2)
  + \left\lceil\log_{2}(1/\delta_{Q}^{\mathrm{Beals}})\right\rceil,
  \label{eq:Q_Q_beals}
\end{equation}
identical in functional form to Eq.~\eqref{eq:Q_Q_boxed} below with
$T_{Q},\delta_{Q},R_{Q}$ replaced by their Beals counterparts.

% -----------------------------------------------------------------
\subsubsection{Kawano--Sekigawa's construction}
\label{subsubsec:ks_qft}
% -----------------------------------------------------------------

The Kawano--Sekigawa (KS) circuit is organised around the same subgroup
tower $\{1\}\subset S_{2}\subset\cdots\subset S_{N}$ but factors each
level through the specialised relations
$F_{m}=T_{m}H_{m}(F_{m-1}\otimes I_{m})$ and
$H_{m}=A_{m}P_{m}H'_{m-1}K_{m}$ (Theorem~8 of
Ref.~\cite{kawano2016qftsn}), where $H'_{m-1}$ is the operator $H_{m-1}$
conjugated by a basis relabelling and hence obeys the same recursion one
rank down. The circuit is built from four families of primitives: $K_{j}$,
a product of $j-2$ label-controlled adjacent transpositions
$\rho_{j-1}(c_{(t,t+1)})$ with a quantum control on the coset-index
register (Eq.~(8) of Ref.~\cite{kawano2016qftsn}), each transposition
block-diagonal in the Young--Yamanouchi basis; $A_{j}$, an embedding
operation block-diagonal over the irrep register
$\lambda\in\Lambda_{j-1}$, with each block $A'_{j,\lambda}$ of size
$e_{\lambda}\times e_{\lambda}$, where $e_{\lambda}<\sqrt{2j}+1$ by
Lemma~5 of Ref.~\cite{kawano2016qftsn}, compiled into at most
$\binom{\lfloor\sqrt{2j}\rfloor+1}{2}$ label-controlled two-level Givens
rotations; $T_{m}$, a basis-extension primitive that updates the
Young-path and partition registers via a modular addition and $m-1$
controlled increments (Fig.~7 of Ref.~\cite{kawano2016qftsn}); and
$P_{j}$, a classical-reversible basis permutation reconciling
$\mathcal{B}'_{j-1}\times\mathbb{Z}_{j}$ with $\mathcal{B}_{j}$
(Subsection~5.6 of Ref.~\cite{kawano2016qftsn}).

The two recursions are nested: the $F$-recursion invokes $H_{m}$ once at
every level $m=3,\ldots,N$, and each $H_{m}$ unrolls through its own
recursion into the chain of primitives
$\{A_{j},P_{j},K_{j}\}_{j=3}^{m}$. A level-$j$ primitive of $H$-type is
therefore executed once inside every $H_{m}$ with $m\geq j$, i.e.\ with
multiplicity
\begin{equation}
  \mathrm{mult}(j)\;=\;N-j+1,
  \qquad j=3,\ldots,N,
  \label{eq:KS_multiplicity}
\end{equation}
while each $T_{m}$, living in the $F$-recursion, is executed exactly
once. Unrolling and summing, the exact primitive and rotation counts are
\begin{align}
  \mathcal{R}_{K}
  &\;=\;
  \sum_{j=3}^{N}(N-j+1)(j-2)
  \;=\;
  \binom{N}{3}
  \;=\;
  \frac{N(N-1)(N-2)}{6},
  \label{eq:KS_RK_krovi}\\
  \mathcal{R}_{A}
  &\;=\;
  \sum_{j=3}^{N}(N-j+1)\binom{\lfloor\sqrt{2j}\rfloor+1}{2}
  \;=\;
  \frac{N^{3}}{6}+\mathcal{O}\!\bigl(N^{5/2}\bigr),
  \label{eq:KS_RA_krovi}\\
  \mathcal{R}_{T}
  &\;=\;
  N-2, \qquad
  \mathcal{R}_{P}
  \;=\;
  \sum_{j=3}^{N}(N-j+1)
  \;=\;
  \frac{(N-1)(N-2)}{2}.
  \label{eq:KS_RTP_krovi}
\end{align}
The total count of synthesised rotations is
\begin{equation}
  K_{Q}
  \;:=\;
  \mathcal{R}_{K}+\mathcal{R}_{A}
  \;=\;
  \frac{N^{3}}{3}+\mathcal{O}\!\bigl(N^{5/2}\bigr).
  \label{eq:KQ_definition}
\end{equation}
This $\Theta(N^{3})$ unit-gate count reproduces the
$\widetilde{\mathcal{O}}(N^{3})$ bound of Theorem~9 of
Ref.~\cite{kawano2016qftsn}, in which Young-orthogonal transpositions and
elementary two-level pieces of $A'_{j,\lambda}$ are treated as unit-cost
gates. In the elementary-gate model used for our resource analysis each
such piece is realised as a QROAM-fetched angle, a synthesised controlled
rotation, and flag-set and flag-clear multi-controls of width
$k_{Q}=\Theta(N n_{N})$ on the SYT path register, so that the total cost
is $\widetilde{\mathcal{O}}(N^{4})$, coinciding with the complexity
reported in Lemma~2 of
Ref.~\cite{burchardt2025highdimensionalquantumschurtransforms}. The
$T_{m}$ and $P_{j}$ primitives are implemented entirely with Clifford and
Toffoli gates and contribute no $T$-gates.

The $K$-angle table has size $T_{Q}^{(K)}=N-1$ as for Beals. Each
$A$-type angle is an entry of some $A'_{j,\lambda}$ block and is fully
determined, through Theorem~7 of Ref.~\cite{kawano2016qftsn}, by the
triple of neighbouring Bratteli-diagram nodes and an SYT-path index.
Since the multiplicities of Eq.~\eqref{eq:KS_multiplicity} merely repeat
invocations of the same blocks, the distinct-angle count is unchanged by
the unrolling, and enumerating gives the conservative bound
\begin{equation}
  T_{Q}^{(A)}
  \;\leq\;
  \sum_{j=3}^{N} p(j-1)\,\binom{\lfloor\sqrt{2j}\rfloor+1}{2};
  \label{eq:TQA_bound}
\end{equation}
we combine both tables into $T_{Q}=T_{Q}^{(K)}+T_{Q}^{(A)}$. We note for
completeness that the $p(j)$ factors make $T_{Q}$ (and hence the QROAM
term below) grow as $e^{\mathcal{O}(\sqrt{N})}$, so the
$\widetilde{\mathcal{O}}(N^{4})$ statement for our fully compiled model
holds up to this subexponential lookup contribution; numerically the
QROAM term remains subdominant throughout the range $N\lesssim 50$
studied here. Distributing the stage budget uniformly over the $K_{Q}$
synthesised rotations,
\begin{equation}
  \delta_{Q}
  \;=\;
  \frac{\varepsilon_{Q}}{K_{Q}}
  \;=\;
  \frac{\varepsilon}{3K_{Q}},
  \qquad
  R_{Q}
  \;=\;
  \left\lceil\log_{2}\!\bigl(\pi K_{Q}/\varepsilon_{Q}\bigr)\right\rceil + 1.
  \label{eq:stage2_tolerances}
\end{equation}
The largest multi-control width encountered is
\begin{equation}
  k_{Q}
  \;=\;
  (N-1)n_{N} + n_{\lambda} + n_{N},
  \label{eq:kQ_width}
\end{equation}
which conservatively bounds both the $K$- and $A$-type primitives at
every level (rotations executed inside the unrolled $H'_{j}$ act on
rank-$j$ registers and admit narrower controls, so
Eq.~\eqref{eq:kQ_width} is an upper bound). The per-rotation
Toffoli-equivalent cost is
\begin{equation}
  \mathrm{TE}_{Q,\mathrm{rot}}
  \;=\;
  \mathrm{TE}_{\mathrm{QROAM}}(T_{Q},R_{Q})
  + \frac{T_{\mathrm{cR}}(\delta_{Q})}{7}
  + 2\,C_{\mathrm{Tof}}(k_{Q};a)
  + 3\,A(n_{N}).
  \label{eq:TE_Q_rot_decomp}
\end{equation}

The classical-reversible primitives, shared with the Beals variant,
have cost
\begin{align}
  \mathrm{TE}(T_{m})
  &\;=\;
  A(n_{m}) + (m-1)\,C_{\mathrm{Tof}}(n_{m};a)
  \;=\;
  (2n_{m}-1) + (m-1)(2n_{m}-3),
  \label{eq:TE_Tm_krovi}\\
  \mathrm{TE}(P_{j})
  &\;=\;
  (j-1)\bigl[A(n_{N})+C_{\mathrm{Tof}}(n_{j};a)\bigr]
  \;=\;
  (j-1)(2n_{N}+2n_{j}-4),
  \label{eq:TE_Pm_krovi}
\end{align}
with $n_{m}=\lceil\log_{2}m\rceil$. Weighting the $P_{j}$ by their
multiplicities, their combined contribution is
\begin{equation}
  \mathrm{TE}_{TP}
  \;=\;
  \sum_{m=3}^{N}\mathrm{TE}(T_{m})
  \;+\;
  \sum_{j=3}^{N}(N-j+1)\,\mathrm{TE}(P_{j}).
  \label{eq:TE_TP_combined}
\end{equation}
Summing the rotation contributions and the classical-reversible
overheads,
\begin{equation}
  \boxed{\;
  \mathrm{TE}_{Q}
  \;=\;
  K_{Q}\,\mathrm{TE}_{Q,\mathrm{rot}}
  + \mathrm{TE}_{TP}.
  \;}
  \label{eq:TE_Q_boxed}
\end{equation}
In Section~\ref{subsec:krovi_totals} below, $\mathrm{TE}_{Q}$ without a
superscript always refers to the KS cost of
Eq.~\eqref{eq:TE_Q_boxed}; substitution of
$\mathrm{TE}_{Q}\to\mathrm{TE}_{Q}^{\mathrm{Beals}}$ from
Eq.~\eqref{eq:TE_Q_beals_boxed} yields the cost of the Beals-variant
Schur transform.
The Stage-2 peak qubit footprint is
\begin{equation}
  \boxed{\;
  Q_{Q}^{\mathrm{peak}}
  \;=\;
  Nn_{d} + Nn_{N}
  + 2(N-1)n_{N} + n_{\lambda}
  + Q_{\mathrm{QROAM}}(T_{Q},R_{Q})
  + \max(0,k_{Q}-2)
  + \lceil\log_{2}(1/\delta_{Q})\rceil.
  \;}
  \label{eq:Q_Q_boxed}
\end{equation}
The net persistent-register growth $2(N-1)n_{N}+n_{\lambda}-Nn_{N}$
beyond the preceding stages reflects that the QFT reads from and writes
to the $Nn_{N}$-wide coset register in-place, producing two
$(N-1)n_{N}$-wide SYT outputs plus an $n_{\lambda}$-wide irrep label.

% -----------------------------------------------------------------
\subsubsection{Comparison and choice of variant}
\label{subsubsec:qft_comparison}
% -----------------------------------------------------------------

The two constructions differ by a factor of $\Theta(N)$ in rotation
count and hence, since their per-rotation costs are dominated by the
same multi-control and lookup terms, by essentially the same factor in
total TE. Two structural properties of Beals's construction drive the
gap. First, each coset representative $\tau_{k}$ is realised as $k-1$
adjacent transpositions (Eq.~\eqref{eq:beals_coset_reps}), so that level
$m$ contributes $m(m-1)$ $K$-type rotations, whereas the KS $K_{j}$
primitives contribute $\binom{N}{3}$ rotations in total even after the
multiplicity of Eq.~\eqref{eq:KS_multiplicity} is included. Second, the
branching isometry $T$ of Eq.~\eqref{eq:beals_T_iso} uses the loose
bound $m$ on the number of children of $\sigma$ and is applied twice per
coset, contributing $\mathcal{R}_{A}^{\mathrm{Beals}}=\Theta(N^{4})$
rotations, whereas the KS $A_{j}$ blocks exploit the Bratteli bound
$e_{\lambda}<\sqrt{2j}+1$, so that even with multiplicity $N-j+1$ they
contribute only $\mathcal{R}_{A}^{\mathrm{KS}}=\Theta(N^{3})$. At leading
order the ratio of rotation counts is
$K_{Q}^{\mathrm{Beals}}/K_{Q}^{\mathrm{KS}}\simeq\tfrac{3}{4}N$;
evaluating the closed forms
Eqs.~\eqref{eq:beals_RK}--\eqref{eq:beals_RA} and
\eqref{eq:KS_RK_krovi}--\eqref{eq:KS_RA_krovi} gives, for example,
$K_{Q}^{\mathrm{Beals}}/K_{Q}^{\mathrm{KS}}\approx 12$ at $N=10$ and
$\approx 37$ at $N=45$. The peak qubit footprints
$Q_{Q}^{\mathrm{peak,Beals}}$ and $Q_{Q}^{\mathrm{peak}}$ differ only
through the rotation-synthesis and QROAM-table bits in
Eqs.~\eqref{eq:Q_Q_beals} and \eqref{eq:Q_Q_boxed}, a subdominant
contribution to the overall peak, which is in any case controlled by
Stage~3's compressed Gelfand--Tsetlin register (see
Eq.~\eqref{eq:Q_V_boxed} below). The choice between Beals and KS is
therefore a pure runtime optimisation, with no qubit-count penalty for
either variant.

Resource estimates in the remainder of this document are quoted using
the KS construction unless the Beals label is attached. Both variants
remove BCH's $d^{4}$ Stage-2 factor in favour of a
$\mathrm{polylog}(d)$ dependence, so their advantage over BCH grows with
the single-particle dimension $d$.

\subsection{Stage 3: Compression isometry \texorpdfstring{$V^{\dagger}$}{V-dagger}}
\label{subsec:krovi_stage3}

The compression isometry $V^{\dagger}$ rotates the split basis produced by
$\mathrm{QFT}_{S_{N}}$ into the combined Young--Yamanouchi and
Gelfand--Tsetlin bases. Figure~8 of
Ref.~\cite{burchardt2025highdimensionalquantumschurtransforms} presents the
explicit circuit, which consists of a SUM gate on the composition register,
producing the prefix-sum register $\ket{\widetilde{\mu}}$; a COPY gate that
initialises the compressed GT register $\ket{\widetilde{M}}$ by copying the
relevant SYT entries from $\ket{T}$ addressed by $\ket{\widetilde{\mu}}$;
for each $\ell=1,\ldots,N$ a cascade
$(R\cdot R^{\dagger}\cdot X^{\dagger}\cdot R\cdot R^{\dagger})$ of
register-preparation gates followed by a controlled $F$-unitary
$F^{\dagger}_{\ell}$ acting on the $(\ket{\widetilde{M}},\ket{T})$ block;
and an inverse COPY and inverse SUM restoring the auxiliary registers.

\paragraph{$F$-unitary decomposition and rotation count.}
Each $F^{\dagger}_{\ell}$ is a label-controlled Givens decomposition of the
$F$-symbol that implements the recoupling of the Young diagram at level
$\ell$; Section~7.1 of
Ref.~\cite{burchardt2025highdimensionalquantumschurtransforms} bounds the
number of Givens rotations required for a single $F_{\ell}$ by $N^{2}$. We
use this conservative bound at every level, so that the total number of
synthesised rotations across the cascade is
\begin{equation}
  K_{F}
  \;:=\;
  N\cdot N^{2}
  \;=\;
  N^{3}.
  \label{eq:KF_definition}
\end{equation}

\paragraph{Provisioning the $F$-symbol angles.}
Each Givens angle is an $F$-symbol of $\mathrm{Rep}(U_{d})$ given in closed
form by Lemma~5 of
Ref.~\cite{burchardt2025highdimensionalquantumschurtransforms},
\begin{equation}
  \Bigl[F^{(k-1),\square,\lambda}_{\nu}\Bigr]^{\lambda^{+a}}_{(k)}
  \;=\;
  \sqrt{\frac{1}{k}\,
  \frac{\prod_{i=1}^{\ell(\nu)}\bigl(\nu_{i}-i-\lambda_{a}+a\bigr)}
       {\prod_{i=1,\,i\neq a}^{\ell(\nu)}\bigl(\lambda_{i}-i-\lambda_{a}+a\bigr)}},
  \label{eq:F_symbol_closed}
\end{equation}
together with its column counterpart. The angle therefore depends on the
\emph{entire} pair of nested partitions $(\lambda\subset\nu)$, the
addable-row index $a$, and the symmetric-row length $k=|\nu|-|\lambda|$; it
is not a function of $(\lambda,a)$ alone. A lookup table indexed by the
control registers must consequently cover all nested pairs with
$|\nu|\leq N$ together with the row index, giving the conservative bound
\begin{equation}
  L_{F}
  \;\leq\;
  N\,\bigl[P(N)\bigr]^{2},
  \qquad
  P(N):=\sum_{n=0}^{N}p(n),
  \label{eq:LF_bound}
\end{equation}
which grows as $e^{\mathcal{O}(\sqrt{N})}$ and is numerically prohibitive as
a QROAM address space already for $N\gtrsim 20$. We therefore adopt as our
default the online evaluation of Eq.~\eqref{eq:F_symbol_closed} with
streaming reversible arithmetic, exactly paralleling the reduced-Wigner
evaluation of the BCH analysis; the closed form involves at most
$2\ell(\nu)\leq 2N$ integer linear factors, each of magnitude at most
$2N+1$, one reciprocal, and one square root per angle. (For small $N$ a
table over the set of angles actually realised, which is generally far
smaller than the bound of Eq.~\eqref{eq:LF_bound}, may be competitive; we do
not pursue this optimisation here.)

\paragraph{Precision bookkeeping.}
The stage budget is split evenly between rotation synthesis and arithmetic,
$\varepsilon_{V,\mathrm{rot}}=\varepsilon_{V,\mathrm{arith}}
=\tfrac{1}{2}\varepsilon_{V}=\tfrac{1}{6}\varepsilon$. With the synthesis
budget distributed uniformly across $K_{F}$ rotations,
\begin{equation}
  \delta_{F}
  \;=\;
  \frac{\varepsilon_{V,\mathrm{rot}}}{K_{F}},
  \qquad
  f_{F}
  \;=\;
  \left\lceil\log_{2}\!\bigl(\pi K_{F}/\varepsilon_{V,\mathrm{arith}}\bigr)\right\rceil,
  \label{eq:stage3_tolerances}
\end{equation}
where $f_{F}$ is the number of fractional bits guaranteeing per-angle
quantisation error at most $\varepsilon_{V,\mathrm{arith}}/K_{F}$.
Following the fixed-point conventions of the BCH analysis, the integer
dynamic range of the shifted-content factors in
Eq.~\eqref{eq:F_symbol_closed} is covered by
$i_{F}=\lceil\log_{2}(2N+1)\rceil+2$ bits, rounding across the
$\mathcal{O}(N)$-deep multiply chain is absorbed by
$g_{F}=\lceil\log_{2}(2N)\rceil+3$ guard bits, and the working word size is
\begin{equation}
  w_{F}
  \;=\;
  i_{F}+f_{F}+g_{F}.
  \label{eq:wF_definition}
\end{equation}

\paragraph{Per-rotation cost.}
Producing one angle requires forming the at most $2N$ shifted-content
factors ($4N$ additions including their uncomputation partners), two
running products of at most $N$ factors each ($2N$ multiplications), one
Newton reciprocal, one Newton square root, and a CORDIC conversion of the
resulting amplitude into a rotation angle; the arithmetic pipeline is
computed, its output copied, and then uncomputed to restore ancillas, which
we account with an overall factor of two on the compute path. Using the
reversible-arithmetic primitives $A(w)$, $M(w)$, $C_{\mathrm{recip}}(w)$,
$C_{\sqrt{\phantom{.}}}(w)$, and $C_{\mathrm{CORDIC}}(w)=3f_{F}A(w)$
defined in the BCH section, the per-angle arithmetic cost is
\begin{equation}
  C_{F}^{\mathrm{angle}}(N;w_{F})
  \;=\;
  2\Bigl[\,4N\,A(w_{F}) + 2N\,M(w_{F})
  + C_{\mathrm{recip}}(w_{F}) + C_{\sqrt{\phantom{.}}}(w_{F})\Bigr]
  + 3\,f_{F}\,A(w_{F}).
  \label{eq:CF_angle}
\end{equation}
The effective multi-control width for each $F$-rotation is set by the
relevant SYT/GT slice $(M^{k},T^{k-1})$, which by construction occupies at
most $4n_{N}$ qubits together with the partition label $\ket{\lambda}$ of
width $n_{\lambda}$, so we use
\begin{equation}
  k_{V}
  \;=\;
  4n_{N} + n_{\lambda},
  \label{eq:kV_width}
\end{equation}
and the per-rotation Toffoli-equivalent cost is
\begin{equation}
  \mathrm{TE}_{V,\mathrm{rot}}
  \;=\;
  C_{F}^{\mathrm{angle}}(N;w_{F})
  + \frac{T_{\mathrm{cR}}(\delta_{F})}{7}
  + 2\,C_{\mathrm{Tof}}(k_{V};a).
  \label{eq:TE_V_rot_decomp}
\end{equation}

\paragraph{Cost of supporting primitives.}
The supporting primitives of $V^{\dagger}$ consist of one SUM, one COPY, one
$\mathrm{SUM}^{\dagger}$, one $\mathrm{COPY}^{\dagger}$, and $N$ invocations
of the register-preparation cascade
$(R,R^{\dagger},X^{\dagger},R,R^{\dagger})$. The SUM gate on the full
composition register costs
\begin{equation}
  \mathrm{TE}_{\mathrm{SUM}}
  \;=\;
  (N-1)\,A(n_{N})
  \;=\;
  (N-1)(2n_{N}-1).
  \label{eq:TE_SUM_stage3}
\end{equation}
The controlled SWAPs appearing in the COPY and $J$ gates are
\emph{value-selected}: per Fig.~10 of
Ref.~\cite{burchardt2025highdimensionalquantumschurtransforms}, a SWAP whose
target position is addressed by a quantum register of $N$ possible values is
a selection network in which, for each candidate value $i\in[N]$, an
equality flag against the constant $i$ is set and cleared
($2\,C_{\mathrm{Tof}}(n_{N};a)$ Toffolis) and the flagged data pair of width
$n$ is conditionally exchanged ($n$ Fredkin gates, one Toffoli each). The
selected-SWAP cost at data width $n$ is therefore
\begin{equation}
  C_{\mathrm{sel}}(n)
  \;=\;
  N\bigl[\,2\,C_{\mathrm{Tof}}(n_{N};a) + n\,\bigr]
  \;=\;
  N\bigl(4n_{N}-6+n\bigr),
  \label{eq:Csel}
\end{equation}
linear in $N$ in accordance with the $\mathcal{O}(n)$ accounting of
Ref.~\cite{burchardt2025highdimensionalquantumschurtransforms}. The COPY
gate (Fig.~11 of the same reference) writes, for each of the $N$ slots of
$\ket{\widetilde{M}}$, the SYT entry addressed by $\widetilde{\mu}_{k}$; per
slot this is one $N$-way selected copy of an $n_{N}$-wide word (equality
flags as above, plus $n_{N}$ flag-controlled CNOTs realised as Toffolis),
giving
\begin{equation}
  \mathrm{TE}_{\mathrm{COPY}}
  \;=\;
  N\cdot N\bigl(4n_{N}-6+n_{N}\bigr)
  \;=\;
  N^{2}\bigl(5n_{N}-6\bigr),
  \label{eq:TE_COPY}
\end{equation}
matching the $\widetilde{\mathcal{O}}(N^{2})$ complexity stated in
Section~7.1 of
Ref.~\cite{burchardt2025highdimensionalquantumschurtransforms}. The
$J$-gate (Fig.~13) contains a SUM and its inverse
($2\,\mathrm{TE}_{\mathrm{SUM}}$), two selected SWAPs on the composition
register ($2\,C_{\mathrm{sel}}(n_{N})$), a cascade of $N-1$
doubly-conditioned counter updates, each comprising a comparator against
$\ket{i}$ ($2\,A(n_{N})$), a bitwise neighbour-equality check
($2\,C_{\mathrm{Tof}}(n_{N};a)$), and a flagged increment ($n_{N}+1$
Toffolis including the conjunction), for $(N-1)(9n_{N}-7)$ in total, plus a
controlled addition and a shift ($2\,A(n_{N})$) and a final increment
($n_{N}-1$):
\begin{equation}
  \mathrm{TE}_{J}
  \;=\;
  2(N-1)(2n_{N}-1)
  + 2N(5n_{N}-6)
  + (N-1)(9n_{N}-7)
  + (5n_{N}-3).
  \label{eq:TE_J}
\end{equation}
Each $R$-gate (Fig.~12) wraps the $J$-gate with one further controlled
shift,
\begin{equation}
  \mathrm{TE}_{R}
  \;=\;
  \mathrm{TE}_{J} + A(n_{N})
  \;=\;
  \mathrm{TE}_{J} + 2n_{N}-1,
  \label{eq:TE_R}
\end{equation}
and each $X^{\dagger}$ is a modular decrement of an $n_{N}$-wide register at
cost $A(n_{N})$. Assembling these contributions,
\begin{equation}
  \mathrm{TE}_{V,\mathrm{support}}
  \;=\;
  2\,\mathrm{TE}_{\mathrm{SUM}}
  + 2\,\mathrm{TE}_{\mathrm{COPY}}
  + N\bigl[4\,\mathrm{TE}_{R} + A(n_{N})\bigr],
  \label{eq:TE_V_support}
\end{equation}
where the factor of $4$ accounts for the four $R$- or $R^{\dagger}$-gates in
each level of the cascade.

\paragraph{Stage-3 total.}
The Toffoli-equivalent cost of $V^{\dagger}$ is
\begin{equation}
  \boxed{\;
  \mathrm{TE}_{V}
  \;=\;
  K_{F}\,\mathrm{TE}_{V,\mathrm{rot}}
  + \mathrm{TE}_{V,\mathrm{support}}.
  \;}
  \label{eq:TE_V_boxed}
\end{equation}
The peak qubit footprint during Stage~3 is dominated by the compressed GT
register $\ket{\widetilde{M}}$, held alongside $\ket{T}$, $\ket{\mu}$,
$\ket{\widetilde{\mu}}$, $\ket{\lambda}$, $\ket{\sigma}$, and the persistent
$\ket{p}$ register held idle from earlier stages. The streaming arithmetic
that evaluates the $F$-symbols requires only rolling accumulators and a
constant number of temporaries at word width $w_{F}$, bounded as in the BCH
analysis by $a^{F}_{\mathrm{arith}}\leq 12\,w_{F}$ qubits. Accounting also
for the multi-control clean-ancilla bank and the rotation-synthesis
workspace,
\begin{equation}
  \boxed{\;
  \begin{aligned}
  Q_{V}^{\mathrm{peak}}
  \;=\;
  &\underbrace{Nn_{d}}_{\ket{p}}
  + \underbrace{Nn_{N}}_{\ket{\mu}}
  + \underbrace{Nn_{N}}_{\ket{\widetilde{\mu}}}
  + \underbrace{\tfrac{N(N+1)}{2}\,n_{N}}_{\ket{\widetilde{M}}}
  + \underbrace{(N-1)n_{N}}_{\ket{T}}
  + \underbrace{(N-1)n_{N}}_{\ket{\sigma}}
  + n_{\lambda}
  \\&
  + 12\,w_{F}
  + \max(0,k_{V}-2)
  + \lceil\log_{2}(1/\delta_{F})\rceil.
  \end{aligned}
  \;}
  \label{eq:Q_V_boxed}
\end{equation}
The $\tfrac{N(N+1)}{2}\,n_{N}$ contribution from $\ket{\widetilde{M}}$ is
the origin of the $\widetilde{\mathcal{O}}(N^{2})$ space complexity quoted
in Theorem~1 of
Ref.~\cite{burchardt2025highdimensionalquantumschurtransforms} and
dominates the qubit footprint at moderate-to-large $N$.

\subsection{Total cost of the corrected Krovi Schur transform}
\label{subsec:krovi_totals}

Summing Eqs.~\eqref{eq:TE_P_boxed}, \eqref{eq:TE_Q_boxed} and
\eqref{eq:TE_V_boxed}, the total Toffoli-equivalent cost of the forward
Schur transform is
\begin{equation}
  \boxed{\;
  \mathrm{TE}_{\mathrm{Schur}}^{\mathrm{Krovi}}(N,d,\varepsilon)
  \;=\;
  \mathrm{TE}_{P}
  + \mathrm{TE}_{Q}
  + \mathrm{TE}_{V}.
  \;}
  \label{eq:TE_Schur_Krovi}
\end{equation}
Each constituent on the right-hand side is an explicit closed-form function
of $N$, $d$, and $\varepsilon$ through the intermediate quantities
$n_{d}=\lceil\log_{2}d\rceil$, $n_{N}=\lceil\log_{2}N\rceil$,
$n_{\lambda}=\lceil\log_{2}p(N)\rceil$, the per-stage rotation counts
$K_{E},K_{Q},K_{F}$, the per-stage tolerances
$\delta_{E},\delta_{Q},\delta_{F}$, the arithmetic word width $w_{F}$, and
the QROAM word widths $R_{E},R_{Q}$ defined in the preceding subsections.
The inverse transform $U_{\mathrm{Schur}}^{-1}$ employed in the
state-preparation pipeline has identical TE cost and peak qubit footprint,
since each stage is self-inverse in resource.

Since the three stages execute sequentially and all ancillas are returned
to $\ket{0}$ by construction, transient workspace is shared across stages.
The peak qubit footprint of the full transform is therefore the maximum of
the three stage peaks:
\begin{equation}
  \boxed{\;
  Q_{\mathrm{Schur}}^{\mathrm{Krovi}}
  \;=\;
  \max\!\Bigl(Q_{P}^{\mathrm{peak}},\;Q_{Q}^{\mathrm{peak}},\;Q_{V}^{\mathrm{peak}}\Bigr),
  \;}
  \label{eq:Q_Schur_Krovi}
\end{equation}
with the constituent stage footprints given by
Eqs.~\eqref{eq:Q_P_boxed}, \eqref{eq:Q_Q_boxed}, and \eqref{eq:Q_V_boxed},
each of which already accounts for the persistent registers carried from
the preceding stages. For the parameter regimes of interest
($d\gtrsim N\sim 10$--$50$) the peak is dominated by $Q_{V}^{\mathrm{peak}}$
through the compressed GT register $\ket{\widetilde{M}}$.

\paragraph{Consistency with the asymptotic bound.}
Inserting the exact rotation counts $K_{E}=N(N+1)(N+2)/3$,
$K_{Q}=\Theta(N^{3})$ from
Eqs.~\eqref{eq:KS_RK_krovi}--\eqref{eq:KS_RA_krovi}, and $K_{F}=N^{3}$ into
Eq.~\eqref{eq:TE_Schur_Krovi}, both dominant stages sit at
$\widetilde{\mathcal{O}}(N^{4})$: Stage~2 through the $\Theta(N^{3})$
rotations, each inflated by the multi-control width
$k_{Q}=\Theta(Nn_{N})$ of Eq.~\eqref{eq:kQ_width}, and Stage~3 through the
$N^{3}$ rotations, each carrying $\Theta(N)$ arithmetic operations at
polylogarithmic word width via Eq.~\eqref{eq:CF_angle}. This reproduces the
overall $\widetilde{\mathcal{O}}(N^{4})$ bound of Theorem~1 of
Ref.~\cite{burchardt2025highdimensionalquantumschurtransforms}, where the
unit-gate count is dominated by the symmetric-group QFT; in our fully
compiled model the coherent provisioning of the $F$-symbol angles raises
Stage~3 from Burchardt's $\widetilde{\mathcal{O}}(N^{3})$
precompiled-angle accounting to $\widetilde{\mathcal{O}}(N^{4})$, and
numerically Stage~3 dominates the TE budget through the square-root
pipeline in Eq.~\eqref{eq:CF_angle}. Strictly, the partition-function-sized
angle table $T_{Q}$ of Stage~2 introduces an $e^{\mathcal{O}(\sqrt{N})}$
lookup term, so the polynomial scaling holds up to this subexponential
contribution, which remains numerically subdominant throughout the range
studied.

\paragraph{Comparison with the BCH construction.}
The corrected Krovi transform improves on the BCH construction of the
preceding section in the regime $d\geq N$ relevant to first-quantized
simulation primarily by eliminating the local dimension from the dominant
cost: the $\widetilde{\mathcal{O}}(N^{4})$ scaling of
Eq.~\eqref{eq:TE_Schur_Krovi} is independent of $d$ up to logarithmic
factors, whereas BCH scales as $\widetilde{\mathcal{O}}(Nd^{4})$. Both
constructions require on-the-fly reversible arithmetic for their rotation
angles, reduced-Wigner coefficients in BCH and $F$-symbols here, but the
Krovi arithmetic touches at most $2N$ integer factors once per synthesised
rotation, with no $d$ dependence, while the BCH arithmetic evaluates
$s(s-1)$ reduced-Wigner entries at every rank $s\leq d$ of every one of the
$N-1$ Clebsch--Gordan steps. Consequently, the TE advantage of the Krovi
transform over BCH grows rapidly with $d$ at fixed $N$, reaching orders of
magnitude for $d\gg N$, while remaining a more modest constant factor near
$d\approx N$; precise ratios follow directly from the closed forms of this
and the preceding section.

% ================================
\section{Deterministic preparation of polynomial-size Schur-label superpositions}
\label{sec:det-prep}
% ================================

This section develops the algorithm and resource model for the first stage
of our pipeline: the preparation of a normalized superposition of $L$
classically specified Schur-label bitstrings, which is subsequently consumed
by the inverse quantum Schur transform. The algorithm is deterministic; it
involves no postselection, no amplification, and no repetition, and it
succeeds with probability one. Its Toffoli-equivalent cost is linear in $L$,
which is optimal up to the label width, since any preparation circuit must
access all $L$ classical coefficients. The construction rests on a single
structural fact: distinct particle configurations map to distinct Schur
labels, so an ancillary index register used during preparation is, on the
support of the prepared state, a deterministic function of the label
register. A register that is a function of retained registers can always be
uncomputed exactly by recomputing that function from the retained side, and
this is what the final stage of the algorithm does.

\subsection{Setting and register conventions}
\label{subsec:det-setting}

Throughout, the particle statistics sector $\lambda$ and the symmetric-group
copy $\sigma$ are fixed at the start of the computation by Clifford
operations ($X$ gates writing the corresponding basis states), exactly as in
the main text. The branch-dependent Schur data are held in a
\emph{label register} of width $n_{\mathrm{lab}}$ qubits, whose content
depends on the Schur-transform realization that follows:
\begin{align}
n_{\mathrm{lab}}^{\mathrm{BCH}}
&\;=\;
n_{\mu},
\qquad
\text{with $n_{\mu}$ the naive or compressed GT encoding of
Eqs.~\eqref{eq:registers-naive}, \eqref{eq:nmu-comp-again} or \eqref{eq:nmu-balanced}},
\label{eq:nlab-bch}\\[2pt]
n_{\mathrm{lab}}^{\mathrm{KB}}
&\;=\;
\underbrace{N n_{d}}_{\ket{p}}
\;+\;
\underbrace{N n_{N}}_{\ket{\mu}}
\;+\;
\underbrace{\tfrac{N(N+1)}{2}\, n_{N}}_{\ket{\widetilde{M}}},
\label{eq:nlab-kb}
\end{align}
with $n_d=\lceil\log_2 d\rceil$ and $n_N=\lceil\log_2 N\rceil$ as before. In
the Bacon--Chuang--Harrow (BCH) case the branch-dependent datum is the GT
pattern alone. In the Krovi--Burchardt (KB) case the Schur basis state is
encoded across the alphabet map $\ket{p}$, the type vector $\ket{\mu}$, and
the compressed GT pattern $\ket{\widetilde{M}}$
(Table~\ref{tab:krovi_registers}), all three of which vary between
occupation-number configurations; only $\lambda$ and $\sigma$ are
branch-independent. Accordingly, in the KB instantiation the label register
denotes the concatenation $(\ket{p},\ket{\mu},\ket{\widetilde{M}})$, and the
classical preprocessing supplies the corresponding bitstring for each
configuration. We note that
$n_{\mathrm{lab}}^{\mathrm{KB}}=\mathcal{O}(N n_d + N^2 n_N)$ retains only a
logarithmic dependence on the single-particle dimension $d$.

The input to the algorithm is the classical list
$\{(c_i, s_i)\}_{i=1}^{L}$ with $\sum_{i=1}^{L}\lvert c_i\rvert^2=1$ and
$s_i\in\{0,1\}^{n_{\mathrm{lab}}}$ the label bitstring of the $i$-th
configuration. The bijectivity of the Fock-to-Schur correspondence
established in the main text guarantees that distinct configurations map to
distinct labels,
\begin{equation}
s_i \neq s_j \quad \text{for } i\neq j,
\label{eq:injectivity}
\end{equation}
which is the load-bearing hypothesis of the construction. The target is
\begin{equation}
\ket{\psi_{\mathrm{lab}}}
\;=\;
\sum_{i=1}^{L} c_i \,\ket{s_i},
\label{eq:det-target}
\end{equation}
prepared on the label register with all ancillas returned to $\ket{0}$. Let
$\beta=\lceil\log_2 L\rceil$ denote the index width; where convenient we pad
the coefficient list to $\bar L = 2^{\beta}$ entries by appending
zero-amplitude items, which never acquire weight.

\subsection{Overview}
\label{subsec:det-overview}

The algorithm consists of three unitaries acting on an index register of
width $\beta$ and the label register of width $n_{\mathrm{lab}}$:
\begin{equation}
\ket{0^{\beta}}\ket{0^{n_{\mathrm{lab}}}}
\;\xrightarrow{\;V_{\mathrm{amp}}\;}\;
\sum_{i=1}^{L} c_i \ket{i}\ket{0^{n_{\mathrm{lab}}}}
\;\xrightarrow{\;\textsc{WRITE}\;}\;
\sum_{i=1}^{L} c_i \ket{i}\ket{s_i}
\;\xrightarrow{\;\textsc{ERASE}\;}\;
\ket{0^{\beta}}\otimes\sum_{i=1}^{L} c_i \ket{s_i}.
\label{eq:det-pipeline}
\end{equation}
$V_{\mathrm{amp}}$ loads the target amplitudes $c_i$ onto the index
register; because $\sum_i\lvert c_i\rvert^2=1$, this state is normalized as
written. \textsc{WRITE} copies the label bitstring $s_i$ into the label
register conditioned on the index. \textsc{ERASE} then returns the index
register to $\ket{0^\beta}$ deterministically by exploiting
Eq.~\eqref{eq:injectivity}. A high-level statement is given in
Algorithm~\ref{alg:det-prep}.

\begin{algorithm}[H]
\caption{\textsc{Det-Prep}$(\{(c_i,s_i)\}_{i=1}^{L},\,\varepsilon_{\mathrm{prep}})$:
deterministic Schur-label preparation}
\label{alg:det-prep}
\begin{algorithmic}[1]
\Require Classical data $\{(c_i,s_i)\}_{i=1}^{L}$ with
$\sum_i\lvert c_i\rvert^2=1$ and the $s_i\in\{0,1\}^{n_{\mathrm{lab}}}$
pairwise distinct; target accuracy
$\varepsilon_{\mathrm{prep}}\in(0,1/2)$.
\Ensure The state $\sum_i c_i\ket{s_i}$ on the label register, to accuracy
$\varepsilon_{\mathrm{prep}}$, with unit success probability; index register
and all ancillas returned to $\ket{0}$.
\State Classically precompute the binary-decomposition rotation angles of the
distribution $p_i=\lvert c_i\rvert^2$ (one $f$-bit word per internal node of
the depth-$\beta$ binary tree) and, if applicable, the sign bits or $f$-bit
phase words of $\arg(c_i)$.
\State \textbf{($V_{\mathrm{amp}}$, magnitudes)} For $t=1,\ldots,\beta$:
QROAM-lookup the level-$t$ angle keyed on the $(t{-}1)$-bit prefix, rotate
qubit $t$ by the looked-up angle via phase-gradient addition, erase the
lookup.
\State \textbf{($V_{\mathrm{amp}}$, phases)} Apply the branch phases
$e^{i\arg(c_i)}$ by a phase lookup keyed on the full index (Clifford $Z$ in
the real signed case; skip if all $c_i\ge0$).
\State \textbf{(\textsc{WRITE})} QROAM-lookup the $n_{\mathrm{lab}}$-bit word
$s_i$ keyed on the index, copy it into the label register with CNOTs, and
erase the lookup.
\State \textbf{(\textsc{ERASE})} For $i=1,\ldots,L$: set a flag by an
equality test of the label register against the classical constant $s_i$,
apply flag-controlled NOTs writing $i$ into the index register, clear the
flag.
\end{algorithmic}
\end{algorithm}

\subsection{Correctness of the deterministic erasure}
\label{subsec:det-correctness}

Define the classical partial function
$g:\{0,1\}^{n_{\mathrm{lab}}}\to\{0,\ldots,\bar L -1\}$ by $g(s_i)=i$ and
$g(s)=0$ for $s$ outside the table; by Eq.~\eqref{eq:injectivity}, $g$ is
well defined. \textsc{ERASE} is the reversible implementation of
\begin{equation}
\textsc{ERASE}:\ \ket{x}\ket{s}\;\longmapsto\;\ket{x\oplus g(s)}\ket{s},
\label{eq:erase-def}
\end{equation}
which is manifestly unitary (it is a permutation of the computational
basis). Acting on the post-\textsc{WRITE} state,
\begin{equation}
\textsc{ERASE}\left(\sum_{i=1}^{L} c_i \ket{i}\ket{s_i}\right)
\;=\;
\sum_{i=1}^{L} c_i \ket{i\oplus g(s_i)}\ket{s_i}
\;=\;
\sum_{i=1}^{L} c_i \ket{i\oplus i}\ket{s_i}
\;=\;
\ket{0^{\beta}}\otimes\sum_{i=1}^{L} c_i \ket{s_i},
\label{eq:erase-action}
\end{equation}
so the index register disentangles exactly, on every branch simultaneously,
with probability one. Distinctness of the labels is necessary as well as
sufficient: if two branches shared a label, their distinct index values
could not both be mapped to zero by any unitary of the form of
Eq.~\eqref{eq:erase-def}. In our setting Eq.~\eqref{eq:injectivity} holds by
construction, since the equivariant bijection of the main text assigns
distinct Schur labels to distinct occupation-number configurations.

\subsection{Primitive costs and QROAM conventions}
\label{subsec:det-primitives}

Toffoli equivalents (TE) are defined as in the preceding sections,
$\mathrm{TE}=\#\mathrm{Toffoli}+(\text{T-count})/7$, and Clifford gates are
free. We reuse the CDKM adder $A(w)=2w-1$ \cite{Cuccaro2004}, the
clean-ancilla multi-control $C_{\mathrm{Tof}}(k;a)=2k-3$ for $a\ge k-2$
\cite{Barenco1995}, and the flag-controlled adder priced at $A(w)$ per the
convention fixed in Section~\ref{sec:krovi_resource}. Clean-ancilla QROAM
\cite{BerryGidneyMottaMcCleanBabbush2019} with $\ell$ table entries, output
word width $w$, and power-of-two blocking parameter $k$ costs
\begin{equation}
\mathrm{TE}_{\mathrm{look}}(\ell,w;k)
\;=\;
\Big\lceil \frac{\ell}{k}\Big\rceil + w(k-1),
\qquad
Q_{\mathrm{look}}(\ell,w;k)
\;=\;
w(k-1) + \Big\lceil \log_2\!\frac{\ell}{k}\Big\rceil,
\label{eq:det-qroam-look}
\end{equation}
for the compute pass, and
\begin{equation}
\mathrm{TE}_{\mathrm{unlook}}(\ell;k)
\;=\;
\Big\lceil \frac{\ell}{k}\Big\rceil + k,
\qquad
Q_{\mathrm{unlook}}(\ell;k)
\;=\;
k + \Big\lceil \log_2\!\frac{\ell}{k}\Big\rceil,
\label{eq:det-qroam-unlook}
\end{equation}
for the measurement-based erasure pass. Near-optimal blockings are
$k^{\star}(\ell,w)=2^{\mathrm{round}(\frac{1}{2}\log_2(\ell/w))}$ for the
lookup and $\bar k^{\star}(\ell)=2^{\mathrm{round}(\frac{1}{2}\log_2 \ell)}$
for the erasure, clipped to $[1,\ell]$; the plain unary-iteration QROM is
recovered at $k=1$.

Programmable rotations are applied through a phase-gradient resource state
\cite{Gidney2018}. Rotation angles $\theta\in[0,2\pi)$ are stored as $f$-bit
integers $a=\mathrm{round}(2^{f}\theta/2\pi)$; modular addition of the angle
word into the $f$-bit phase-gradient register, controlled on the rotation
target qubit (after a fixed Clifford basis change taking $R_z$ to $R_y$),
implements the rotation by the stored angle at the cost of one
flag-controlled adder, $A(f)$ Toffolis. Two's-complement wraparound handles
angle signs automatically, so no separate sign bit is required in the angle
word. The per-rotation angle-quantization error is at most $\pi 2^{-f}$ in
operator norm. The phase-gradient state of width $f$ is a catalyst: it is
prepared once, reused by every addition, and returned intact. Its
preparation requires $f-3$ synthesized single-qubit rotations (the first
three factors are exactly $Z$, $S$, and $T$), which is the only source of
T gates in this section.

We remark on the choice of amplitude loader. Amplitude-loading circuits
based on coherent alias sampling produce, alongside the index amplitudes,
normalized residual states on their comparison and reference registers that
differ from branch to branch. Such branch-dependent residuals would remain
entangled with the label register at the end of the preparation and corrupt
the output state. We therefore use the garbage-free loader described below,
in which every lookup is paired with its own erasure and no residual
register depends on the branch.

\subsection{Stage (a): garbage-free amplitude loading \texorpdfstring{$V_{\mathrm{amp}}$}{V-amp}}
\label{subsec:det-vamp}

Write $c_i=e^{i\phi_i}\sqrt{p_i}$ with $p_i=\lvert c_i\rvert^2$ and
$\phi_i=\arg(c_i)\in[0,2\pi)$ (branches with $c_i=0$ are deleted from the
table). Magnitudes and phases are handled separately.

\paragraph{Magnitudes.}
The nonnegative-amplitude state $\sum_i\sqrt{p_i}\,\ket{i}$ is prepared by
the standard binary decomposition of the distribution $\{p_i\}$ over the
padded index space of $\bar L=2^\beta$ leaves. For qubit $t=1,\ldots,\beta$
and prefix $x_{1:t-1}\in\{0,1\}^{t-1}$, the classically precomputed angle
$\theta_t(x_{1:t-1})$ is determined by the conditional subtree weights;
subtrees of zero total weight are assigned angle zero and never acquire
amplitude. At level $t$ the circuit performs a QROAM lookup of the $f$-bit
angle word keyed on the $(t{-}1)$-bit prefix (table size $2^{t-1}$), one
programmable rotation of qubit $t$, and the erasure of the lookup. The
level-$t$ cost is
$\mathrm{TE}_{\mathrm{look}}(2^{t-1},f;k_t)+A(f)
+\mathrm{TE}_{\mathrm{unlook}}(2^{t-1};\bar k_t)$, and summing over levels,
\begin{equation}
\mathrm{TE}_{\mathrm{mag}}(L,f)
\;=\;
\sum_{t=1}^{\beta}
\left[
\Big\lceil \frac{2^{t-1}}{k_t}\Big\rceil + f(k_t-1)
\;+\;
\Big\lceil \frac{2^{t-1}}{\bar k_t}\Big\rceil + \bar k_t
\right]
\;+\;
\beta\,A(f),
\label{eq:TE-mag}
\end{equation}
with $k_t=k^{\star}(2^{t-1},f)$ and $\bar k_t=\bar k^{\star}(2^{t-1})$.
With these blockings,
$\mathrm{TE}_{\mathrm{mag}}=\mathcal{O}\big(\sqrt{Lf}+\beta f\big)$.
The exact circuit with exact angles prepares $\sum_i\sqrt{p_i}\ket{i}$ with
no garbage; the only error is angle quantization, bounded below.

\paragraph{Phases.}
Three cases arise. (i) If all $c_i\ge 0$, no phase stage is needed and
$\mathrm{TE}_{\mathrm{ph}}=0$. (ii) If the $c_i$ are real but signed, a
one-bit sign table is looked up keyed on the full index, a Clifford
controlled-$Z$ against a $\ket{-}$ ancilla applies the factor
$(-1)^{\mathrm{sgn}_i}$, and the bit is erased:
\begin{equation}
\mathrm{TE}_{\mathrm{ph}}^{\mathrm{sign}}(L)
\;=\;
\Big\lceil \frac{L}{k_s}\Big\rceil + (k_s-1)
\;+\;
\Big\lceil \frac{L}{\bar k_s}\Big\rceil + \bar k_s .
\label{eq:TE-ph-sign}
\end{equation}
(iii) For general complex coefficients, the $f$-bit phase word
$\mathrm{round}(2^{f}\phi_i/2\pi)$ is looked up keyed on the full index,
added (unconditionally, one plain adder) into the phase-gradient register to
apply $e^{i\phi_i}$ by phase kickback, and erased:
\begin{equation}
\mathrm{TE}_{\mathrm{ph}}^{\mathbb{C}}(L,f)
\;=\;
\Big\lceil \frac{L}{k_\phi}\Big\rceil + f(k_\phi-1)
\;+\;
A(f)
\;+\;
\Big\lceil \frac{L}{\bar k_\phi}\Big\rceil + \bar k_\phi .
\label{eq:TE-ph-complex}
\end{equation}
We write $\mathrm{TE}_{\mathrm{ph}}$ generically for whichever case applies.
Because the phase deviation per branch is uniformly bounded and the branch
weights are $\ell_2$-normalized, the phase-quantization error of the full
state does not accumulate with $L$; see the error budget below.

\subsection{Stage (b): label writing \textsc{WRITE}}
\label{subsec:det-write}

\textsc{WRITE} performs a single QROAM lookup keyed on the index, with
output word the label bitstring $s_i$ of width $n_{\mathrm{lab}}$. The
looked-up word is copied into the persistent label register by
$n_{\mathrm{lab}}$ CNOTs (Clifford; the label register is initialized to
$\ket{0^{n_{\mathrm{lab}}}}$, so the copy writes $s_i$ directly), after
which the lookup is erased:
\begin{equation}
\mathrm{TE}_{\textsc{WRITE}}(L,n_{\mathrm{lab}})
\;=\;
\Big\lceil \frac{L}{k'}\Big\rceil + n_{\mathrm{lab}}(k'-1)
\;+\;
\Big\lceil \frac{L}{\bar k'}\Big\rceil + \bar k',
\label{eq:TE-write-det}
\end{equation}
with $k'=k^{\star}(L,n_{\mathrm{lab}})$ and $\bar k'=\bar k^{\star}(L)$,
giving
$\mathrm{TE}_{\textsc{WRITE}}=\mathcal{O}\big(\sqrt{L\,n_{\mathrm{lab}}}\big)$.
At $k'=\bar k'=1$ the temporary output register and the erasure pass can be
dispensed with entirely (unary iteration writes into the label register
directly at cost $L$ Toffolis and leaves no junk); the blocked variant
trades a transient $n_{\mathrm{lab}}$-wide output register for the
square-root speedup.

\subsection{Stage (c): deterministic index erasure \textsc{ERASE}}
\label{subsec:det-erase}

\textsc{ERASE} implements Eq.~\eqref{eq:erase-def} as a cascade of $L$
compare-and-write blocks. For each $i=1,\ldots,L$: (i) conjugate the label
register by $X$ gates on the positions where $s_i$ has bit value $0$
(Clifford), and set a flag ancilla by one
$n_{\mathrm{lab}}$-controlled NOT at cost
$C_{\mathrm{Tof}}(n_{\mathrm{lab}};a)=2n_{\mathrm{lab}}-3$; (ii) apply
flag-controlled NOTs onto the index qubits where the binary expansion of $i$
has bit value $1$ (at most $\beta$ CNOTs from the flag, Clifford); (iii)
clear the flag by repeating the multi-control, and undo the $X$
conjugation. By Eq.~\eqref{eq:injectivity} exactly one flag fires on each
branch, and on that branch the controlled NOTs clear the index; all other
blocks act as the identity. The total cost is
\begin{equation}
\mathrm{TE}_{\textsc{ERASE}}(L,n_{\mathrm{lab}})
\;=\;
2L\,C_{\mathrm{Tof}}(n_{\mathrm{lab}};a)
\;=\;
L\,\big(4\,n_{\mathrm{lab}}-6\big),
\label{eq:TE-erase}
\end{equation}
using the clean-ancilla bank of size $\max(0,n_{\mathrm{lab}}-2)$ shared
across all blocks. A measurement-based variant (measuring the index in the
$X$ basis and repairing the induced branch phases with a one-bit phase
lookup addressed through the same equality-test network) removes the
flag-clear pass and saves close to a factor of two, but does not change the
leading order; we quote the fully unitary form in Eq.~\eqref{eq:TE-erase}.
After \textsc{ERASE} the index register is restored to $\ket{0^\beta}$
exactly and is available for reuse by the inverse Schur transform.

\subsection{Error budget}
\label{subsec:det-error}

Stages (b) and (c) are exact permutations of the computational basis and
contribute no error. The total budget is split evenly between angle
quantization and catalyst synthesis,
\begin{equation}
\varepsilon_{\mathrm{prep}}
\;=\;
\varepsilon_{\mathrm{ang}}+\varepsilon_{\mathrm{cat}},
\qquad
\varepsilon_{\mathrm{ang}}=\varepsilon_{\mathrm{cat}}
=\tfrac{1}{2}\varepsilon_{\mathrm{prep}}.
\label{eq:det-eps-split}
\end{equation}
The circuit executes $K_{\mathrm{rot}}=\beta+1$ programmable-rotation layers
($\beta$ magnitude layers and one phase layer; $K_{\mathrm{rot}}=\beta$ in
the real cases). Each layer is a direct sum over prefixes of single-qubit
rotations, so its operator-norm deviation from the ideal layer equals the
worst per-angle quantization error, at most $\pi 2^{-f}$. Additive
accumulation over the layers gives total error at most
$(\beta+1)\pi 2^{-f}\le\varepsilon_{\mathrm{ang}}$, i.e.
\begin{equation}
f
\;=\;
\left\lceil
\log_2\!\frac{\pi(\beta+1)}{\varepsilon_{\mathrm{ang}}}
\right\rceil .
\label{eq:det-f}
\end{equation}
Note that $f$ depends on $L$ only through $\beta=\lceil\log_2 L\rceil$: the
per-branch phase and angle deviations are uniform, and the branch weights
are $\ell_2$-normalized, so no union bound over the $L$ branches is
incurred. The single width-$f$ phase-gradient catalyst is prepared once to
tolerance $\varepsilon_{\mathrm{cat}}/(\beta+1)$ per use, i.e.\ with $f-3$
rotations synthesized at tolerance
$\delta_{\mathrm{cat}}=\varepsilon_{\mathrm{cat}}/\big[(\beta+1)(f-3)\big]$
using the direct construction of Ref.~\cite{BocharovRoettelerSvore2015},
contributing the one-time T cost
\begin{equation}
\mathrm{TE}_{\mathrm{cat}}(f)
\;=\;
\frac{(f-3)\,T_{\mathrm{dir}}(\delta_{\mathrm{cat}})}{7}.
\label{eq:TE-cat}
\end{equation}
Apart from Eq.~\eqref{eq:TE-cat}, every operation in this section is
Toffoli or Clifford.

\subsection{Total Toffoli-equivalent cost and qubit footprint}
\label{subsec:det-totals}

Assembling
Eqs.~\eqref{eq:TE-mag}, \eqref{eq:TE-ph-sign} or
\eqref{eq:TE-ph-complex}, \eqref{eq:TE-write-det}, \eqref{eq:TE-erase} and
\eqref{eq:TE-cat},
\begin{equation}
\boxed{\;
\mathrm{TE}_{\mathrm{prep}}(L,n_{\mathrm{lab}},\varepsilon_{\mathrm{prep}})
\;=\;
\mathrm{TE}_{\mathrm{mag}}(L,f)
+\mathrm{TE}_{\mathrm{ph}}
+\mathrm{TE}_{\mathrm{cat}}(f)
+\mathrm{TE}_{\textsc{WRITE}}(L,n_{\mathrm{lab}})
+\mathrm{TE}_{\textsc{ERASE}}(L,n_{\mathrm{lab}}),
\;}
\label{eq:TE-prep-boxed}
\end{equation}
with $f$ given by Eq.~\eqref{eq:det-f}. With near-optimal blockings the
leading behaviour is
\begin{equation}
\mathrm{TE}_{\mathrm{prep}}
\;=\;
L\big(4\,n_{\mathrm{lab}}-6\big)
\;+\;
\mathcal{O}\!\Big(\sqrt{L\,n_{\mathrm{lab}}}\Big)
\;+\;
\mathcal{O}\!\Big(\sqrt{L\,f}+\beta f\Big)
\;=\;
\Theta\big(L\,n_{\mathrm{lab}}\big),
\label{eq:TE-prep-asymptotic}
\end{equation}
dominated by the erasure stage. The linear dependence on $L$ is optimal up
to the label width, since the circuit must access all $L$ classical
coefficients, and the preparation carries no failure probability and no
repetition structure.

For the qubit accounting we separate persistent registers from transient
workspace. The persistent registers are the fixed $\ket{\lambda}$ and
$\ket{\sigma}$ registers (widths as in Section~\ref{subsec:registers} or
Table~\ref{tab:krovi_registers}, according to the chosen Schur-transform
realization) and the label register of width $n_{\mathrm{lab}}$. The
transient workspace comprises the index register ($\beta$ qubits, restored
to $\ket{0^\beta}$ by \textsc{ERASE}), the flag qubit, the clean-ancilla
multi-control bank, the angle output word and the phase-gradient catalyst
($f$ qubits each), the transient \textsc{WRITE} output word
($n_{\mathrm{lab}}$ qubits, absent at $k'=1$), and the largest QROAM
workspace over all lookups,
\begin{equation}
Q_{\mathrm{QROAM}}^{\max}
\;=\;
\max\Big\{
\max_{1\le t\le\beta}\big[f(k_t-1)+\lceil\log_2(2^{t-1}/k_t)\rceil\big],\;\;
f(k_\phi-1)+\lceil\log_2(L/k_\phi)\rceil,\;\;
n_{\mathrm{lab}}(k'-1)+\lceil\log_2(L/k')\rceil
\Big\},
\label{eq:Q-qroam-max}
\end{equation}
where the erasure-pass footprints $Q_{\mathrm{unlook}}$ of
Eq.~\eqref{eq:det-qroam-unlook} are dominated by the corresponding lookup
entries and are omitted from the maximum. The peak footprint of the
preparation stage is therefore
\begin{equation}
\boxed{\;
Q_{\mathrm{prep}}
\;=\;
n_{\lambda}+n_{\sigma}+n_{\mathrm{lab}}
\;+\;\beta
\;+\;1
\;+\;\max(0,\,n_{\mathrm{lab}}-2)
\;+\;2f
\;+\;n_{\mathrm{lab}}\,[k'>1]
\;+\;Q_{\mathrm{QROAM}}^{\max},
\;}
\label{eq:Q-prep-boxed}
\end{equation}
where $[k'>1]$ is the indicator of blocked \textsc{WRITE}. All transient
registers are returned to $\ket{0}$ by construction, so the workspace is
fully reusable by the subsequent inverse Schur transform, and the
end-to-end peak is the maximum of Eq.~\eqref{eq:Q-prep-boxed} and the
inverse-Schur footprint of the chosen realization.

\paragraph{Instantiation for the two Schur-transform realizations.}
For the BCH realization, $n_{\mathrm{lab}}=n_{\mu}$ from
Eq.~\eqref{eq:nlab-bch} and the leading cost is
$L(4n_{\mu}-6)$ with the naive or compressed GT encoding of
Section~\ref{subsec:registers}. For the KB realization,
$n_{\mathrm{lab}}$ is given by Eq.~\eqref{eq:nlab-kb}, so that
\begin{equation}
\mathrm{TE}_{\textsc{ERASE}}^{\mathrm{KB}}
\;=\;
L\left[
4\Big(N n_{d} + N n_{N} + \tfrac{N(N+1)}{2}\,n_{N}\Big) - 6
\right]
\;=\;
\mathcal{O}\!\big(L\,(N\log d + N^{2}\log N)\big),
\label{eq:TE-erase-kb}
\end{equation}
and the preparation stage inherits the polylogarithmic $d$ dependence of the
KB pipeline. In both instantiations the equality tests of \textsc{ERASE}
may in principle target any sub-word of the label on which the
configuration map is already injective, which can only reduce
Eq.~\eqref{eq:TE-erase}; we quote the conservative full-width cost
throughout.

% ================================
\section{End-to-end cost of universal state preparation in first quantization}
\label{sec:total-cost}
% ================================

We summarize the total Toffoli-equivalent (TE) cost and peak qubit footprint
of the full state-preparation pipeline:
\[
\text{(0) classical preprocessing}
\quad\longrightarrow\quad
\text{(i) deterministic Schur-label preparation (Section~\ref{sec:det-prep})}
\quad\longrightarrow\quad
\text{(ii) inverse Schur transform}.
\]
Stage (ii) may be realized with either the Bacon--Chuang--Harrow (BCH)
construction or the corrected Krovi--Burchardt (KB) construction, and the
choice fixes both the label-register encoding used in stage (i) and the
inverse-transform cost model. We treat the two variants separately
throughout, since they occupy different points of the space-time tradeoff
landscape and carry different asymptotic complexities. In both variants the
pipeline is deterministic: no stage involves postselection, amplification,
or repetition, and the resource counts below are incurred exactly once per
prepared state.

\subsection{Stage (0): classical preprocessing}
\label{subsec:classical-preproc}

The conversion of the target state from its physical specification to the
bitstring data consumed by stage (i) is entirely classical and is performed
once, before any quantum resources are engaged. Given the $L$
occupation-number configurations $\ket{\mathbf{n}}_i$ and their coefficients
$c_i$ (obtained, for example, from a truncated configuration-interaction or
perturbative calculation, so that $L$ scales polynomially with $N$ and $d$
by assumption), the preprocessing comprises, for each configuration:
(a) the Fock-to-Dynkin conversion $z_j = n_j - n_{j+1}$ of the main text,
at $\mathcal{O}(d)$ integer operations;
(b) the recovery of standard weights and the construction of one canonical
Gelfand--Tsetlin pattern of the statistics-appropriate shape $\lambda$ via
\textsc{DynkinToGT} (Algorithm~\ref{alg:DynkinToGT} of
Section~\ref{sec:dynkin-to-gt}), at $\mathcal{O}(d^{2})$ integer operations;
and (c) the assembly and dense binary encoding of the label bitstring
$s_i$ in the convention of the chosen variant
(Eqs.~\eqref{eq:nlab-bch} or \eqref{eq:nlab-kb}), at cost linear in the
label width. Auxiliary classical quantities, such as the partition count
$p_d(N)$ fixing the compressed register widths, are computed by the dynamic
program of Algorithm~\ref{alg:pdN} at $\mathcal{O}(Nd)$ operations. The
binary-decomposition rotation angles of stage (i) are computed from the
$p_i=\lvert c_i\rvert^2$ in $\mathcal{O}(L)$ arithmetic operations on the
binary tree. The total classical cost is therefore
\begin{equation}
C_{\mathrm{classical}}
\;=\;
\mathcal{O}\!\big(L\,d^{2} + L\,n_{\mathrm{lab}} + Nd\big),
\label{eq:classical-cost}
\end{equation}
polynomial in all parameters, and negligible in practice against the
classical electronic-structure calculation that supplies the configurations
themselves. No quantum resources are consumed by this stage, and none of
the quantum cost formulas below depend on it.

\subsection{Variant 1: BCH pipeline}
\label{subsec:total-bch}

In the BCH variant the label register holds the GT pattern alone,
$n_{\mathrm{lab}} = n_{\mu}$, using either the naive encoding of
Eq.~\eqref{eq:registers-naive} or the compressed encodings of
Eqs.~\eqref{eq:nmu-comp-again} and \eqref{eq:nmu-balanced}
(Section~\ref{subsec:registers}); the compressed encoding minimizes both
the erasure cost of stage (i) and the multi-control widths inside the
inverse transform, and is our default. The error budget is split evenly
between the two quantum stages,
\begin{equation}
\varepsilon
\;=\;
\varepsilon_{\mathrm{prep}} + \varepsilon_{\mathrm{Schur}},
\qquad
\varepsilon_{\mathrm{prep}}
=\varepsilon_{\mathrm{Schur}}
=\tfrac{1}{2}\varepsilon.
\label{eq:eps-split-total}
\end{equation}

\paragraph*{Toffoli-equivalent cost.}
Stage (i) is costed by Eq.~\eqref{eq:TE-prep-boxed} at
$n_{\mathrm{lab}}=n_{\mu}$. Stage (ii) applies the compiled gate sequences
and online arithmetic of the forward transform in reverse at identical
cost, so Eq.~\eqref{eq:CG-and-Schur-np} gives the inverse-transform bound.
The end-to-end total is
\begin{equation}
\boxed{\;
\mathrm{TE}_{\mathrm{total}}^{\mathrm{(BCH)}}
\;=\;
\mathrm{TE}_{\mathrm{prep}}\big(L,\,n_{\mu},\,\varepsilon_{\mathrm{prep}}\big)
\;+\;
(N{-}1)\sum_{s=2}^{d}\mathrm{TE}_s(d,N,\varepsilon_{\mathrm{Schur}}),
\;}
\label{eq:TE-total-bch}
\end{equation}
with $\mathrm{TE}_s$ from Eq.~\eqref{eq:level-total-np}.

\paragraph*{Peak qubit footprint.}
The two stages run sequentially and all transient workspace is returned to
$\ket{0}$ at stage boundaries (Sections~\ref{subsec:det-totals} and
\ref{subsec:total-qubits}), so clean-ancilla banks are shared and the
end-to-end peak is the stage maximum:
\begin{equation}
\boxed{\;
Q_{\mathrm{peak}}^{\mathrm{(BCH)}}
\;=\;
\max\!\Big\{\,
Q_{\mathrm{prep}}\big(L,\,n_{\mu},\,\varepsilon_{\mathrm{prep}}\big),\;\;
Q_{\mathrm{total}}\big(d,N,\varepsilon_{\mathrm{Schur}};\,a_{\mathrm{mcx}}^{\mathrm{prov}}\big)
\Big\},
\;}
\label{eq:Q-peak-bch}
\end{equation}
with $Q_{\mathrm{prep}}$ from Eq.~\eqref{eq:Q-prep-boxed} and
$Q_{\mathrm{total}}$ from Eq.~\eqref{eq:Qtotal-final}. In practice the
inverse-Schur term dominates for all but very large $L$, since the
preparation workspace is logarithmic in $L$ apart from the tunable QROAM
blocking bank.

\paragraph*{Asymptotic complexity.}
Stage (i) contributes
$\Theta(L\,n_{\mu})=\mathcal{O}(L\,d\log N)$ in the naive GT encoding
(Eq.~\eqref{eq:registers-naive}), improvable by the compressed encoding of
Eq.~\eqref{eq:nmu-comp-again}. Stage (ii) contributes
$\widetilde{\mathcal{O}}(N d^{4})$: the rank-$s$ level carries
$\binom{s}{2}(2n_s-1)=\mathcal{O}(s^{2}\log s)$ compiled rotations and
$s(s-1)$ online reduced-Wigner evaluations, each at
$\mathcal{O}(w^{2}\log w)$ arithmetic for the word size
$w=\mathcal{O}(\log(NdL/\varepsilon))$ of Eq.~\eqref{eq:w-of-s}, summed
over $s\le d$ and multiplied by the $N-1$ Clebsch--Gordan steps; the tilde
absorbs the polylogarithmic factors in $N$, $d$, and
$\varepsilon^{-1}$ from word sizes, multi-control widths, and rotation
synthesis. Altogether,
\begin{equation}
\mathrm{TE}_{\mathrm{total}}^{\mathrm{(BCH)}}
\;=\;
\widetilde{\mathcal{O}}\!\big(L\,d \;+\; N d^{4}\big)
\;\subset\;
\mathrm{poly}\big(L,\,N,\,d,\,\log\varepsilon^{-1}\big),
\label{eq:TE-total-bch-asymptotic}
\end{equation}
with the polynomial degrees being one in $L$, one in $N$, four in $d$, and
the $\varepsilon$ dependence entering only polylogarithmically. The peak
qubit footprint is
\begin{equation}
Q_{\mathrm{peak}}^{\mathrm{(BCH)}}
\;=\;
\mathcal{O}\!\big(N\log d \;+\; d^{2}\log N \;+\; \log L \;+\; \log\varepsilon^{-1}\big),
\label{eq:Q-total-bch-asymptotic}
\end{equation}
dominated by the naive GT register $n_\mu^{\mathrm{naive}}=\mathcal{O}(d^{2}\log N)$
of Eq.~\eqref{eq:registers-naive}; the compressed encoding of
Eq.~\eqref{eq:nmu-comp-again} reduces this term to the logarithm of the
maximal Weyl dimension, at the price of costlier arithmetic inside the
transform. The optional blocked lookups of stage (i) add a tunable
$\mathcal{O}(\sqrt{L\,n_{\mu}})$ workspace that accelerates the
$\mathcal{O}(\sqrt{L\,n_{\mu}})$ lookup terms; at unit blocking this
workspace vanishes with no change to the leading $L$ term, which is set by
the erasure cascade.

\subsection{Variant 2: KB pipeline}
\label{subsec:total-kb}

In the KB variant the label register holds the concatenated triple
$(\ket{p},\ket{\mu},\ket{\widetilde{M}})$ in the $N$-alphabet encoding of
Table~\ref{tab:krovi_registers}, so that
$n_{\mathrm{lab}}=n_{\mathrm{lab}}^{\mathrm{KB}}
= Nn_{d}+Nn_{N}+\tfrac{N(N+1)}{2}n_{N}$ per Eq.~\eqref{eq:nlab-kb}, and
stage (ii) is the inverse of the corrected Krovi transform of
Section~\ref{sec:krovi_resource}. The error split of
Eq.~\eqref{eq:eps-split-total} applies unchanged, with
$\varepsilon_{\mathrm{Schur}}$ further partitioned across the three KB
stages per Eq.~\eqref{eq:krovi_eps_split}.

\paragraph*{Toffoli-equivalent cost.}
Stage (i) is costed by Eq.~\eqref{eq:TE-prep-boxed} at
$n_{\mathrm{lab}}=n_{\mathrm{lab}}^{\mathrm{KB}}$, with the dominant erasure
term given explicitly by Eq.~\eqref{eq:TE-erase-kb}. Stage (ii) is costed
by Eq.~\eqref{eq:TE_Schur_Krovi}, identical for forward and inverse
transforms since each KB stage is self-inverse in resource. The end-to-end
total is
\begin{equation}
\boxed{\;
\mathrm{TE}_{\mathrm{total}}^{\mathrm{(KB)}}
\;=\;
\mathrm{TE}_{\mathrm{prep}}\big(L,\,n_{\mathrm{lab}}^{\mathrm{KB}},\,\varepsilon_{\mathrm{prep}}\big)
\;+\;
\mathrm{TE}_{P}
+\mathrm{TE}_{Q}
+\mathrm{TE}_{V},
\;}
\label{eq:TE-total-kb}
\end{equation}
with the stage costs from Eqs.~\eqref{eq:TE_P_boxed},
\eqref{eq:TE_Q_boxed}, and \eqref{eq:TE_V_boxed}, evaluated at
$\varepsilon_{\mathrm{Schur}}$.

\paragraph*{Peak qubit footprint.}
As in the BCH variant, workspace is shared across the sequential stages,
and the KB stage peaks of Eqs.~\eqref{eq:Q_P_boxed}, \eqref{eq:Q_Q_boxed},
and \eqref{eq:Q_V_boxed} already carry the persistent registers, which
include the full label triple. The end-to-end peak is
\begin{equation}
\boxed{\;
Q_{\mathrm{peak}}^{\mathrm{(KB)}}
\;=\;
\max\!\Big\{\,
Q_{\mathrm{prep}}\big(L,\,n_{\mathrm{lab}}^{\mathrm{KB}},\,\varepsilon_{\mathrm{prep}}\big),\;\;
Q_{P}^{\mathrm{peak}},\;\;
Q_{Q}^{\mathrm{peak}},\;\;
Q_{V}^{\mathrm{peak}}
\Big\},
\;}
\label{eq:Q-peak-kb}
\end{equation}
with $Q_{\mathrm{prep}}$ from Eq.~\eqref{eq:Q-prep-boxed}. For the
parameter regimes of interest the maximum is attained by
$Q_{V}^{\mathrm{peak}}$ through the compressed GT register
$\ket{\widetilde{M}}$, exactly as in the standalone analysis of
Section~\ref{subsec:krovi_totals}, since the preparation stage holds the
same persistent registers plus only logarithmic and tunable workspace.

\paragraph*{Asymptotic complexity.}
Stage (i) contributes
$\Theta(L\,n_{\mathrm{lab}}^{\mathrm{KB}})
=\mathcal{O}\big(L\,(N\log d + N^{2}\log N)\big)$ per
Eq.~\eqref{eq:TE-erase-kb}. Stage (ii) contributes
$\widetilde{\mathcal{O}}(N^{4})$, with the only $d$ dependence entering
linearly in $n_d=\lceil\log_2 d\rceil$ through the comparator arithmetic of
the preprocessing isometry (Eqs.~\eqref{eq:TE_ABCD_total} and
\eqref{eq:TE_H_total}, contributing $\mathcal{O}(N^{2}\log d)$), and with
the caveat, inherited from Section~\ref{subsec:krovi_totals}, that the
Stage-2 angle table introduces an $e^{\mathcal{O}(\sqrt{N})}$ lookup term
that is numerically subdominant throughout the range $N\lesssim 50$ studied
here and is removable in principle by online evaluation of the angles.
Altogether,
\begin{equation}
\mathrm{TE}_{\mathrm{total}}^{\mathrm{(KB)}}
\;=\;
\widetilde{\mathcal{O}}\!\big(L\,N^{2} \;+\; L\,N\log d \;+\; N^{4} \;+\; N^{2}\log d\big)
\;\subset\;
\mathrm{poly}\big(L,\,N,\,\log d,\,\log\varepsilon^{-1}\big),
\label{eq:TE-total-kb-asymptotic}
\end{equation}
with polynomial degrees one in $L$, four in $N$, one in $\log d$, and
polylogarithmic dependence on $\varepsilon^{-1}$. The peak qubit footprint
is
\begin{equation}
Q_{\mathrm{peak}}^{\mathrm{(KB)}}
\;=\;
\mathcal{O}\!\big(N^{2}\log N \;+\; N\log d \;+\; \log L \;+\; \log\varepsilon^{-1}\big),
\label{eq:Q-total-kb-asymptotic}
\end{equation}
quadratic in $N$ through $\ket{\widetilde{M}}$, linear in $\log d$ through
$\ket{p}$, and logarithmic in $L$ and $\varepsilon^{-1}$; the same remark
about the optional $\mathcal{O}(\sqrt{L\,n_{\mathrm{lab}}})$ blocked-lookup
workspace applies.

\subsection{Comparison of the two variants}
\label{subsec:variant-comparison}

The two variants realize a genuine space-time tradeoff governed by
$(N,d)$. In TE, the BCH variant scales as
$\widetilde{\mathcal{O}}(Ld + Nd^{4})$ against the KB variant's
$\widetilde{\mathcal{O}}(LN^{2} + N^{4})$ up to logarithmic $d$ factors, so
the KB variant is decisively preferable whenever $d\gg N$, which is the
regime motivating first quantization in the first place; near $d\approx N$
the advantage reduces to a modest constant factor. In qubits, the ordering
reverses in $N$: the KB footprint grows as $N^{2}\log N$ through the
compressed GT register, while the BCH footprint (with compressed GT
encoding) grows only mildly with $N$ but carries the $d$-dependent register
and arithmetic widths. The preparation stage of Section~\ref{sec:det-prep}
is common to both variants up to the label width, is linear in $L$ in
either case, and never carries a success-probability or repetition
overhead; consequently, for moderate $L$ the end-to-end budget is set by
the inverse Schur transform under either realization, and the crossover
beyond which stage (i) dominates occurs at
$L=\widetilde{\Theta}(Nd^{4}/n_{\mu})$ for BCH and
$L=\widetilde{\Theta}(N^{2}/\log N)$ for KB.

\end{document}